\documentclass{article}

\usepackage[utf8]{inputenc}
\usepackage[margin=1.25in]{geometry}
\usepackage{authblk}
\usepackage{setspace}
\usepackage{graphicx}
\graphicspath{{./Images/}}
\usepackage{subcaption}
\usepackage{amsmath}
\usepackage{booktabs}
\usepackage{hyperref}
\usepackage{comment}
\usepackage{enumitem}
\usepackage{float}
\usepackage{amsfonts}

\usepackage[switch]{lineno}      
\usepackage[
  style=nejm, 
  citestyle=numeric-comp,
  sorting=nyt
]{biblatex}
\usepackage{appendix}
\usepackage{tabularx}
\usepackage{array}

\title{Fixed Rank co-Kriging: a model for multivariate spatial prediction}

\author{Gaia Caringi}
\author{Piercesare Secchi}

\affil[]{MOX - Department of Mathematics, Politecnico di Milano, Piazza Leonardo da Vinci 32, 20133, Milano, Italy}

\date{}
\begin{document}

\maketitle

\begin{abstract}

This work develops a multivariate extension of the Fixed Rank Kriging (FRK) framework for spatial prediction in settings where multiple spatial processes may provide complementary information. The goal is to preserve the computational efficiency, the ability to operate without assuming stationarity over the domain, and the spatial support flexibility of FRK, while incorporating cross-process dependence. To this end, we employ a multiresolution coregionalization structure for the latent spatial effects, in which spatial basis functions are combined with Gaussian Markov Random Field coefficients. An estimation procedure based on the expectation-maximization algorithm is developed, designed to exploit the multiresolution latent structure. 
Through simulation studies, we examine when the proposed joint modeling is beneficial. We consider cases in which one process is observed more sparsely or is entirely unobserved in a subregion and find that the multivariate formulation is able to borrow information from the more densely observed process, producing coherent and accurate predictions even where direct observations are limited or absent.
Finally, the model is applied to the analysis of PM\textsubscript{10} concentrations in Northern Italy, illustrating its applicability in a real environmental context.

\end{abstract}
\vspace{1em}

\section{Introduction}

Georeferenced data are becoming increasingly prevalent. Modern satellites, sensor networks, and monitoring systems continuously produce large amounts of spatially referenced information. These data are often multivariate, providing measurements for several variables at each spatial location.  Therefore, efficiently handling and modeling such large multivariate spatial datasets has become a fundamental challenge in modern statistics.

One of the best-known approaches to make predictions on spatial data is \textit{kriging} (\cite{Cressie1993}). Despite its theoretical optimality, kriging is computationally demanding, as it requires the inversion of the covariance matrix, an operation whose computational cost grows cubically with the sample size, making classical kriging prohibitive for large spatial datasets. 
Fixed Rank Kriging (FRK), a spatial model developed by Cressie et al. in  
\cite{CressieJohannesson2008}, addresses this problem through a low-rank representation of the spatial process. The model decomposes the spatial process into a large-scale trend, modeled via spatial covariates, a low-rank spatial random effect expressed as a combination of basis functions, and a fine-scale residual term, capturing small-scale variability. This formulation imposes a fixed-rank structure on the covariance matrix, allowing its inversion to be reduced to that of an $R$-rank matrix, with \(R \ll n\), where \(R\) is the fixed number of basis functions employed. As a result, the computational cost is substantially reduced, providing improvements for large spatial datasets.
Moreover, through its basis–function expansion, the covariance structure is allowed to vary across the spatial domain, enabling nonstationary spatial modeling. This allows the model to represent spatial processes whose dependence structure changes over the domain, an essential feature for large and heterogeneous spatial regions, where assuming stationarity is often unrealistic.
Finally, FRK also offers flexibility in handling different spatial supports. It naturally accommodates both point and areal data through the discretization of the spatial domain into fine-scale regions known as Basic Areal Units (BAUs) (\cite{Nguyen2012}).
This feature is particularly important in modern environmental and remote sensing applications, where measurements are often collected over spatial footprints rather than individual locations, and predictions are frequently required for aggregated regions. 

While FRK provides an efficient solution for univariate spatial prediction, it does not directly address multivariate spatial processes.
Many real-world problems involve several spatial variables. For example, in environmental and climate science, a satellite measurement can jointly provide information on temperature, pressure, wind speed, and various pollutants. Empirical studies have demonstrated that taking into account the dependence between variables significantly improves the predictive performance in numerous fields, such as environmental sciences (\cite{Goovaerts2000}), urban and socio-economic spatial analysis (\cite{Helbich}), mining and geoscience applications (\cite{Wackernagel1994}, \cite{Dowd2023}). 

When considering multiple variables, the problem of spatial prediction is commonly handled through \textit{co-kriging}. The main challenge posed by co-kriging is to define proper cross-covariance functions, in order to obtain a valid covariance matrix. This requires modeling both the spatial dependence of each variable across locations and the cross-dependence among different variables, making the multivariate extension far from trivial. 

Several modeling strategies have been proposed in the literature to construct valid multivariate covariance structures for co-kriging. A classical and widely used approach is the Linear Model of Coregionalization (LMC) (\cite{ChilesDelfiner2012, Wackernagel1998, GoulardVoltz1992}), which writes each process as a linear combination of independent spatial latent processes. Other approaches define cross-covariance functions through kernel smoothing (\cite{CressieRonald}), where each variable is obtained by smoothing a common latent process with its own kernel function; this provides a flexible and nonparametric way of modeling spatial dependence. Alternatively, parametric models, such as the Matérn class, have been extended to the multivariate setting, introducing parameters that directly control correlation across variables, providing both interpretability and theoretical validity (\cite{Gneiting2010, Apanasovich2012}). A detailed overview of existing approaches for multivariate spatial modeling and co-kriging can be found in \cite{GentonKleiber2015}. 

In this work, we propose a multivariate version of Fixed Rank Kriging that bridges the gap between low-rank spatial models and multivariate spatial covariance construction. This extension requires addressing two main challenges. The first concerns the specification of cross-covariance functions that ensure positive definiteness of the resulting multivariate covariance structure. The second is the preservation of the fixed-rank representation, essential in order to retain the computational advantages of the corresponding univariate model. Additionally, the implemented model maintains the BAU-based spatial discretization, allowing prediction and aggregation across both point- and areal-level supports.
To address these challenges, we build upon the multivariate multiresolution lattice model introduced by Kleiber et al. in \cite{Kleiber2019} to model the covariance structure of the spatial random effects coefficients within the FRK framework. The spatial random effect component is represented through a basis-function expansion, whose functions are grouped by resolution levels and whose coefficients are modeled as a Gaussian Markov Random Field (GMRF). This construction captures cross-process dependence that varies with spatial scale, allowing both large- and fine-scale interactions to be represented within a unified framework. As a result, the model can improve predictive performance while maintaining interpretability. This multiresolution parameterization guarantees a valid multivariate covariance structure and preserves the computational efficiency characteristic of FRK.

Beyond the methodological contribution, we investigate the performance of the proposed framework through simulation studies. These experiments aim to determine under which conditions the joint model yields more accurate predictions than fitting each spatial variable independently, thereby identifying scenarios in which the additional modeling complexity introduced by the multivariate construction is justified. 

The work is organized as follows.
Section~2 introduces the proposed multivariate extension of FRK, referred to as the Fixed Rank co-Kriging model (coFRK). 
Section~3 details the estimation procedure and presents the Expectation–Maximization (EM) algorithm used for inference.
Section~4 reports simulation studies that evaluate the predictive performance of the multivariate model under different scenarios.
Section~5 applies the model to a dataset of PM\textsubscript{10} concentrations in Northern Italy, illustrating its practical applicability in a real-world context.
Finally, Section~6 concludes the work and outlines directions for future research.

\section{The coFRK model}
\label{sec:coFRK}
We consider a continuous multivariate spatial process defined over a domain \( D \subset \mathbb{R}^2 \): 
\( \{
\mathbf{Y}(\mathbf{s}) = \big(Y_1(\mathbf{s}), \ldots, Y_p(\mathbf{s})\big), 
\quad \mathbf{s} \in D\}.
\)
For each component \(j=1,\dots,p\), we adopt the standard FRK decomposition(\cite{CressieJohannesson2008}):
\[
Y_j(\mathbf{s}) 
= \mathbf{f}(\mathbf{s})^\top \boldsymbol{\beta}_j 
+ w_j(\mathbf{s}) 
+ \xi_j(\mathbf{s}),
\]
where \(\mathbf{f}(\mathbf{s})\) is the vector of spatial covariates, which we assume to be shared across all processes (although this assumption can be easily relaxed), and \(\boldsymbol{\beta}_j\) is the corresponding vector of regression coefficients. 
The error component is decomposed into a spatially correlated error
\(w_j(\mathbf{s})\), capturing medium- and large-scale process variation, and spatially uncorrelated noise \(\xi_j(\mathbf{s})\) that represents small-scale variation. 
Cross-dependence among different processes is introduced through the joint covariance structure of the spatially correlated vector
$\mathbf{w}(\mathbf{s}) = (w_1(\mathbf{s}), \ldots, w_p(\mathbf{s}))^\top$, 
while the fine-scale terms $\xi_j(\mathbf{s})$ are assumed to be mutually independent across processes.

\subsection{Multivariate spatial random effects model}

In the proposed multivariate framework, the spatially correlated component \(\mathbf{w}(\mathbf{s})\) captures both the spatial variation and the cross-dependencies across different processes. The purpose of this subsection is to describe the formulation adopted for \(\mathbf{w}(\mathbf{s})\), which is based on the multivariate multiresolution model developed by Kleiber et al. in \cite{Kleiber2019}. We show how this formulation inherently guarantees a valid covariance structure and models the cross-dependencies by exploiting the multiresolution framework, still preserving the computational efficiency of the univariate FRK.

\subsubsection{Basis functions}
We employ the same set of basis functions for all processes. 
This choice, besides being common in multivariate spatial modeling, is not restrictive. 
Indeed, it is widely recognized (e.g., see \cite{CressieJohannesson2008}, \cite{Nychka2002}) that when basis functions are used to approximate the spatial covariance structure, they should be capable of capturing variation across multiple spatial scales. In this way, distinct covariance structures can be represented: smoother processes will be mainly captured by lower-resolution levels, whereas rougher variables will be better described by higher-resolution ones. Moreover, adopting a common set of basis functions facilitates the definition of cross-dependencies, entirely modeled through the covariance structure of the random coefficients \(c_{\ell jr}\).  

The bases are defined as translations and dilations of the same parent function. Typical choices for parent functions include bisquare, Gaussian, exponential or Matérn functions. 
A single basis function is thus represented as
\[
\phi_{\ell r}(\mathbf{s}) = \phi\!\left(\frac{\mathbf{s}-\mathbf{x}_{\ell r}}{s_{\ell r}}\right),
\]
where \(\mathbf{x}_{\ell r}\) denotes the centroid and \(s_{\ell r}\) the scale parameter. 
Within each resolution level, the centers are arranged on a regular grid, with progressively finer spacing at higher resolutions.

\subsubsection{Gaussian Markov Random Field representation}

Following Kleiber et al. (\cite{Kleiber2019}), we propose to model the coefficients \(\mathbf{c}_{\ell} = (\mathbf{c}_{\ell 1}^T,...,\mathbf{c}_{\ell p}^T)^T,\) within each level \(\ell=1,...,L,\) as a multivariate lattice process.
For a given process \(j\in \{1,...,p\}\), 
a univariate lattice model (as first introduced in \cite{Nychka2015}) is employed, whose nodes correspond to the centroids of the basis functions. 
Specifically, the vector of coefficients \(\mathbf{c}_{\ell j} = (c_{\ell j 1},...,c_{\ell j R_{\ell}})^T\) is modeled as a \textit{Gaussian Markov Random Field} (GMRF). 

There are two main reasons for relying on a GMRF formulation. 
First, its Markov property implies conditional independence between non-neighboring nodes, leading to a sparse precision matrix. This sparsity significantly decreases the computational cost of matrix operations, making GMRFs an appealing and scalable strategy. 
Second, GMRFs can be interpreted as discrete approximations of Gaussian random fields with Matérn covariance, as demonstrated in \cite{LindgrenLindstromRue2011}. This link combines the interpretability and flexibility of the Matérn class with the computational efficiency of GMRFs, making the latter a powerful tool for large-scale spatial modeling.

In the proposed framework, GMRF is defined through the specification of the Spatial Autoregressive (SAR) matrix \(\mathbf{B}_\ell\) that encodes the neighborhood structure. Specifically, following \cite{Kleiber2019}, we set the diagonal elements to \((\mathbf{B}_\ell)_{ii} =4+\kappa_{\ell}^2\) and the others to \((\mathbf{B}_\ell)_{ij} =-1\) if \(i\) and \(j\) are neighbors, while \((\mathbf{B}_\ell)_{ij} =0\) if they are not. In this formulation, each vector \(\mathbf{c}_{\ell j}\) is expressed as the linear transformation of a Gaussian white noise vector \(\mathbf{e}_{\ell j}\):
\[
\mathbf{c}_{\ell j} = \mathbf{B}_\ell^{-T} \, \mathbf{e}_{\ell j},
\]
This representation ensures that spatial dependence arises directly from the specified neighborhood structure 
while maintaining computational tractability through the sparse form of $\mathbf{B}_\ell$. 

In order to define the multivariate model, a lattice model for the vector \(\mathbf{c}_{\ell} = (\mathbf{c}_{\ell 1}^T,...,\mathbf{c}_{\ell p}^T)^T \) is employed. Here, the covariance is defined as: \[
\mathrm{Var}(\mathbf{c}_\ell) = \boldsymbol{\Sigma}_\ell \otimes (\mathbf{B}_\ell \mathbf{B}_\ell^{T})^{-1}.
\] Therefore, within each level, the covariance structure is assumed separable. The typical co-kriging problem of being able to model two different kinds of dependence is here addressed as follows: the \(\boldsymbol{\Sigma}_\ell\) term models the cross-dependencies among processes at the same location, while \((\mathbf{B}_\ell \mathbf{B}_\ell^{T})^{-1}\) captures the spatial covariance structure. 
\(\boldsymbol{\Sigma}_\ell\) is a covariance matrix of entries: \((\boldsymbol{\Sigma}_\ell)_{ii}= \sigma_{si}^2\alpha_{\ell i}\), \((\boldsymbol{\Sigma}_\ell)_{ij}= \rho_{\ell i j}\sigma_{si}\sigma_{sj}\sqrt{\alpha_{\ell i}\alpha_{\ell j}}\). Here, \(\sigma_{sj}^2\) controls the variance of the \(j^{th}\) process across levels, while \(\alpha_{\ell j}\) expresses the proportion of variance of process \(j\) explained at level \(\ell\). Finally, \(\rho_{\ell i j}\) measures the correlation between process \(i\) and process \(j\) at that resolution level.
A more detailed discussion about the parametrization of these components is reported in Section 3.

In order to integrate this construction into our model,  we specify the ordering of the coefficient vector, that involves both process and resolution indices. We adopt a process-first ordering to define its covariance structure:
\[
\mathbf{c} = (\mathbf{c}_1^\top, \ldots, \mathbf{c}_p^\top)^\top,
\quad
\mathbf{c}_j = (\mathbf{c}_{j1}^\top, \ldots, \mathbf{c}_{jL}^\top)^\top.
\]
where $\mathbf{c}_{j\ell}$ denotes the vector of random coefficients associated with process $j$ and resolution level $\ell$. 
Under this construction: 
\begin{enumerate}[label=(\roman*), leftmargin=*]
  \item Coefficients associated with different resolution levels are independent;
  \item Within a given level $\ell$, coefficients follow the multivariate GMRF structure described above. 
\end{enumerate}

Consequently, the covariance matrix $\mathbf{K} = \mathrm{Cov}(\mathbf{c})$ is sparse and exhibits a block structure, with \(R_\ell\times R_\ell\) blocks, \(\ell=1,\dots,L.\)
The generic $(i,j,\ell)$ block of $\mathbf{K}$ is defined as
\[
\mathbf{K}_{ij\ell} = \mathrm{Cov}(\mathbf{c}_{j\ell}, \mathbf{c}_{i\ell}).
\]
Within a fixed level $\ell$, the cross–process covariance structure follows from the multivariate GMRF formulation:
\[
\mathbf{K}_{ij\ell}
= (\boldsymbol{\Sigma}_\ell)_{ij} \, (\mathbf{B}_\ell \mathbf{B}_\ell^\top)^{-1}
=
\begin{cases}
\sigma_{si}^{2}\,\alpha_{\ell i}\,(\mathbf{B}_\ell \mathbf{B}_\ell^\top)^{-1}, & \text{if } i = j, \\[6pt]
\rho_{\ell ij}\,\sigma_{si}\sigma_{sj}\sqrt{\alpha_{\ell i}\alpha_{\ell j}}\,(\mathbf{B}_\ell \mathbf{B}_\ell^\top)^{-1}, & \text{if } i \neq j.
\end{cases}
\]
This formulation preserves the fixed-rank structure of \(\mathbf{K}\), whose dimensions scale with the number of basis functions and processes (\(pR\)), rather than with the total number of spatial observations (\(n\)).

\subsubsection{Covariance Validity and Properties}

The multiresolution formulation of Kleiber et al. (\cite{Kleiber2019}) models spatial and cross-process dependence at the level of latent coefficients. Consistent with FRK, spatial dependence is represented through a fixed-rank structure, rather than directly over the observation domain.
This representation allows addressing the challenge of defining a valid covariance matrix in co-kriging. By transferring the problem from an infinite-dimensional spatial domain to a finite-dimensional latent space, the covariance structure is specified for the latent coefficients 
\(\mathbf{c}_{\ell}\), which do not depend explicitly on spatial location.  
This approach transforms the task of modeling continuous cross-covariance functions \(C_{ij}(\mathbf{s}_1, \mathbf{s}_2)\) into the simpler problem of ensuring that the covariance matrix \(\operatorname{Var}(\mathbf{c}_{\ell})=
\boldsymbol{\Sigma}_{\ell} \otimes (\mathbf{B}_{\ell}\mathbf{B}_{\ell}^{\top})^{-1}\) is positive-definite.

We parametrize the cross-process covariance matrix as
\(
\boldsymbol{\Sigma}_\ell 
= \mathbf{D}_\ell \mathbf{R}_\ell \mathbf{D}_\ell,
\)
where \(\mathbf{D}_\ell\) contains level-specific standard deviations and \(\mathbf{R}_\ell\) is a correlation matrix. 
Covariance validity follows from the positive definiteness of \(\mathbf{R}_\ell\) and from standard regularity conditions of the SAR specification. 
Since the Kronecker product of positive definite matrices is positive definite, the resulting multivariate covariance structure is well defined.

Beyond covariance validity, the multiresolution formulation admits a rigorous interpretation based on the spectral representation of Gaussian processes (\cite{GentonKleiber2015}). 
Each resolution level can be viewed as capturing a distinct band of spatial frequencies, from large-scale smooth variation at coarse levels to fine-scale structure at higher resolutions. 
Within this framework, the cross-process correlation parameters act as discrete analogues of spectral coherence, describing how processes co-vary across frequency bands. 
Under suitable parameterizations, this construction asymptotically reproduces the spectral behavior of multivariate Matérn models \cite{Kleiber2019}.

Finally, as in the original FRK framework, the proposed multiresolution representation does not assume spatial stationarity. 
Because the covariance structure is induced by basis functions explicitly defined over the domain, dependence between two locations depends not only on their separation but also on their positions in space. 
This construction therefore accommodates nonstationary spatial behavior.

\subsection{Basic Areal Units and Change of Support}
As in \cite{ZammitMangion_Cressie_2024_FRKintro}, the spatial domain is discretized into fine-scale, non-overlapping regions called Basic Areal Units (BAUs) that define a common spatial support for modeling and prediction. 
This construction enables the integration of data observed over heterogeneous spatial supports by mapping all measurements to the BAU level.
The BAU construction extends naturally to the multivariate setting. 
For completeness, we briefly summarize the main modeling components, which also serve to establish notation. 

\subsubsection{Definition of BAU-level processes}
We consider a partition of the spatial domain \(D\) into a collection of 
Basic Areal Units (BAUs),
\(
D^{\mathrm{BAU}} = \{A_1, \ldots, A_B\},
\)
such that the BAUs are mutually disjoint and together cover the entire domain:
\(
A_i \cap A_j = \emptyset \) for all \( i \neq j\)
and 
\(\bigcup_{i=1}^{B} A_i = D.\)

The true spatial processes $\{Y_1(\mathbf{s}),...,Y_p(\mathbf{s})\}, \mathbf{s} \in D$, are aggregated at BAU level and generate BAU--averaged processes: 
\[
Y_{j b} \;=\; \frac{1}{|A_b|}\int_{A_b} Y_j(\mathbf{s}) \, d\mathbf{s}, 
\qquad b = 1,\ldots,B,\qquad j = 1,\ldots,p.
\]
The model therefore takes the form 
\[
Y_{j b} \;=\; \mathbf{f}_{ b}^\top \boldsymbol{\beta}_j + w_{j b} + \xi_{j b}, 
\qquad b = 1,\ldots,B, \qquad j = 1,\ldots,p,
\]
where all terms are understood to represent averages over the corresponding BAU, defined analogously to the univariate FRK construction. 

The $\mathbf{f}_b$ above denotes the vector of covariates associated with BAU $A_b$. Stacking these vectors across BAUs yields the design matrix
\(
\mathbf{F} = (\mathbf{f}_1^\top, \ldots, \mathbf{f}_B^\top)^\top \in \mathbb{R}^{B \times q}
\), where $q$ is the number of covariates.

For \(j=1,...,p,\) the spatial random effect $w_j$ averaged over a BAU yields
\[
w_{j b} 
\;=\; \frac{1}{|A_b|} \int_{A_b} 
\sum_{\ell=1}^L \sum_{r=1}^{R_\ell} \phi_{\ell r}(\mathbf{s}) \, c_{\ell j r} \, d\mathbf{s}
\;=\; \sum_{\ell=1}^L \sum_{r=1}^{R_\ell} 
\left( \frac{1}{|A_b|} \int_{A_b} \phi_{\ell r}(\mathbf{s}) \, d\mathbf{s} \right) c_{\ell j r},
\; b=1,\ldots,B, \; j = 1,\ldots,p.
\]

Finally, we define the matrix of BAU-averaged basis function evaluations as
\[
\boldsymbol{\Phi} \;=\; 
\left( \frac{1}{|A_b|} \int_{A_b} \boldsymbol{\phi}(\mathbf{s}) \, d\mathbf{s} 
:\; b=1,\ldots,B \right)^\top,
\qquad \Phi \in \mathbb{R}^{B \times R},
\]
where $\boldsymbol{\phi}(\mathbf{s}) = (\phi_{11}(\mathbf{s}),\ldots,\phi_{1R_1}(\mathbf{s}),
\ldots,\phi_{L1}(\mathbf{s}),\ldots,\phi_{L R_L}(\mathbf{s}))^\top$. 
In practice, this averaging is typically approximated by evaluating the basis functions at the centroid of each BAU, 
so that
\[
\boldsymbol{\Phi} \;\approx\; \big( \boldsymbol{\phi}(\mathbf{s}_b) : b=1,\ldots,B \big)^\top,
\]
with $\mathbf{s}_b$ denoting the centroid of BAU $A_b$.
The matrix is organized in blocks according to the resolution levels,
\(
\boldsymbol{\Phi} =
[\boldsymbol{\Phi}_1 \mid \ldots \mid \boldsymbol{\Phi}_L],
\)
where $\boldsymbol{\Phi}_{\ell} \in \mathbb{R}^{B \times R_\ell}$ contains the basis
functions associated with resolution level $\ell$.

\subsubsection{From observations to BAUs}
We denote by \textit{observational domain} the collection of areas, called footprints, over which the processes have been measured: \(D^O=\{B^O_m:m=1,...,N_{obs}\}\). For simplicity of notation, we assume that all processes are observed on the same footprints; this assumption can be easily relaxed.
We denote the observed processes at footprint \(B^O_m\) as \((Z_{m1},...,Z_{mp})\). The goal is to express these measurements as a function of the true process defined at the BAU level in the previous step. The assignment of BAUs to footprints is based on a centroid inclusion rule: a BAU \(A_b\) is considered to belong to footprint \(B^O_m\) if the centroid of \(A_b\) lies within \(B^O_m\). Accordingly, we introduce the aggregation matrix: 
\[
\mathbf{C}_Z = 
\left( 
\frac{\omega_{bm}}
     {\sum_{l=1}^{B} \omega_{lm}} 
\; : \;
b = 1, \ldots, B,\;
m = 1, \ldots, N_{obs}
\right),
\]
where $\omega_{bm}$ is a weight linking BAU $b$ to footprint $m$ (typically $\omega_{bm}=|A_b|$ if $A_b \subseteq B^O_m$ and zero otherwise). 
The observation model for process $j$ at footprint $m$ can be written as
\[
Z_{mj} 
= \frac{1}{\sum_{b=1}^B \omega_{bm}} 
\left( \sum_{b=1}^B\omega_{bm} Y_{j b}
 \right) + \epsilon_{mj},
 \qquad m=1,...,N_{obs},
 \qquad j=1,...,p.
\]
The term $\epsilon_{mj}$ represents a measurement error specific to footprint $B_m^O$ and process $j$. It is modeled as a zero-mean Gaussian variable, independent across footprints and processes.

\subsubsection{From BAUs to prediction supports}
The prediction phase aims to obtain estimates of the spatial process over a set of regions of interest, denoted as \textit{prediction domain}: \(D^P=\{B^P_k: k=1,...,N_{pred}\}\). Each prediction region \(B^P_k\) is the union of one or more BAUs. Analogously to the observation case, the prediction over region \(B^P_k\) is obtained by aggregating the BAU-level processes with proportional weights:
\[
Y_{jk}^{P}
= 
\frac{\sum_{b=1}^{B} \tilde{\omega}_{bk} \, Y_{jb}}
       {\sum_{b=1}^{B} \tilde{\omega}_{bk}},
       \qquad k=1,...,N_{pred},
       \qquad j=1,...,p,
\]
where the weights $\tilde{\omega}_{bk}$ are 
defined in the same way as those used for the observations.
This expression simply states that the prediction for each region is a weighted average of the predictions available at the BAU level. 
Accordingly, we define the aggregation matrix
\[
\mathbf{C}_P = 
\left( 
\frac{\tilde{\omega}_{bk}}
     {\sum_{l=1}^{B} \tilde{\omega}_{lk}} 
\; : \;
b = 1, \ldots, B,\;
k = 1, \ldots, N_{pred}
\right),
\]
which maps predictions from the BAU level to the desired prediction supports.

\subsection{Fixed Rank co-Kriging model}
\label{subsec:coFRKFinalFormulation}
Each process is measured over the observational domain \(D^O\), resulting in $N_{\text{obs}}$ observations collected in the vector:
\[\mathbf{Z}_j = \mathbf{C}_Z\mathbf{Y}_j + \boldsymbol{\epsilon_j}
\qquad j=1,...,p
\]
where \({\mathbf{Y}}_j\) is the vector of the BAU--averaged process for process \(j\) and \({\boldsymbol{\epsilon}}_j\) is the corresponding footprint-level measurement error vector.
Substituting the latent process representation into the observation model yields
\[
{\mathbf{Z}}_j = \mathbf{C}_Z \mathbf{F}\boldsymbol{\beta}_j + \mathbf{C}_Z \boldsymbol{\Phi} \mathbf{c}_j + \mathbf{C}_Z {\boldsymbol{\xi}}_j + {\boldsymbol{\epsilon}}_j,
\].

The fine-scale variation term $\boldsymbol{\xi}_j$ captures small-scale, spatially uncorrelated variability:
\(
\boldsymbol{\xi}_j \sim \mathcal{N}\bigl(0, \, \sigma_{\xi_j}^2 \mathbf{V}_{\xi_j}\bigr),
\)
where $\mathbf{V}_{\xi_j} = \mathrm{diag}(v_{\xi_j1},\ldots,v_{\xi_jB})$ is a known diagonal matrix that accounts for potential heteroscedasticity across BAUs.  
Since this component represents microscale variation not explained by the basis functions, it is assumed independent across BAUs and across processes. Similarly, the measurement error term \( {\boldsymbol{\epsilon}}_j\) is also modeled as \(
{\boldsymbol{\epsilon}}_j \sim \mathcal{N}\bigl(0, \, \sigma_{\epsilon_j}^2 \mathbf{V}_{\epsilon_j}\bigr)
\), where \( \mathbf{V}_{\epsilon_j} = \mathrm{diag}(v_{\epsilon_j1}, \ldots, v_{\epsilon_jN_{\text{obs}}}) \) contains known observation-specific error variances, for example derived from instrument uncertainty or retrieval error estimates. Measurement errors are assumed independent across footprints, processes, and from all latent components $(\mathbf{c}_j, \boldsymbol{\xi}_j)$.

Stacking all $p$ processes jointly, the observational model can be written as
\[
{\mathbf{Z}} = 
\begin{pmatrix}
{\mathbf{Z}}_1 \\
{\mathbf{Z}}_2 \\
\vdots \\
{\mathbf{Z}}_p
\end{pmatrix}
=
\begin{pmatrix}
\mathbf{C}_Z \mathbf{F} \boldsymbol{\beta}_1 \\
\mathbf{C}_Z \mathbf{F} \boldsymbol{\beta}_2 \\
\vdots \\
\mathbf{C}_Z \mathbf{F} \boldsymbol{\beta}_p
\end{pmatrix}
+
\begin{pmatrix}
\mathbf{C}_Z \boldsymbol{\Phi} \mathbf{c}_1 \\
\mathbf{C}_Z \boldsymbol{\Phi} \mathbf{c}_2 \\
\vdots \\
\mathbf{C}_Z \boldsymbol{\Phi} \mathbf{c}_p
\end{pmatrix}
+
\begin{pmatrix}
\mathbf{C}_Z {\boldsymbol{\xi}}_1 \\
\mathbf{C}_Z {\boldsymbol{\xi}}_2 \\
\vdots \\
\mathbf{C}_Z {\boldsymbol{\xi}}_p
\end{pmatrix}
+
\begin{pmatrix}
{\boldsymbol{\epsilon}}_1 \\
{\boldsymbol{\epsilon}}_2 \\
\vdots \\
{\boldsymbol{\epsilon}}_p
\end{pmatrix}.
\]
The expectation of the full stacked observed vector \({\mathbf{Z}}\) is then given by
\[
\mathbb{E}[{\mathbf{Z}}] = 
\begin{pmatrix}
\mathbf{C}_Z \mathbf{F} \boldsymbol{\beta}_1 \\
\vdots \\
\mathbf{C}_Z \mathbf{F}\boldsymbol{\beta}_p
\end{pmatrix}.
\]
Its covariance structure is given by
\[
\mathrm{Var}({\mathbf{Z}}) = (\mathbf{C}_Z \boldsymbol{\Phi}) \mathbf{K} (\mathbf{C}_Z \boldsymbol{\Phi})^\top 
+ \text{block-diag}(\sigma^2_{\xi_1} \mathbf{C}_Z \mathbf{V}_{\xi_1} \mathbf{C}_Z^\top, \ldots, \sigma^2_{\xi_p} \mathbf{C}_Z \mathbf{V}_{\xi_p} \mathbf{C}_Z^\top)
 + \text{block-diag}(\sigma^2_{\epsilon_1} \mathbf{V}_{\epsilon_1}, \ldots, \sigma^2_{\epsilon_p} \mathbf{V}_{\epsilon_p}).
\]
The first term $\mathbf{K} = \mathrm{Cov}(\mathbf{c})$ denotes the covariance matrix of the stacked random coefficients. 
The block-diagonal terms correspond, respectively, to the fine-scale spatial variability and the footprint-level measurement errors.

A summary of the key similarities and differences between the univariate and multivariate FRK models is reported in Table~\ref{tab:FRKcomparison}.
It is important to note that the proposed coFRK formulation can be interpreted as an extension of FRK in the sense that it preserves its main structural components. However, the two models are not equivalent, even in the case $p=1$. 
Indeed, coFRK induces a different parametrization of the covariance structure, specifically of the covariance matrix of the latent coefficients $\mathbf{K}$. While FRK directly specifies $\mathbf{K}$ as the covariance matrix of the basis coefficients, modeled as a zero-mean Gaussian vector, the proposed formulation derives $\mathbf{K}$ through the Gaussian Markov random field representation described above.

Further details on the behavior of coFRK in the univariate setting, including a qualitative comparison of the induced covariance structures and additional simulation results, are provided in Appendix~\ref{AppendixA}.

\begin{table}[h!]
\centering
\renewcommand{\arraystretch}{1.4}
\caption{Comparison between the univariate FRK and the proposed multivariate FRK formulation.}
\label{tab:FRKcomparison}
\begin{tabular}{@{}p{4cm}p{5cm}p{6cm}@{}}
\toprule
\textbf{} & \textbf{Univariate FRK } & \textbf{Multivariate FRK} \\ 
\midrule
Spatial domain 
& \(D = \bigcup_{b=1}^{B} A_b\), discretized into BAUs 
& Same BAU discretization used for all processes\\

Observation model
& \(\mathbf{Z} = \mathbf{C}_Z \mathbf{Y} + \mathbf{C}_Z \boldsymbol{\xi} + \boldsymbol{\epsilon}\) 
& \(\mathbf{Z}_j = \mathbf{C}_Z \mathbf{Y}_j + \mathbf{C}_Z \boldsymbol{\xi}_j + \boldsymbol{\epsilon}_j\); all stacked in \(\mathbf{Z}=(\mathbf{Z}_1^\top,\ldots,\mathbf{Z}_p^\top)^\top\) 
 \\

Spatial random effect 
& \(w(s) = \sum_{r=1}^R \phi_r(s)\, \eta_r\) 
& \(w_j(s) = \sum_{\ell=1}^{L}\sum_{r=1}^{R_\ell} \phi_{\ell r}(s)\, c_{\ell  jr}\) \\

Latent coefficients
& {\raggedright Gaussian vector: $\boldsymbol{\eta}\sim\mathcal{N}(\mathbf{0},\mathbf{K})$ \par}
& Gaussian Random Markov Field per resolution level:
\(
\mathrm{Var}(\mathbf{c}_\ell) = \boldsymbol{\Sigma}_\ell \otimes (\mathbf{B}_\ell \mathbf{B}_\ell^{T})^{-1}.
\) \\

Cross-process dependency
& Not modeled (single process)
& Modeled through \(\boldsymbol{\Sigma}_\ell\), which defines correlations \(\rho_{\ell ij}\) between processes \(i\) and \(j\) at each resolution level \\

Covariance structure 
& \(\mathrm{Var}(\mathbf{Z}) = (\mathbf{C}_Z\boldsymbol{\Phi})\mathbf{K}(\mathbf{C}_Z\boldsymbol{\Phi})^{\top} + \sigma^2_\xi \mathbf{V}_\xi + \sigma^2_\epsilon \mathbf{V}_\epsilon\)
& \(\mathrm{Var}(\mathbf{Z}) = (\mathbf{C}_Z\boldsymbol{\Phi})\mathbf{K}(\mathbf{C}_Z\boldsymbol{\Phi})^{\top} + 
\text{block-diag}(\sigma^2_{\xi_j}\mathbf{V}_{\xi_j}) +\text{block-diag}(\sigma^2_{\epsilon_j}\mathbf{V}_{\epsilon_j})\) \\

Computational complexity
& Inversion cost: \(\mathcal{O}(R^3)\)
& Inversion cost: \(\mathcal{O}((pR)^3)\) \\

\bottomrule
\end{tabular}
\end{table}

\section{Estimation Procedure}
\label{sec:estimation}
\subsection{Model parameters}
\label{sec:model_parameters}
Before detailing the estimation procedure, we summarize the model parameters and their parameterization, following \cite{Kleiber2019}.

\paragraph{Regression parameters.}
For each process \( j = 1, \ldots, p \), the regression coefficients \( \boldsymbol{\beta}_j \) 
capture the large-scale (mean) spatial trend. 
These parameters are treated as unknown and estimated within the EM algorithm.\\

\paragraph{Spatial random effect.}
The latent coefficients vector 
\(\mathbf{c}\) has covariance
\(\mathrm{Var}(\mathbf{c}) = \mathbf{K}\), structured by resolution level, with blocks of the form \(
\mathbf{K}_\ell = \boldsymbol{\Sigma}_\ell \otimes (\mathbf{B}_\ell \mathbf{B}_\ell^{\top})^{-1},
\;\ell = 1, \ldots, L
\).

The matrix \(\boldsymbol{\Sigma}_\ell\) governs the cross-process dependence at level \(\ell\). Its parameters are summarized below.
\begin{itemize}[leftmargin=1.5em]
    \item Process-specific variances \( \sigma_{sj}^2 \) control the marginal variance of the latent coefficients associated with process \(j\).
    \item Level-specific weights \( \alpha_{\ell j} \) measure the relative contribution for the variance of process \(j\) at level \(\ell\). Intuitively, they control the smoothness of the field: larger weights at coarse levels (small \(\ell\)) emphasize broad-scale variability, whereas larger weights at finer levels (large \(\ell\)) produce more localized spatial detail. Kleiber et al. in \cite{Kleiber2019} suggest the following parameterization: \(\alpha_{\ell j} = 2^{-2\nu_j \ell}\). These weights are normalized: \(\sum_{\ell=1}^{L} \alpha_{\ell j} = 1.\)
Here \(\nu_j\) is a smoothness parameter analogous to that in the Matérn covariance model. This choice implies that the proportion of variance decreases geometrically with increasing resolution level, consistent with a Matérn covariance model of smoothness \(\nu_j\). In practice \(\nu_j\) is typically fixed to a plausible value chosen on the basis of prior knowledge (e.g. \(\nu_j = 0.5\) corresponds to an exponential covariance). Otherwise, the implementation procedure described in the next section allows, if desired, to estimate \(\nu_j\) together with the other parameters. In the implementations described in this work, \(\nu_j\) is always fixed at $0.5$ for all processes.

    \item Cross-process correlation \( \rho_{\ell ij} \) describes how strongly the two processes \(i\) and \(j\) are correlated at a given resolution level \(\ell\). Specifically, it captures the extent to which the two processes share common spatial features at that scale. We adopt an exponential parameterization: \(\rho_{\ell i j} = r_0\exp(-r_1(\ell-1))\). This choice imposes a decreasing correlation at increasing scales, where \(r_0\) represents the correlation at the coarsest level, \(r_1\) controls the rate at which correlation decreases with resolution. This formulation reflects the idea that processes tend to be more strongly linked at broad spatial scales, where they are influenced by common large-scale patterns, while their correlation gradually weakens at finer resolutions as more process-specific variability appears.
 \end{itemize}

The matrix \( \mathbf{B}_\ell \) encodes spatial dependence within each level \(\ell\). Its structure depends on a single parameter \(\kappa_{\ell}\), which governs the strength of spatial dependence. It plays a role analogous to the range parameter in a Matérn covariance function: it determines how quickly spatial correlation decays with distance. Larger values of \(\kappa_{\ell}\) correspond to faster decay (shorter-range correlation), while smaller values imply smoother and more spatially extended dependence. To reflect the fact that higher resolution levels represent finer-scale variation we adopt an exponential parameterization: \(\kappa_{\ell}^2 = \exp(\kappa_0 \ell)\), so that \(\kappa_0\) controls the rate at which correlation range changes across resolution levels.  Therefore, the only parameter that needs to be estimated is \(\kappa_0\).

\paragraph{Fine-scale variation.}
The fine-scale error term \( \boldsymbol{\xi}_j \) is modeled as
\[
\boldsymbol{\xi}_j \sim \mathcal{N}(\mathbf{0}, \sigma^2_{\xi_j}\mathbf{V}_{\xi_j}).
\]
The weights \( \mathbf{V}_{\xi_j} = \text{diag}(v_{\xi_{j1}}, \ldots, v_{\xi_{jB}}) \) encode the heteroscedasticity pattern: they modulate how much uncertainty each BAU contributes relative to the others. They can be derived from prior domain knowledge, such as terrain roughness, measurement reliability, or the local variability of residuals (\cite{ZammitMangion_Cressie_2024_FRKintro}, \cite{ZammitMangion_2015_multivariateST}). In the following, unless otherwise specified, all weights are set to one, implying a homoscedastic fine-scale error structure. 
The variance parameter \( \sigma^2_{\xi_j} \) controls the overall magnitude of fine-scale variation and must be estimated.
\\

\paragraph{Measurement errors.}
The observational errors \( \boldsymbol{\epsilon}_j \) are modeled as zero-mean Gaussian vectors with known variances:
\[
\mathrm{Var}(\boldsymbol{\epsilon}_j) = \sigma^2_{\epsilon_j}\mathbf{V}_{\epsilon_j}.
\]
The weights \(\mathbf{V}_{{\epsilon}_j} = \text{diag}(v_{\epsilon_j 1},...,v_{\epsilon_j N_{obs}}) \) are treated as known and fixed and represent the relative reliability of each observation footprint. The global scale parameter \(\sigma_{\epsilon,j}^2\) controls the overall magnitude of measurement noise. Following \cite{ZammitMangion_Cressie_2024_FRKintro}, it is estimated initially (e.g., using variogram-based methods) 
and then treated as fixed throughout the EM estimation procedure described below.

\subsection{EM algorithm}
Due to the presence of latent coefficients, the model parameters are estimated using an Expectation–Maximization (EM) algorithm, following the approaches proposed in \cite{ZammitMangion_Cressie_2024_FRKintro} and \cite{Zhang2007_MLcoregionalization}.

To simplify notation, we rewrite the model in compact form as 
\[
\mathbf{Z}
= \widetilde{\mathbf{F}}\boldsymbol{\beta}
+ \widetilde{\boldsymbol{\Phi}}\mathbf{c}
+ (\mathbf{C}_Z \otimes \mathbf{I}_p)\,\boldsymbol{\xi}
+ \boldsymbol{\varepsilon},
\]
where \[
\widetilde{\mathbf{F}} = (\mathbf{C}_Z \otimes \mathbf{I}_p)\,\mathbf{F}, 
\qquad
\widetilde{\boldsymbol{\Phi}} = (\mathbf{C}_Z \otimes \mathbf{I}_p)\,\boldsymbol{\Phi},
\]
We denote the covariance of the spatially uncorrelated components as:  
\[
\mathbf{D} = (\mathbf{C}_Z \otimes \mathbf{I}_p)\,
\mathrm{Var}(\boldsymbol{\xi})\,
(\mathbf{C}_Z \otimes \mathbf{I}_p)^{\!\top}
+ \mathrm{Var}(\boldsymbol{\varepsilon}).
\]

The latent coefficients satisfy
\(
\mathbf{c} \sim \mathcal{N}(0,\mathbf{Q}^{-1}),
\)
where the precision matrix admits the multiresolution structure
\[
\mathbf{Q}
=
\mathbf{P}^{\top}
\Bigg(
\bigoplus_{\ell=1}^{L}
\big(\boldsymbol{\Sigma}_{\ell}^{-1}
\otimes \mathbf{B}_{\ell}\mathbf{B}_{\ell}^{\top}\big)
\Bigg)
\mathbf{P}.
\]
Here $\mathbf{P}$ is a permutation matrix that reorders the coefficient vector 
$\mathbf{c}$ so that coefficients are grouped by resolution level rather than by process, 
as in the ordering defined above.
This precision-based formulation is computationally advantageous because it avoids explicit inversion of the covariance matrix $\mathbf{K}$ and fully exploits the sparsity induced by the GMRF representation. Moreover, the permutation matrix $\mathbf{P}$ reveals the block-diagonal multiresolution structure of $\mathbf{Q}$, allowing parameters associated with different resolution levels to be updated independently in the M-step.

For efficient inversion and determinant computation, we use the Sherman–Morrison–Woodbury identity and the matrix determinant lemma, exploiting the fixed-rank formulation: 
\[
\boldsymbol{\Sigma}_{\mathbf{Z}}^{-1}
= 
\mathbf{D}^{-1}
-
\mathbf{D}^{-1}\widetilde{\boldsymbol{\Phi}}
\left(
\mathbf{Q}
+
\widetilde{\boldsymbol{\Phi}}^{\top}\mathbf{D}^{-1}\widetilde{\boldsymbol{\Phi}}
\right)^{-1}
\widetilde{\boldsymbol{\Phi}}^{\top}\mathbf{D}^{-1},
\qquad
|\boldsymbol{\Sigma}_{\mathbf{Z}}|
=
|\mathbf{Q}^{-1}|\,
|\mathbf{D}|\,
\big|
\mathbf{Q}
+
\widetilde{\boldsymbol{\Phi}}^{\top}\mathbf{D}^{-1}\widetilde{\boldsymbol{\Phi}}
\big|,
\]
where $\boldsymbol{\Sigma}_{\mathbf{Z}}$ denotes the covariance matrix of $\mathbf{Z}$.

\subsubsection{E--step}
 
Let the set of parameters be denoted by 
\(\boldsymbol{\theta} = (\boldsymbol{\beta},\boldsymbol{\sigma}_s^2, \kappa_0, r_0, r_1, \boldsymbol{\sigma}_\xi^2).\)
The goal of the E--step is to compute the conditional expectation of the complete--data log--likelihood given the observed data $\mathbf{Z}$ and the current parameter estimates $\boldsymbol{\theta}^{(t)}$:
\[
\mathcal{Q}(\boldsymbol{\theta} \mid \boldsymbol{\theta}^{(t)}) = 
\mathbb{E}_{\mathbf{c} \mid \mathbf{Z}, \boldsymbol{\theta}^{(t)}} 
\left[\log p(\mathbf{Z}, \mathbf{c} \mid \boldsymbol{\theta})\right].
\]

To compute the expected value \(\mathcal{Q}(\boldsymbol{\theta} \mid \boldsymbol{\theta}^{(t)})\), we need to define the conditional distribution of $\mathbf{c}$ given $\mathbf{Z}$.  Exploiting some known results on Gaussian conditioning (see, for example, \cite{RasmussenWilliams2006_GPML}):
\[
\mathbf{c} \mid \mathbf{Z}, \boldsymbol{\theta}^{(t)}
\sim \mathcal{N}\!\big(\boldsymbol{\mu}_c^{(t)}, \boldsymbol{\Sigma}_c^{(t)}\big),
\]
where
\[
\boldsymbol{\Sigma}_c^{(t)} 
= \left(\widetilde{\boldsymbol{\Phi}}^\top ( \mathbf{D}^{(t)} )^{-1} 
\widetilde{\boldsymbol{\Phi}} + \mathbf{Q}^{(t)} \right)^{-1},
\qquad
\boldsymbol{\mu}_c^{(t)} 
= \boldsymbol{\Sigma}_c^{(t)}\, \widetilde{\boldsymbol{\Phi}}^\top 
( \mathbf{D}^{(t)} )^{-1}\big(\mathbf{Z} - \widetilde{\mathbf{F}}\,\boldsymbol{\beta}^{(t)}\big).
\]
Substituting these expressions into the definition of $\mathcal{Q}(\boldsymbol{\theta} \mid \boldsymbol{\theta}^{(t)})$ and using standard results for the moments of the multivariate normal distribution, we obtain
\[
\begin{aligned}
\mathcal{Q}(\boldsymbol{\theta} \mid \boldsymbol{\theta}^{(t)}) &=
-\frac{1}{2}\log|\mathbf{D}|
+ \frac{1}{2}\log|\mathbf{Q}|
-\frac{1}{2}(\mathbf{Z}-\widetilde{\mathbf{F}}\boldsymbol{\beta})^{\top}\mathbf{D}^{-1}(\mathbf{Z}-\widetilde{\mathbf{F}}\boldsymbol{\beta})
\\[4pt]
&\quad
+ (\mathbf{Z}-\widetilde{\mathbf{F}}\boldsymbol{\beta})^{\top}\mathbf{D}^{-1}\widetilde{\boldsymbol{\Phi}}\boldsymbol{\mu}_c^{(t)}
- \frac{1}{2}\operatorname{tr}\!\Big(
\widetilde{\boldsymbol{\Phi}}^{\top}\mathbf{D}^{-1}\widetilde{\boldsymbol{\Phi}}\,
(\boldsymbol{\Sigma}_c^{(t)}+\boldsymbol{\mu}_c^{(t)}{\boldsymbol{\mu}_c^{(t)}}^{\top})
\Big)
\\[4pt]
&\quad
- \frac{1}{2}\operatorname{tr}\!\Big(
\mathbf{Q}\,
(\boldsymbol{\Sigma}_c^{(t)}+\boldsymbol{\mu}_c^{(t)}{\boldsymbol{\mu}_c^{(t)}}^{\top})
\Big)
+ \text{const.}
\end{aligned}
\]

\subsubsection{M--step}
In the M--step, the expected complete-data log-likelihood 
$\mathcal{Q}(\boldsymbol{\theta} \mid \boldsymbol{\theta}^{(t)})$
is maximized with respect to each parameter.
The multiresolution structure allows parameters to be updated level by level, improving computational efficiency.

In what follows, we provide the explicit expressions maximized in the M--step to update each parameter. The detailed derivations of these updates are provided in Appendix A.

\begin{itemize}
    \item Maximizing $\mathcal{Q}(\boldsymbol{\theta} \mid \boldsymbol{\theta}^{(t)})$ with respect to $\boldsymbol{\beta}$ yields the updated
generalized least--squares estimator:
\[
\widehat{\boldsymbol{\beta}}^{(t+1)} = 
\big(\widetilde{\mathbf{F}}^{\top}\mathbf{D}^{-1}\widetilde{\mathbf{F}}\big)^{-1}
\widetilde{\mathbf{F}}^{\top}\mathbf{D}^{-1}
\big(\mathbf{Z} - \widetilde{\boldsymbol{\Phi}}\boldsymbol{\mu}_c^{(t)}\big).
\]

\item  Differentiating $\mathcal{Q}(\boldsymbol{\theta} \mid \boldsymbol{\theta}^{(t)})$ with respect to each $\sigma_{\xi_j}^2$
leads to the equation
\[
\frac{\partial \mathcal{Q}}{\partial \sigma_{\xi_j}^2} =
-\tfrac{1}{2}\operatorname{tr}\!\big(\mathbf{D}^{-1}\tfrac{\partial \mathbf{D}}{\partial \sigma_{\xi_j}^2}\big)
+ \tfrac{1}{2}\operatorname{tr}\!\big(\mathbf{D}^{-1}\tfrac{\partial \mathbf{D}}{\partial \sigma_{\xi_j}^2}
\mathbf{D}^{-1}\boldsymbol{\Omega}\big) = 0,
\]
where the matrix $\boldsymbol{\Omega}$ is:
\[
\boldsymbol{\Omega} = 
\widetilde{\boldsymbol{\Phi}}\,\boldsymbol{\Sigma}_c^{(t)}\,\widetilde{\boldsymbol{\Phi}}^{\top}
+ \widetilde{\boldsymbol{\Phi}}\,\boldsymbol{\mu}_c^{(t)}{\boldsymbol{\mu}_c^{(t)}}^{\top}\widetilde{\boldsymbol{\Phi}}^{\top}
- 2\,\widetilde{\boldsymbol{\Phi}}\,\boldsymbol{\mu}_c^{(t)}(\mathbf{Z}-\widetilde{\mathbf{F}}\boldsymbol{\beta})^{\top}
+ (\mathbf{Z}-\widetilde{\mathbf{F}}\boldsymbol{\beta})(\mathbf{Z}-\widetilde{\mathbf{F}}\boldsymbol{\beta})^{\top}.
\]

Each $\sigma_{\xi_j}^2$ is then updated by numerically solving the above equation.

\item The update for \(\boldsymbol{\sigma}_s^2=(\sigma_{s 1}^2,...,\sigma_{s p}^2)^{\top}\) is obtained by maximizing 
\[
\mathcal{Q}_{\sigma}\!\big(\boldsymbol{\sigma}_s^2\big)
=
\sum_{\ell=1}^{L}
\left[
-\,2\,R_\ell \sum_{i=1}^{p}\log d_{\ell i}
\;-\;
\operatorname{tr}\!\Big(
\big(\boldsymbol{\Sigma}_\ell^{-1}\otimes \mathbf{B}_\ell \mathbf{B}_\ell^{\top}\big)\,
\mathbf{P}\, \boldsymbol{S}_c^{(t)}\,\mathbf{P}^{\top}
\Big)
\right]
\;+\;\mathrm{const},
\]
where \(d_{\ell_i} = \sqrt{\alpha_{\ell i}\,\sigma_{si}^2}\) and 
\(\boldsymbol{S}_c^{(t)} =  \boldsymbol{\Sigma}_c^{(t)} + \boldsymbol{\mu}_c^{(t)}{\boldsymbol{\mu}_c^{(t)}}^{\top}\).

\item The parameter \(\kappa_0\) is updated by maximizing

\[
\mathcal{Q}_{\kappa_0}(\kappa_0)
=
\sum_{\ell=1}^{L}
\left[
\;2p\,\log\big|\,\mathbf{B}_{\ell}\,\big|
\;-\;
\operatorname{tr}\!\Big(
\big(\boldsymbol{\Sigma}_{\ell}^{-1}\otimes \mathbf{B}_{\ell}\mathbf{B}_{\ell}^{\top}\big)\,
\,\mathbf{P}\,\boldsymbol{S}_c^{(t)}\,\mathbf{P}^{\top}
\Big)
\right]
\;+\;\mathrm{const},
\qquad
\kappa_\ell^2 = e^{\kappa_0 \ell}.
\]

\item Finally, the function to maximize for updating the parameters controlling cross--process correlation \(r_0\) and \(r_1\) is: 

\begin{equation*}
\begin{aligned}
    &\mathcal{Q}_{r_0,r_1}(r_0, r_1)
= 
\sum_{\ell=1}^{L}
\Big[
 -R_\ell \big( (p-1)\log(1 - \rho_\ell)
              + \log(1 + (p-1)\rho_\ell) \big)-
\operatorname{tr}\!\Big(
\big(\boldsymbol{\Sigma}_{\ell}^{-1}\otimes \mathbf{B}_{\ell}\mathbf{B}_{\ell}^{\top}\big)\,
\,\mathbf{P}\,\boldsymbol{S}_c^{(t)}\,\mathbf{P}^{\top}
\Big) \Big]
,
\\
&\rho_\ell = r_0\,e^{-r_1(\ell - 1)}.
\end{aligned}
\end{equation*}
\end{itemize}

\subsubsection{Convergence criterion}
To assess convergence of the EM algorithm, we monitor the incomplete-data (marginal) log-likelihood
of the observed multivariate data vector 
$\mathbf{Z}$
at each iteration.
Convergence of the EM algorithm is declared when the relative increase in log-likelihood between two consecutive iterations falls below a predefined tolerance threshold (typically \(10^{-4}\)).

\section{Simulation studies}
In this section, we present a series of simulation experiments with two main aims: to validate the proposed model under controlled conditions and to quantify the potential gains of the multivariate formulation in scenarios where cross-process dependence can be exploited.

The main experiments consider a bivariate spatial field and evaluate both cross-covariance parameter recovery and predictive performance in scenarios where modeling cross-correlation is expected to be beneficial, comparing coFRK to independent univariate FRK fits.

For completeness, we also examine the behavior of the proposed model in a univariate setting and compare its predictive performance with that of the standard FRK model. These results are reported in Appendix~\ref{AppendixA}.

\subsection{Bivariate simulation}
\label{subsec:bivariate_simulation}

We consider a bivariate spatial field \(\{Z_1(s), Z_2(s)\}\), following the model illustrated in Section~(\ref{sec:coFRK}). 
The simulations are conducted over a two-dimensional square domain 
$\mathcal{D} = [0,1]^2$. 
No covariates are included in the model, resulting in a zero-mean spatial process. At each Monte Carlo replication, the latent coefficients, fine-scale variation, and measurement errors are sampled from their respective Gaussian distributions.

A total of $n_{\text{total}} = 1000$ spatial locations are uniformly sampled within the domain. 
Specifically, the dataset is divided into $n_{\text{train}} = 800$ training points and 
$n_{\text{test}} = 200$ test points for out-of-sample validation, selected randomly from the uniformly distributed spatial locations. Locations and the train–test partition are held fixed across 50 Monte Carlo replications.

We adopt a two-resolution system of bisquare basis functions ($L = 2$), with $R_1 = 9$ coarse-scale and $R_2 = 81$ fine-scale bisquare basis functions, for a total of $R = 90$. The basis-function centroids are placed on regular grids over the spatial
domain, namely a $3\times 3$ grid at the coarse level and a $9\times 9$ grid at
the fine level. The corresponding scale parameters are 0.936 and 0.234,
respectively.

\begin{figure}[H]
    \centering
    \includegraphics[width=0.9\columnwidth]{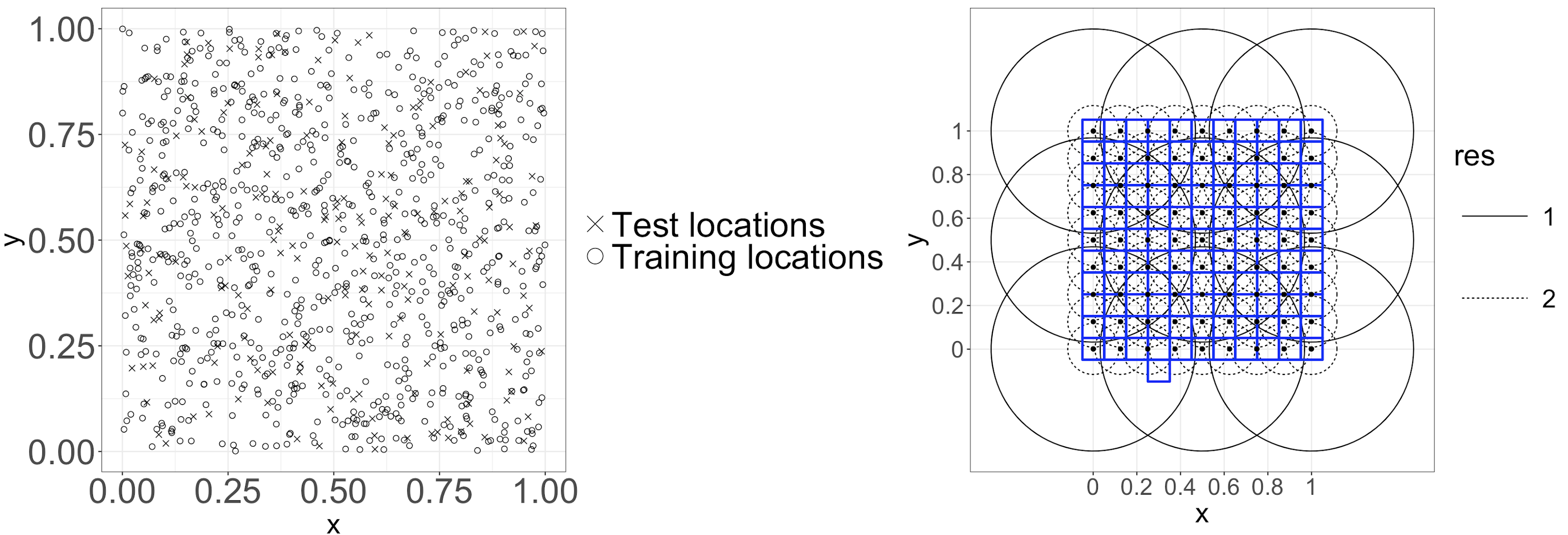}
    \caption{Left: spatial sampling design used in the simulation study, with
training locations shown as circles and test locations as crosses. Right: BAU
grid and bisquare basis functions used in the two-resolution system, showing
basis centers and supports at the coarse and fine levels.}
    \label{fig:BAUsBasis}
\end{figure}

The variance and correlation parameters used to simulate the spatial process are:
\[
\boldsymbol{\sigma}^2_s = (0.7,0.7), \quad \boldsymbol{\sigma}^2_\xi = (0.01,0.01), \quad \boldsymbol{\sigma}^2_\varepsilon = (10^{-4},10^{-4}), \quad \kappa_0^2 = 0.05.
\]
We set $\mathbf{V}_\xi = \mathbf{I}$ and $\mathbf{V}_\varepsilon = \mathbf{I}$, thereby imposing homoscedastic fine-scale variation and identical measurement error variance across all point locations.

Cross-dependence between the two processes is 
introduced at each resolution level through the correlation structure 
\(\rho_{\ell 12} = r_0 \exp[-r_1(\ell - 1)]\). 

\subsubsection{Inspecting cross-covariance structure}

We evaluate the recovery of cross-dependence parameters under three distinct scenarios: 
\begin{enumerate}[label=(\roman*)]
    \item Strong correlation with slow decay: \(\{r_0 = 0.9,\, r_1 = 0.5\}\)
    \item Moderate constant correlation: \(\{r_0 = 0.6,\, r_1 = 0\}\)
    \item Strong correlation with fast decay: \(\{r_0 = 0.9,\, r_1 = 2\}\)
\end{enumerate}

\begin{figure}[H]
  \centering

  \begin{minipage}[t]{0.32\textwidth}
    \centering
    \includegraphics[width=\linewidth]{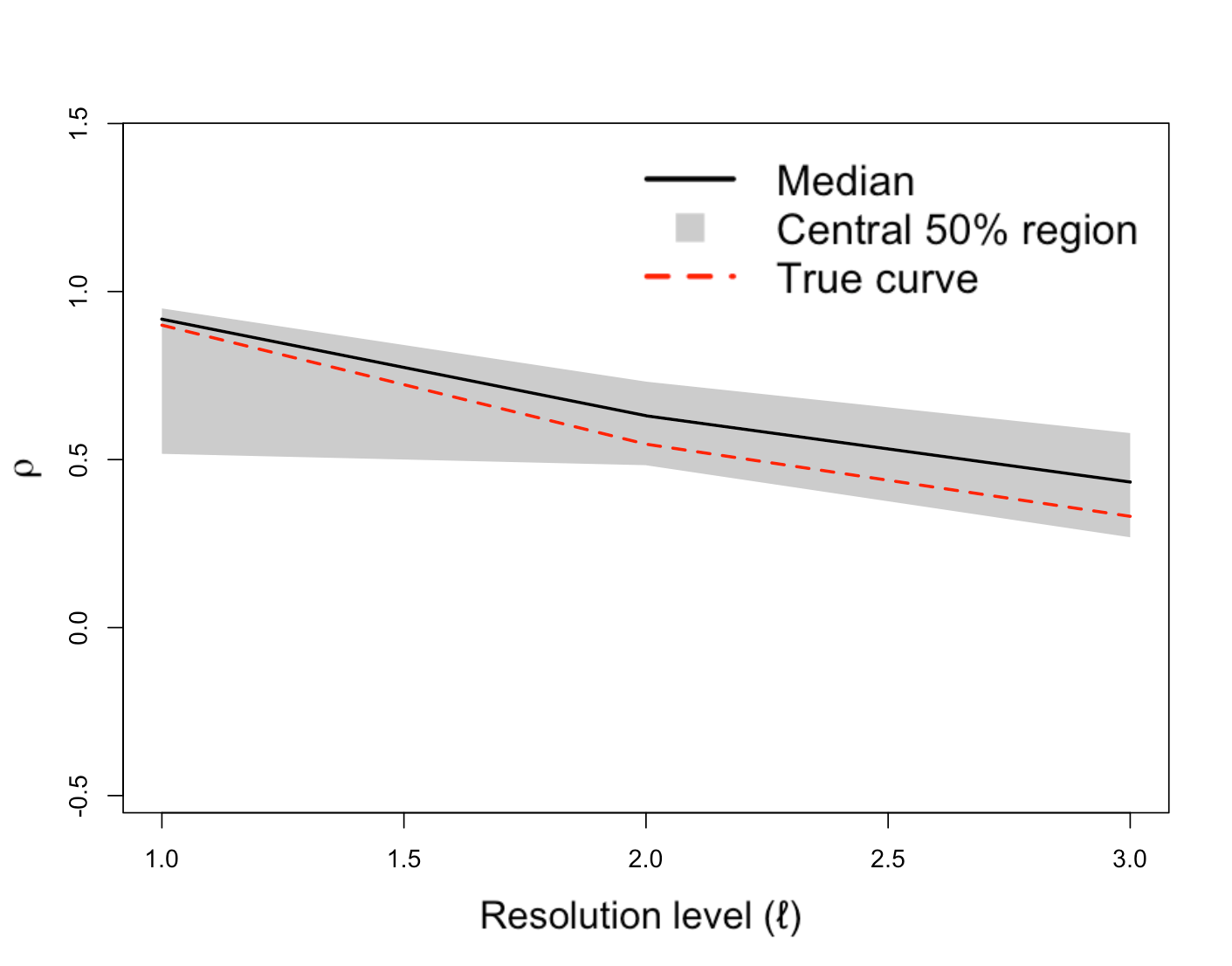}\\[0.1em]
    {\scriptsize Slow Decay}
  \end{minipage}
  \begin{minipage}[t]{0.32\textwidth}
    \centering
    \includegraphics[width=\linewidth]{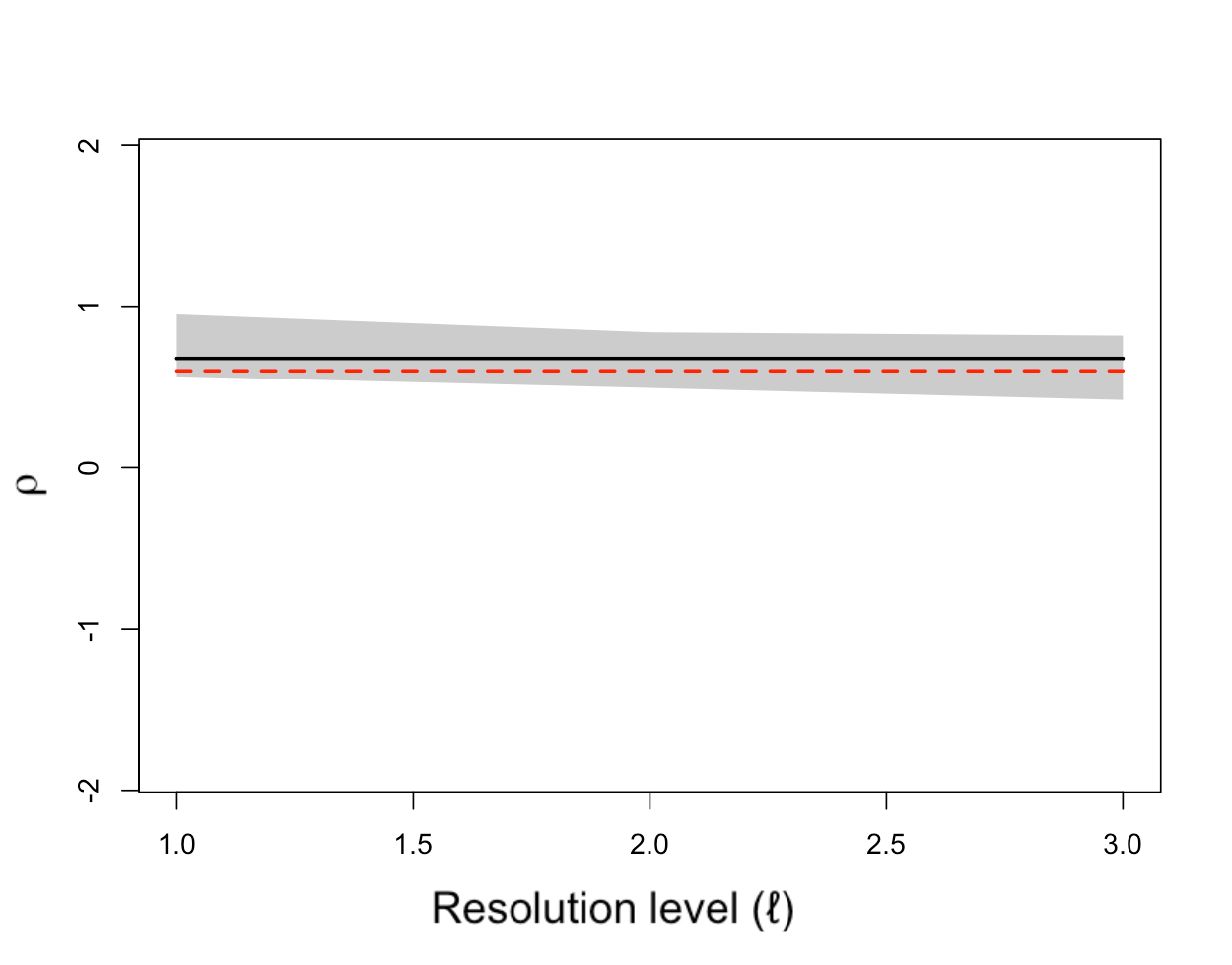}\\[0.1em]
    {\scriptsize No Decay}
  \end{minipage}
  \begin{minipage}[t]{0.32\textwidth}
    \centering
    \includegraphics[width=\linewidth]{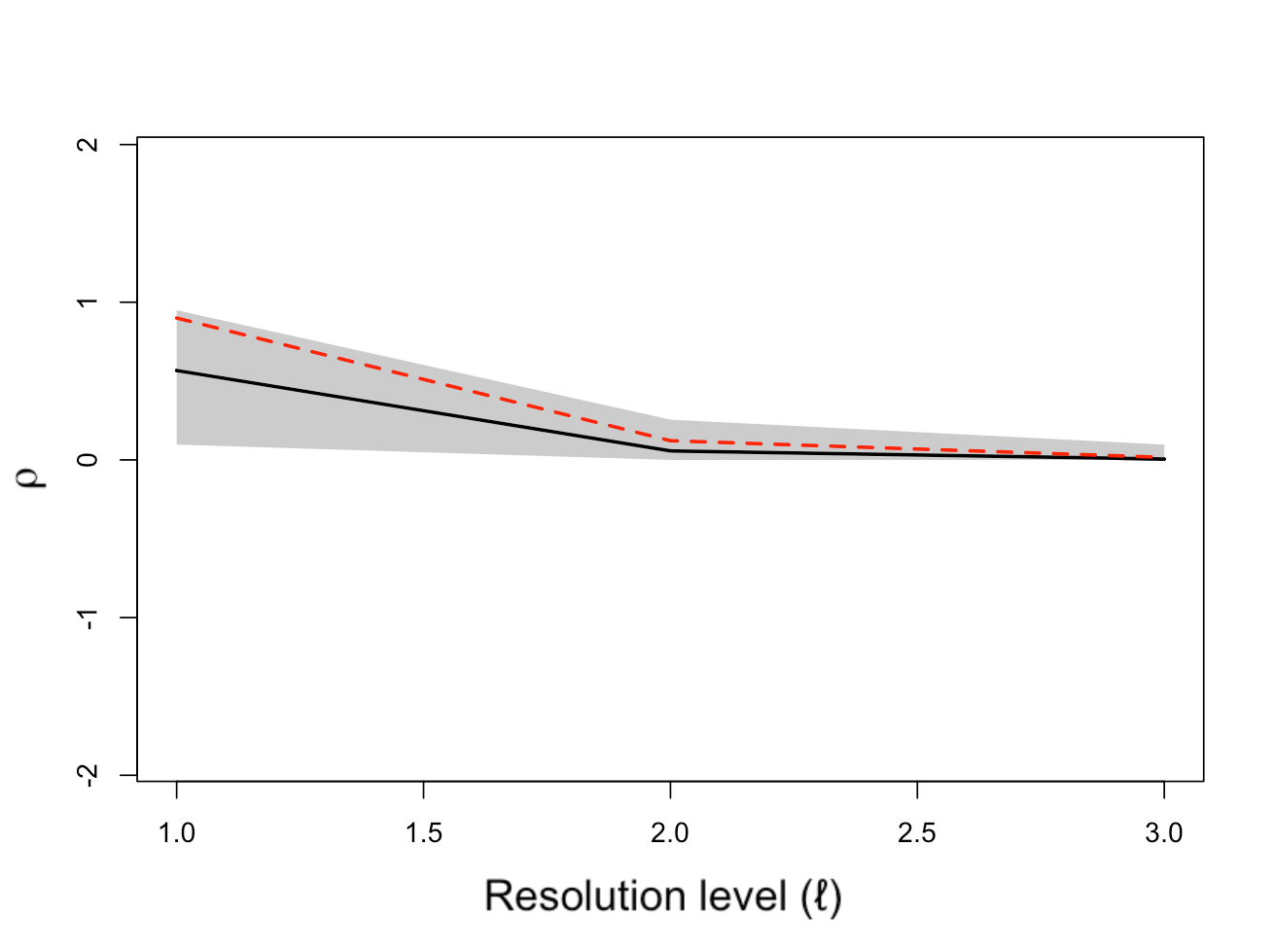}\\[0.1em]
    {\scriptsize Fast Decay}
  \end{minipage}

  \caption{Functional boxplots for the three spatial correlation scenarios.}
  \label{fig:functional_boxplots_three_cases}
\end{figure}

Table~\ref{tab:rho_estimates} reports the mean and standard deviation of the estimated cross-dependence parameters across Monte Carlo replications.
In addition, Figure~\ref{fig:functional_boxplots_three_cases} displays functional boxplots of the estimated correlation functions $\hat{\rho}(\ell)$, constructed following the framework of Sun and Genton (\cite{SunGenton}), using the Median Band Depth (MBD) measure (\cite{PintadoRomo}).

Across the three scenarios, the estimated cross-scale correlation functions closely reproduce the true patterns: the functional median closely follows the true curve and the 50\% central region consistently contains it. The constant--correlation case ($r_1 = 0$) exhibits greater variability, reflecting the intrinsic difficulty of disentangling resolution-level contributions when correlation does not vary with scale, as also noted by \cite{Kleiber2019}. Overall, the results indicate that the proposed parameterization adequately captures the main cross-covariance structure.

\begin{table}[H]
    \centering
    
    \begin{minipage}[t]{0.28\textwidth}
        \centering
        \caption*{\small Slow decay}
        \vspace{0.4em}
        \begin{tabular}{lccc}
            \toprule
            \small Parameter & \small True & \small Mean & \small SD \\
            \midrule
            \(r_0\) & 0.9 & 0.799 & 0.146 \\
            \(r_1\) & 0.5 & 0.329 & 0.256 \\
            \bottomrule
        \end{tabular}
    \end{minipage}
    \hfill

    \begin{minipage}[t]{0.28\textwidth}
        \centering
        \caption*{\small No decay}
        \vspace{0.4em}
        \begin{tabular}{lccc}
            \toprule
            \small Parameter & \small True & \small Mean & \small SD \\
            \midrule
            \(r_0\) & 0.6 & 0.618 & 0.44  \\
            \(r_1\) & 0.0 & 0.56 & 2.22 \\
            \bottomrule
        \end{tabular}
    \end{minipage}
    \hfill
   
    \begin{minipage}[t]{0.28\textwidth}
        \centering
        \caption*{\small Fast decay}
        \vspace{0.4em}
        \begin{tabular}{lccc}
            \toprule
            \small Parameter & \small True & \small Mean & \small SD \\
            \midrule
            \(r_0\) & 0.9 & 0.31 & 0.525 \\
            \(r_1\) & 2.0 & 2.11 & 3.49 \\
            \bottomrule
        \end{tabular}
    \end{minipage}

    \caption{Estimated and true parameter values for \(r_0\) and \(r_1\) under the three spatial correlation scenarios.}
    \label{tab:rho_estimates}
\end{table}

\subsubsection{Co-Kriging gain}

It is well established that co-kriging does not systematically outperform kriging in practice. 
Both theoretical and empirical studies have shown that the predictive gain depends critically on the strength of cross-correlation and on the sampling design 
\cite{GentonKleiber2015,ZhangCai2015,LimWu2022_cokriging_vs_kriging,EldeiryGarcia2010_soilsalinity}. 
In particular, Zhang and Cai \cite{ZhangCai2015} derived the asymptotic relation
\[
\lim_{n \to \infty} 
\frac{\mathrm{MSE}(\widehat{Z}_1^{\,\text{kriging}})}
     {\mathrm{MSE}(\widehat{Z}_1^{\,\text{cokriging}})}
= 1 - \frac{\rho^2}{2},
\]
showing that the improvement is directly driven by the squared cross-correlation $\rho$. 
Hence, limited gains are expected when dependence between variables is weak.

Motivated by these considerations, we design simulation scenarios that explicitly investigate conditions under which co-kriging is theoretically expected to provide benefits, namely:
(i) strong cross-correlation, and 
(ii) unbalanced sampling schemes in which one process is more densely observed than the other.

\paragraph{Experiment 1: Different sampling frequencies.}

To investigate the impact of unbalanced sampling, we consider a design in which the two processes are observed at markedly different frequencies. At each Monte Carlo replication, a fixed test set is held out. Among the remaining locations, \(Z_1(s)\) is observed at 5\% of the sites, whereas \(Z_2(s)\) is observed at 80\%.
Predictions for \(Z_1\) are then computed on the held-out test set using both the bivariate coFRK model and the standard univariate FRK applied to \(Z_1\) alone.

The parameters used for data generation are 
\(\kappa_0 = 0.4\), 
\(\boldsymbol{\sigma}^2_s = (0.7,\,0.7)\), 
\(\boldsymbol{\sigma}^2_\xi = (0.001,\,0.001)\), 
\(\boldsymbol{\sigma}^2_\varepsilon = (0.0002,\,0.0008)\), 
and \((r_0, r_1) = (0.9,\,0.5)\). 
At each replication, random coefficients, fine-scale effects, and measurement errors are drawn according to this specification.

Predictive performance is assessed over fifty Monte Carlo iterations using the Root Mean Squared Error (RMSE) and the coefficient of determination (\(R^2\)). 
Under this unbalanced sampling design, the coFRK consistently attains lower RMSE and higher \(R^2\) values than the univariate FRK applied to \(Z_1\) alone. These results indicate that, when the primary process is sparsely observed, incorporating information from a densely sampled auxiliary variable can lead to improved predictive accuracy. \\

\begin{figure}[H]
    \centering

    \begin{minipage}[t]{0.5\textwidth}
        \centering
        \includegraphics[width=\textwidth]{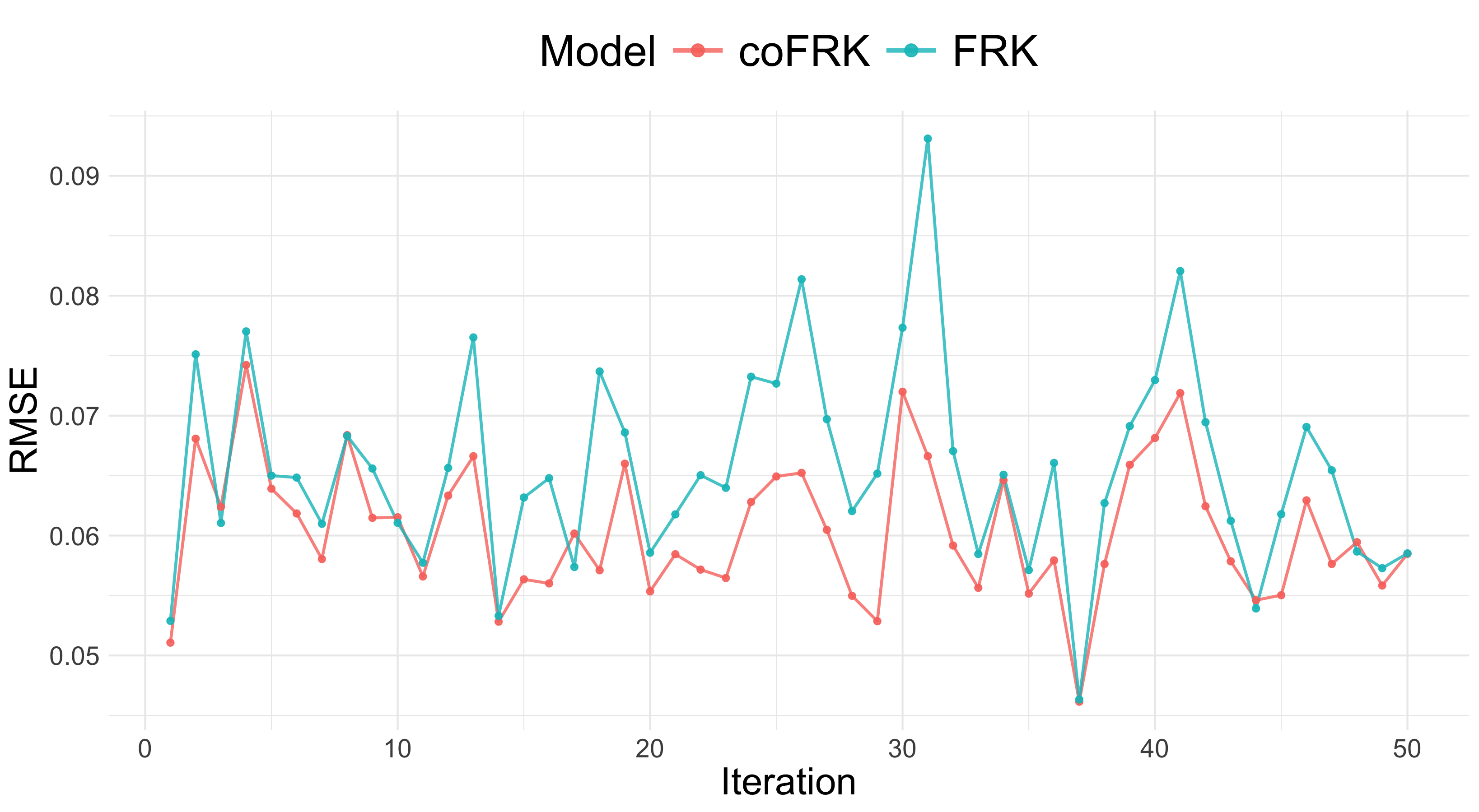}
        \caption{RMSE of \(Z_1\) predictions across Monte Carlo iterations.}
    \end{minipage}
    \hfill
    
    \begin{minipage}[t]{0.5\textwidth}
        \centering
        \includegraphics[width=\textwidth]{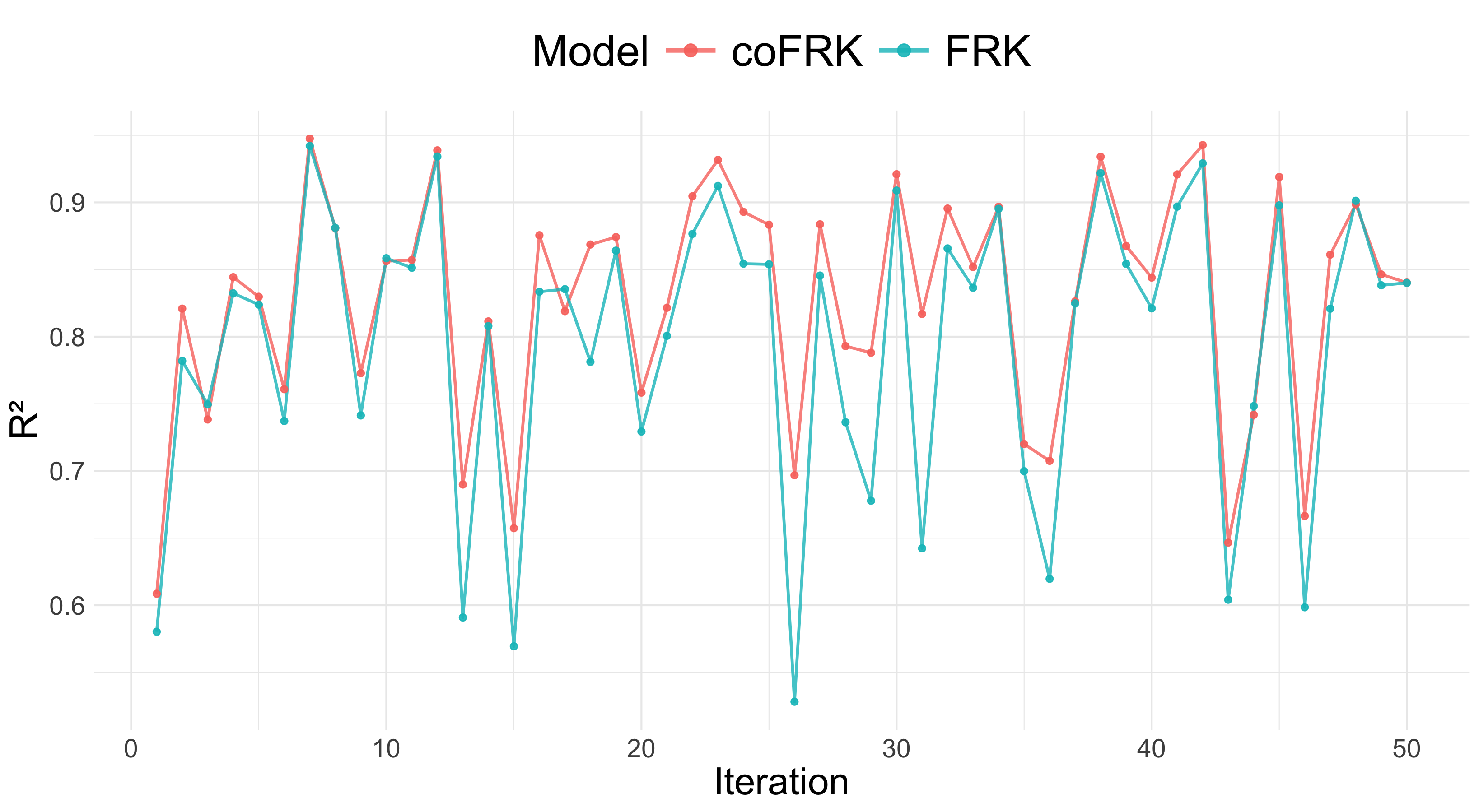}
        \caption{$R^2$ of \(Z_1\) predictions across Monte Carlo iterations.}
    \end{minipage}

    \label{fig:rmse_r2_z1}
\end{figure}

\begin{figure}[H]
    \centering
    \includegraphics[width=0.45\textwidth]{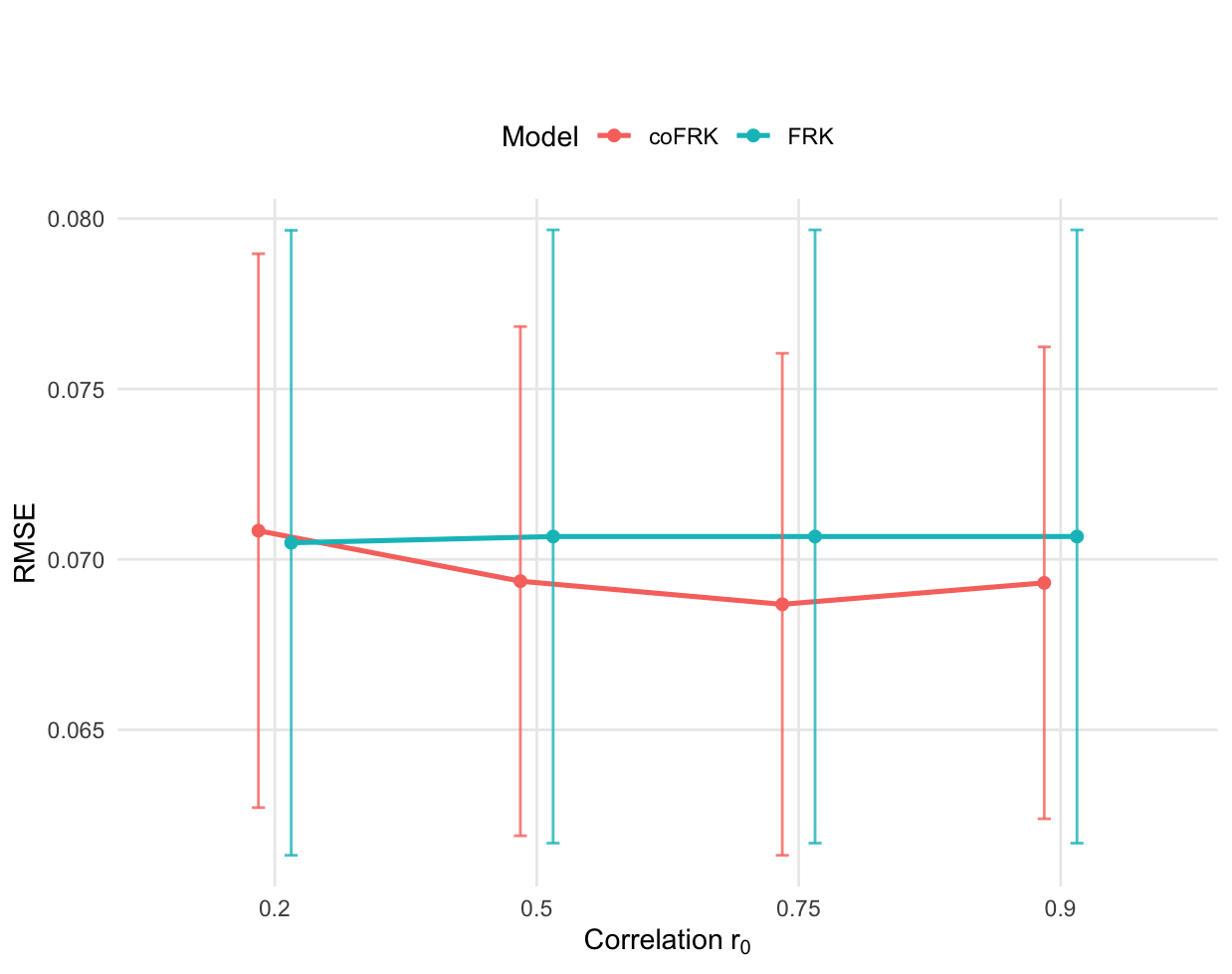}
    \caption{RMSE of coFRK and FRK as a function of the inter-variable correlation parameter \(r_0\). Vertical bars indicate \(\pm 1\) standard deviation across Monte Carlo replications.}
    \label{fig:cofrk_correlation_rmse}
\end{figure}

To further investigate the effect of cross-correlation on predictive performance, we repeated the same experiment under different correlation strengths between the two processes. The parameter \(r_0\), which controls the correlation \(\rho_{\ell 12}\), was varied over the set \(\{0.2,\,0.5,\,0.75,\,0.9\}\), while all other parameters were kept fixed. As illustrated in Figure~\ref{fig:cofrk_correlation_rmse}, an improvement in RMSE can be observed for the coFRK model as \(r_0\) increases.

\paragraph{Experiment 2: Missing spatial subregion}

In this experiment, we assess the predictive performance of the models when one of the two spatial processes is completely unobserved over a portion of the spatial domain. 
Specifically, the process \(Z_1(s)\) is left unobserved within a contiguous subregion of the domain, while \(Z_2(s)\) is observed at all sampled training locations. 
For data generation, the same parameters as in Experiment 1 are employed. 

\begin{figure}[H]
    \centering

    \begin{minipage}[t]{0.42\textwidth}
        \centering
        \includegraphics[width=\textwidth]{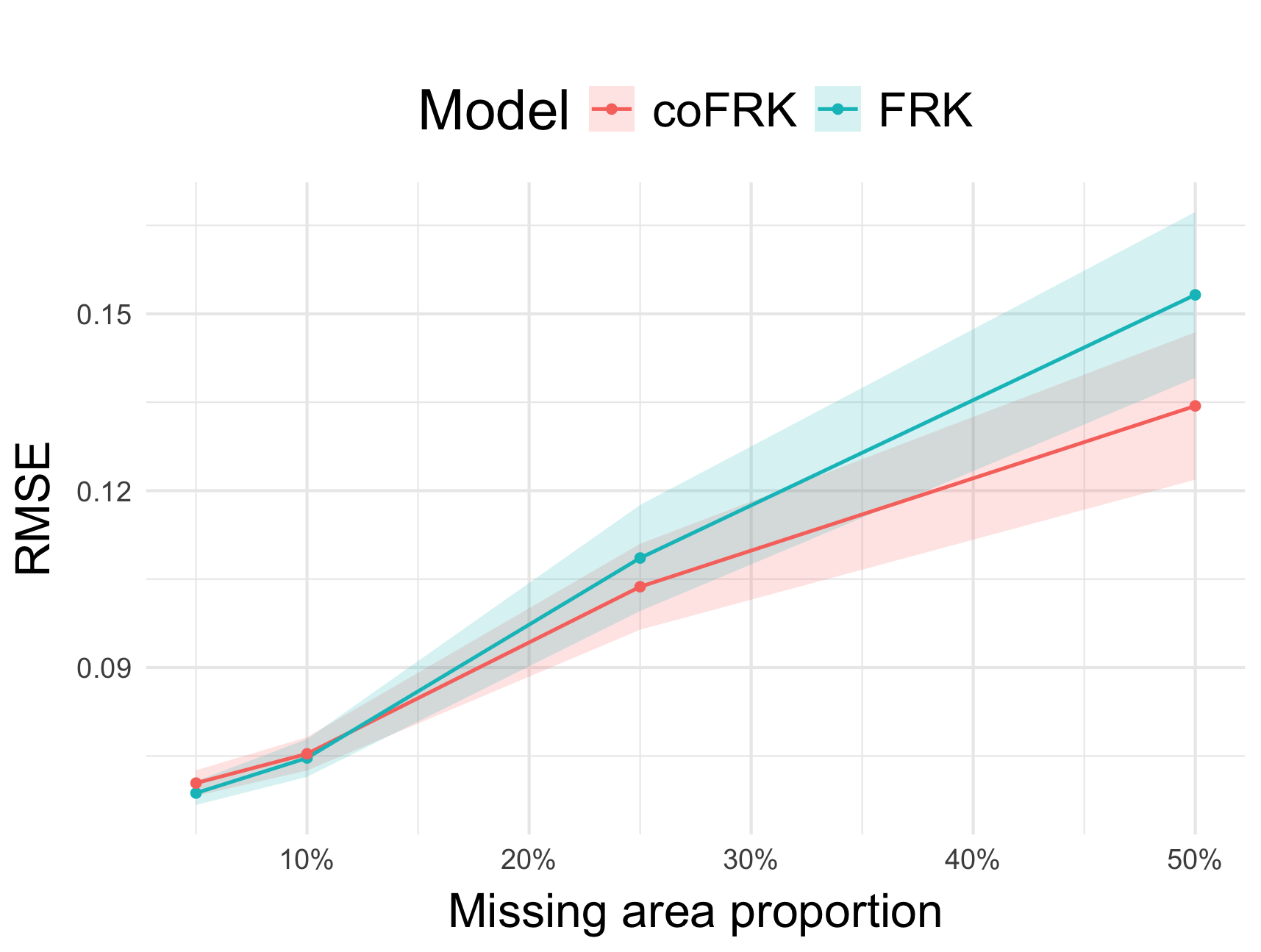}
        \vspace{0.3em}
        {\small }
    \end{minipage}
    \hspace{0.03\textwidth}
    \begin{minipage}[t]{0.42\textwidth}
        \centering
        \includegraphics[width=\textwidth]{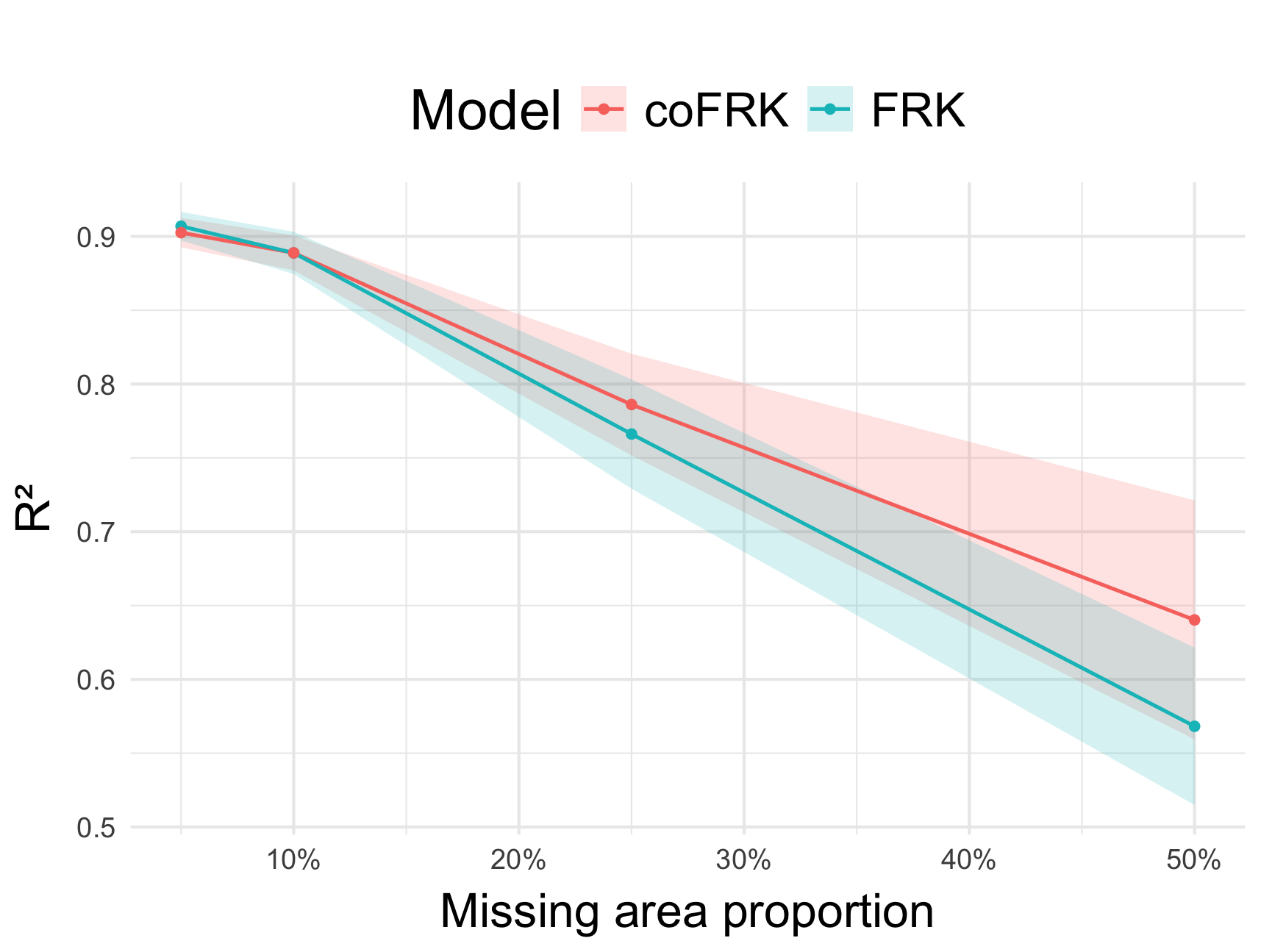}
        \vspace{0.3em}
        {\small }
    \end{minipage}

    \caption{RMSE (left) and R$^2$ (right) for $Z_1$ as a function of the missing-area proportion.}
    \label{fig:rmse_r2_bi}
\end{figure}

Two complementary analyses are carried out. 
In the first analysis, we remove a square region of the domain in which \(Z_1(s)\) is completely unobserved. The position of the missing region is fixed (bottom-left corner), and its size is increased across four proportions of the domain: \(\{0.05,\,0.10,\,0.25,\,0.50\}\) of the total area.
For each proportion, Monte Carlo simulations are performed to assess the predictive performance of coFRK compared to the standard univariate FRK applied separately to \(Z_1\). 
The results, summarized through the evolution of RMSE and \(R^2\) across iterations in Figure~\ref{fig:rmse_r2_bi}, show that coFRK provides consistently better predictions than the univariate model. Both metrics are computed over the full set of 200 test locations.

\begin{figure}[H]
    \centering
    \includegraphics[width=0.9\columnwidth]{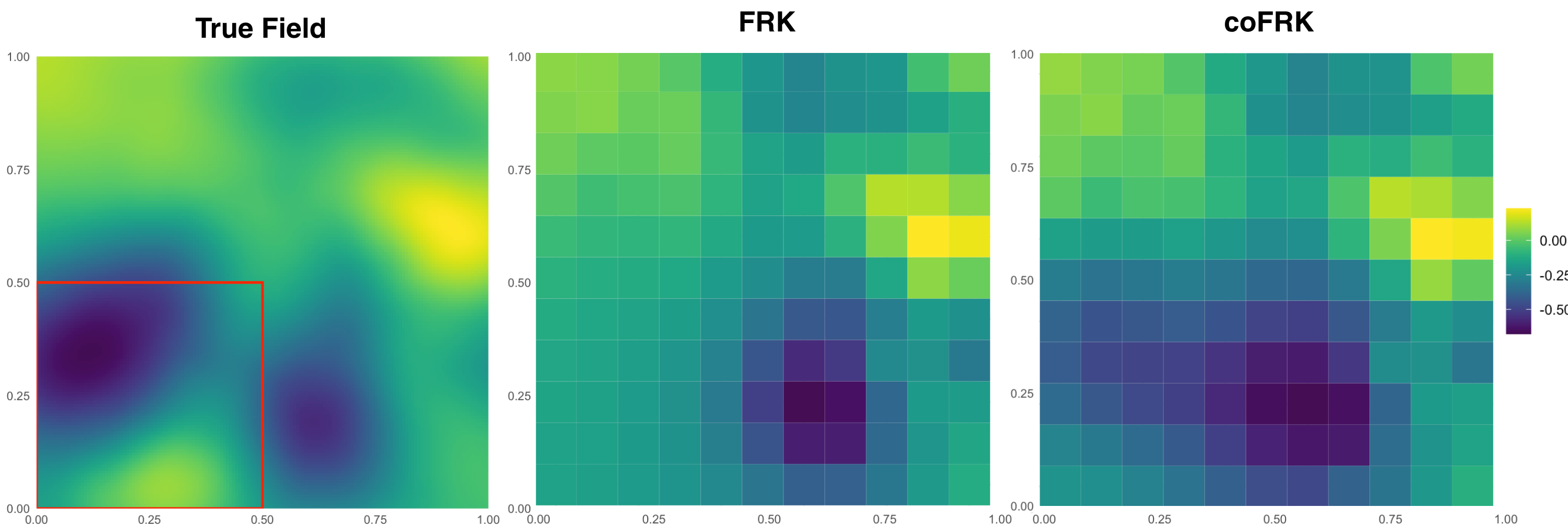}
    \caption{Comparison between the true field \( Z_1(s) \) and the predicted fields from FRK and coFRK. 
The red border in the \textit{True field} panel indicates the region where \( Z_1(s) \) is treated as unobserved in the experiment. 
Predictions are shown at the BAU level.}
    \label{fig:true-frk-cofrk}
\end{figure}

To complement the quantitative assessment, Figure~\ref{fig:true-frk-cofrk} provides a visual comparison between the true field and the predicted fields obtained under FRK and coFRK for a representative replication, highlighting the ability of the multivariate model to recover the missing region.
 
\begin{figure}[H]
  \centering
  \includegraphics[width=0.50\textwidth]{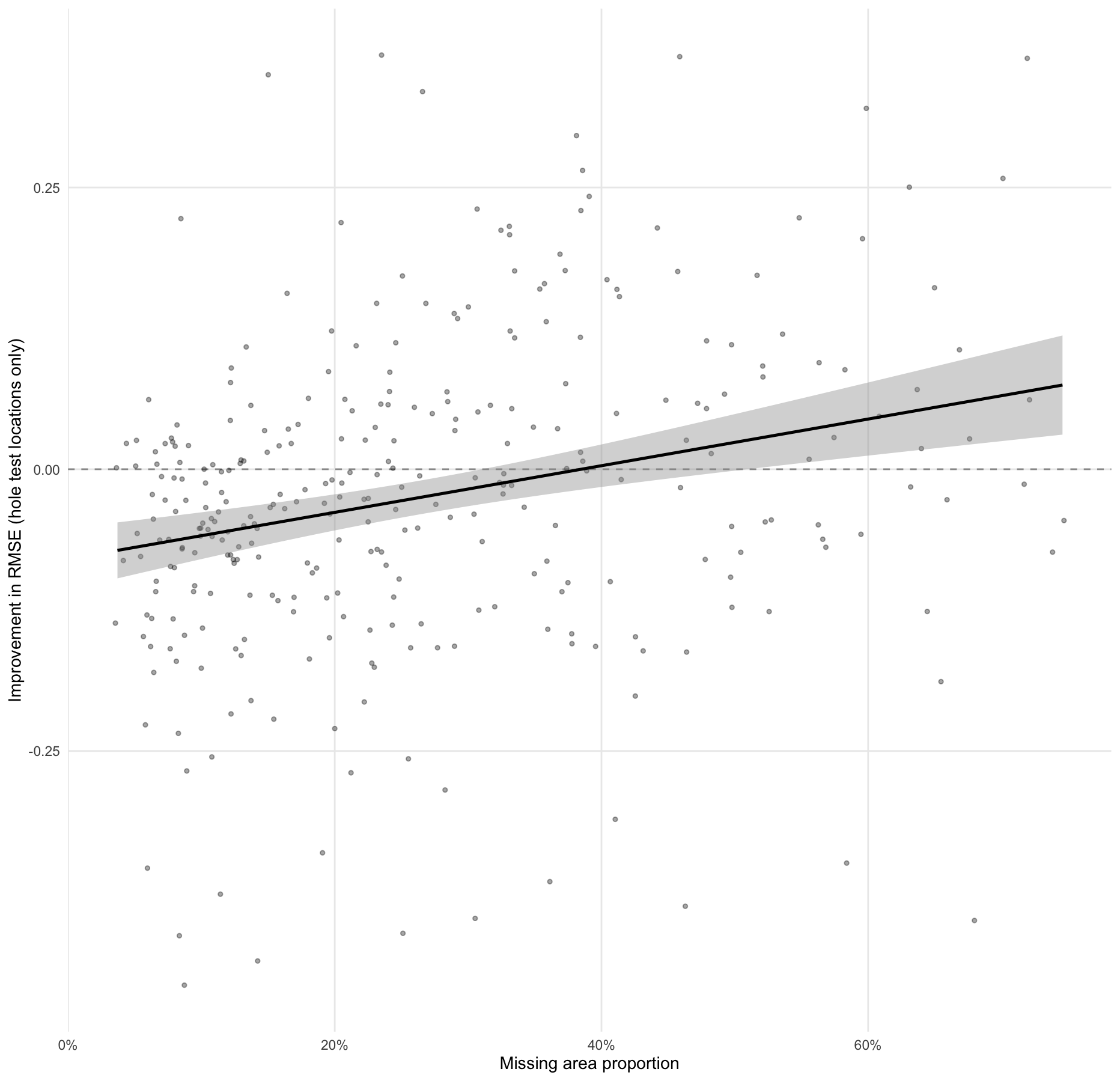}
  \caption{Improvement in RMSE of co-kriging with respect to FRK as a function of the proportion of unobserved area.}
  \label{fig:rmse_improvement}
\end{figure}

\noindent
In the second analysis, the proportion and location of the unobserved region is varied randomly across Monte Carlo replications. 
At each iteration, a different portion of the spatial domain, corresponding to a randomly selected fraction approximately between 0.02 and 0.75 of the total area, was withheld from model fitting. 
This design allows us to investigate the general relationship between the amount of missing area and predictive performance.
As shown in Figure~\ref{fig:rmse_improvement}, a clear positive association emerges between the relative improvement in RMSE of coFRK with respect to FRK and the proportion of unobserved area, indicating that the benefit of co-kriging becomes more pronounced as the information gap for \( Z_1(s) \) widens. 
The improvement was quantified using prediction errors computed only over test locations falling inside the unobserved region, as
\(
\frac{\mathrm{RMSE}_{\text{FRK}} - \mathrm{RMSE}_{\text{coFRK}}}{\mathrm{RMSE}_{\text{FRK}}}
\).
To summarize the relationship between predictive gain and the size of the unobserved region, we added a simple least-squares regression line with its 95\% confidence interval. The positive trend confirms that the benefit of coFRK increases as the missing area grows.

\section{Modeling PM$_{10}$ concentrations over Northern Italy}

In this section, we apply the proposed methodology to a real-world dataset describing PM\textsubscript{10} concentrations across Northern Italy.
This analysis forms part of a broader research framework illustrated in De Sanctis et al.(\cite{DeSanctis2025_DistributionalPM10}).

PM\textsubscript{10} (particulate matter with an aerodynamic diameter smaller than or equal to $10\,\mu\text{m}$) represents a major component of atmospheric pollution. It comprises both fine and coarse particles originating from a wide range of sources. A major one is domestic heating, but other relevant sources include combustion processes, industrial emissions, vehicular traffic, construction activities, and natural phenomena such as soil dust or sea salt.  
According to World Health Organization (\cite{WHO2021_AQG}), PM\textsubscript{10} poses significant risks to human health through both short- and long-term exposure. Owing to their small size, these particles can penetrate the upper respiratory tract and reach the bronchi, where they may cause adverse effects on human health, particularly on the respiratory and cardiovascular systems. For these reasons, ambient concentrations of PM\textsubscript{10} are widely adopted as a key indicator of air quality and are subject to regulation by major international environmental and public health agencies. The European Directive~2008/50/EC on ambient air quality
establishes a limit value for $\mathrm{PM}_{10}$ of 
$50~\mu\mathrm{g/m^3}$ for the daily mean concentration,
allowing up to $35$ exceedance days per calendar year
to account for occasional episodic events (\cite{EEA2024_AirQualityStatus2024}). 

The data analyzed refer to Northern Italy, a region where monitoring air quality is of particular importance. As reported by the European Environment Agency (EEA)(\cite{EEA2024website}), Northern Italy and especially the Po Valley experience some of the worst conditions across Europe. It is, in fact, a highly industrialised and densely populated area, making compliance with air-quality guidelines for pollutants such as PM\textsubscript{10} especially critical.

In \cite{DeSanctis2025_DistributionalPM10}, the same dataset was investigated through a functional perspective, rather than by focusing on summary indicators such as means or exceedance frequencies. In that study, the authors modeled the entire distribution of PM\textsubscript{10} concentrations over space. The work presented in this section is therefore situated within this established framework and provides an additional method to make functional predictions, by applying the proposed multivariate coFRK model.  

\subsection{Data and Modeling Framework}

This section briefly summarizes the modeling framework of \cite{DeSanctis2025_DistributionalPM10}, 
which forms the basis of the present application. 
We directly employ some results from that study, namely the estimated quantiles used to define the trimming thresholds and the spatially smoothed covariate surfaces. Readers are referred to the original paper for a complete methodological description.

\begin{figure}[htbp] \centering \includegraphics[width=0.6\textwidth]{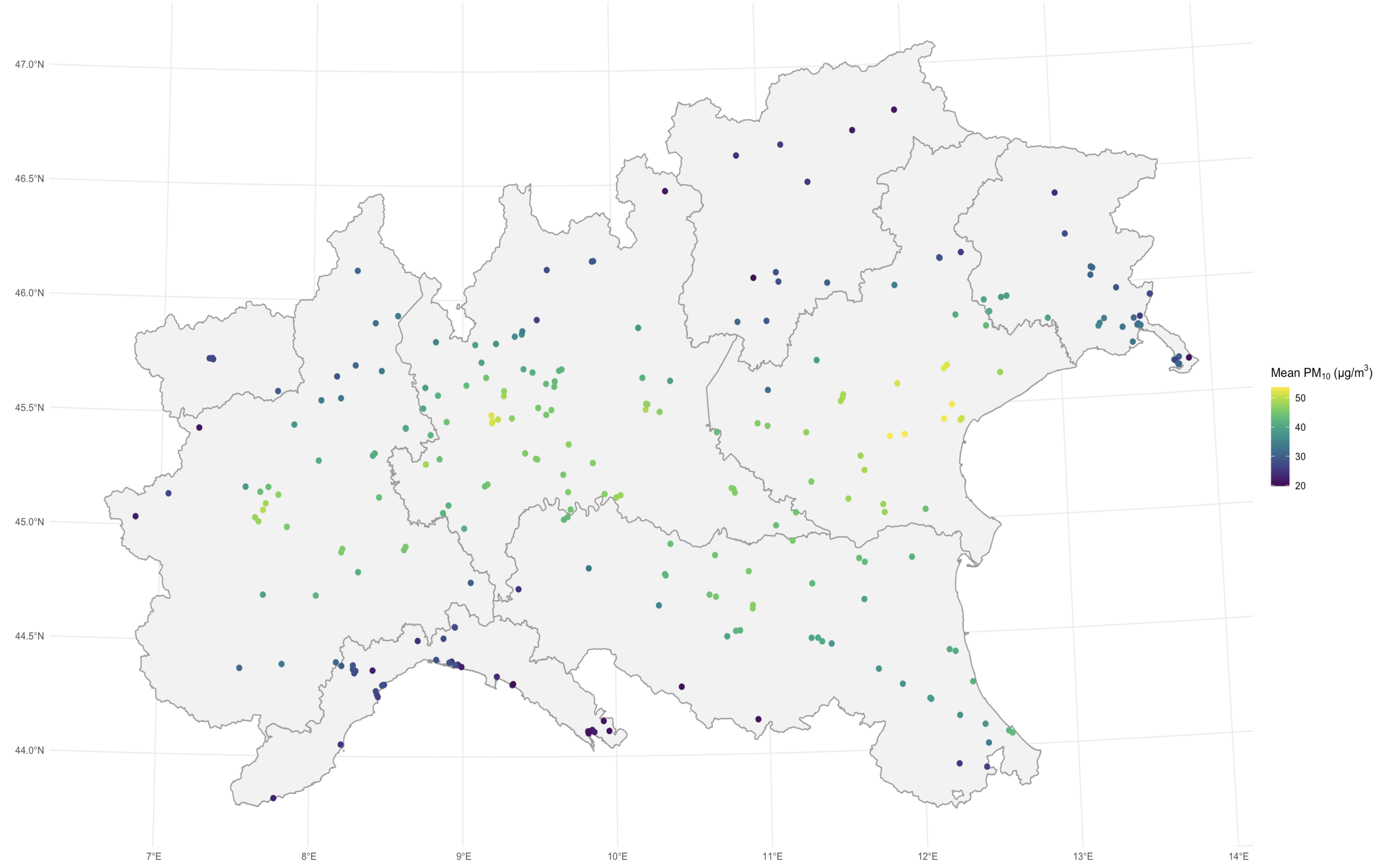} \caption{ Spatial distribution of the average $\mathrm{PM_{10}}$ concentration per monitoring station across Northern Italy (2018--2022). } \label{fig:pm10_mean_map} \end{figure}

The dataset consists of daily average PM\textsubscript{10} concentrations recorded at 266 monitoring stations across Northern Italy during the period 2018--2022. Figure~\ref{fig:pm10_mean_map} shows the monitoring stations across Northern Italy, with each station coloured according to the average PM\textsubscript{10} concentration observed at that location. 
As noted in \cite{DeSanctis2025_DistributionalPM10}, these data are affected by occasional extreme values, not representative of typical PM\textsubscript{10} concentrations. To mitigate the influence of such anomalies, the observations were trimmed, retaining only values between the 1st and 99th empirical quantiles. The quantiles were estimated through the spatial quantile regression method introduced by Castiglione et~al.~(\cite{Castiglione2023_PDEQuantile}). 
Two spatial, real-valued covariates are considered: altitude and population density, both of which are strongly associated with PM\textsubscript{10} concentrations. Higher population density generally reflects more intense anthropogenic activity, while elevation plays a key role in shaping atmospheric circulation. In particular, the Alpine chain to the north and west of the study area limits air exchange and favors thermal inversion phenomena over the Po Valley, thus contributing to the accumulation of pollutants. Both covariate surfaces were spatially smoothed using the FDA–PDE approach (\cite{Sangalli2021_SR_PDE}). 

Finally, predictions are carried out over a municipality-level spatial grid, so that the resulting PM\textsubscript{10} distributions are provided at the municipality level.

\subsection{Functional Representation of PM\textsubscript{10} Distributions}
In order to produce distributional predictions of PM\textsubscript{10} concentrations, two main steps are needed.
The first concerns the construction of an appropriate functional representation of the data starting from raw measurements.
Secondly, these functional data must be embedded into a finite–dimensional 
representation compatible with the multivariate spatial modeling framework introduced in this work.

\paragraph{From Raw Measurements to Functional Densities.} 
We adopt the functional representation of PM\textsubscript{10} distributions 
introduced by \cite{DeSanctis2025_DistributionalPM10}. 
In particular, we use the clr–transformed and smoothed density functions provided in that study 
as the functional inputs to our model.
Their methodology provides smoothed, site–specific estimates of the daily PM\textsubscript{10} 
concentration distributions in a functional form suitable for subsequent statistical modeling.  
In particular, each empirical density is represented within the Bayes space \(B^2(I)\) 
(\cite{VanDenBoogaart2011_BayesLinearSpaces}), which ensures positivity and unit integral, 
and is mapped to the Hilbert space \(L^2(I)\) through the centered log–ratio (clr) transformation.  
This transformation allows standard Functional Data Analysis (FDA) tools to be used while 
preserving the compositional nature of probability densities. 

Because PM\textsubscript{10} concentrations vary across monitoring sites, the supports of the 
densities differ substantially. This issue is addressed by decomposing each density into a 
\emph{support} and a \emph{shape} component.  
For each site \(i,\) the site–specific supports are determined from the empirical quantiles \(Q_1(i)\) and \(Q_{99}(i)\), 
estimated through spatial quantile regression \cite{Castiglione2023_PDEQuantile}.  
The observations are then rescaled to the common domain \([0,1]\) according to
\[
\tilde{y}_{ij} = \frac{y_{ij} - Q_1(i)}{Q_{99}(i) - Q_1(i)}, \qquad j = 1, \dots, n_i,
\]
so that the shapes of all densities become comparable.  
On this aligned domain, the densities are smoothed using penalized cubic B–splines within 
\(B^2([0,1])\).
For complete theoretical and computational details of this construction, we refer to 
\cite{DeSanctis2025_DistributionalPM10}.

\paragraph{Basis expansion and coefficient extraction.}  
To integrate these functional data into the proposed coFRK model, each density function must be expressed in finite–dimensional form.  
We represent every clr–transformed density \( g_i(t) \) on the common domain \([0,1]\) through a cubic B–spline basis expansion
\[
g_i(t) \approx \sum_{k=1}^{K} b_{ik} \, \phi_k(t),
\]
where \(\{\phi_k\}_{k=1}^K\) denotes the set of B–spline basis functions and \(b_{ik}\) their corresponding coefficients.  
In this work, a cubic B–spline basis on \([0,1]\) with a second–derivative smoothing penalty is employed.  This results in a finite–dimensional representation with \(K = 5\) spline coefficients for each site \(i,\)
\[
\mathbf{b}_i = (b_{i1}, b_{i2}, b_{i3}, b_{i4}, b_{i5}),^\top
\]
that constitutes the finite–dimensional representation of the density associated with site \(i\) and serves as multivariate input in the coFRK spatial model, together with the BAU-level covariates, namely altitude and population density, incorporated as spatially aggregated values over each BAU.

After model estimation, predicted densities are reconstructed by combining the estimated coefficients with the same spline basis:
\[
\hat{g}_i(t) = \sum_{k=1}^{K} \hat{b}_{ik} \, \phi_k(t),
\]
and subsequently applying the inverse clr transformation to recover valid density functions in \(B^2([0,1])\).  
Finally, each predicted density is mapped back to its original concentration scale using the site–specific quantiles \(Q_1(i)\) and \(Q_{99}(i)\) provided by \cite{DeSanctis2025_DistributionalPM10}.

\subsection{Results}

As a first step, we summarize the modeled PM\textsubscript{10} distributions at municipality level in terms of their mean concentration and their \(95^{\text{th}}\) percentile (\(Q_{95}\)).  
Both quantities were obtained directly from the reconstructed densities by numerical integration.

\begin{figure}[H]
    \centering
    \begin{minipage}[t]{0.46\textwidth}
        \centering
        \includegraphics[width=\linewidth]{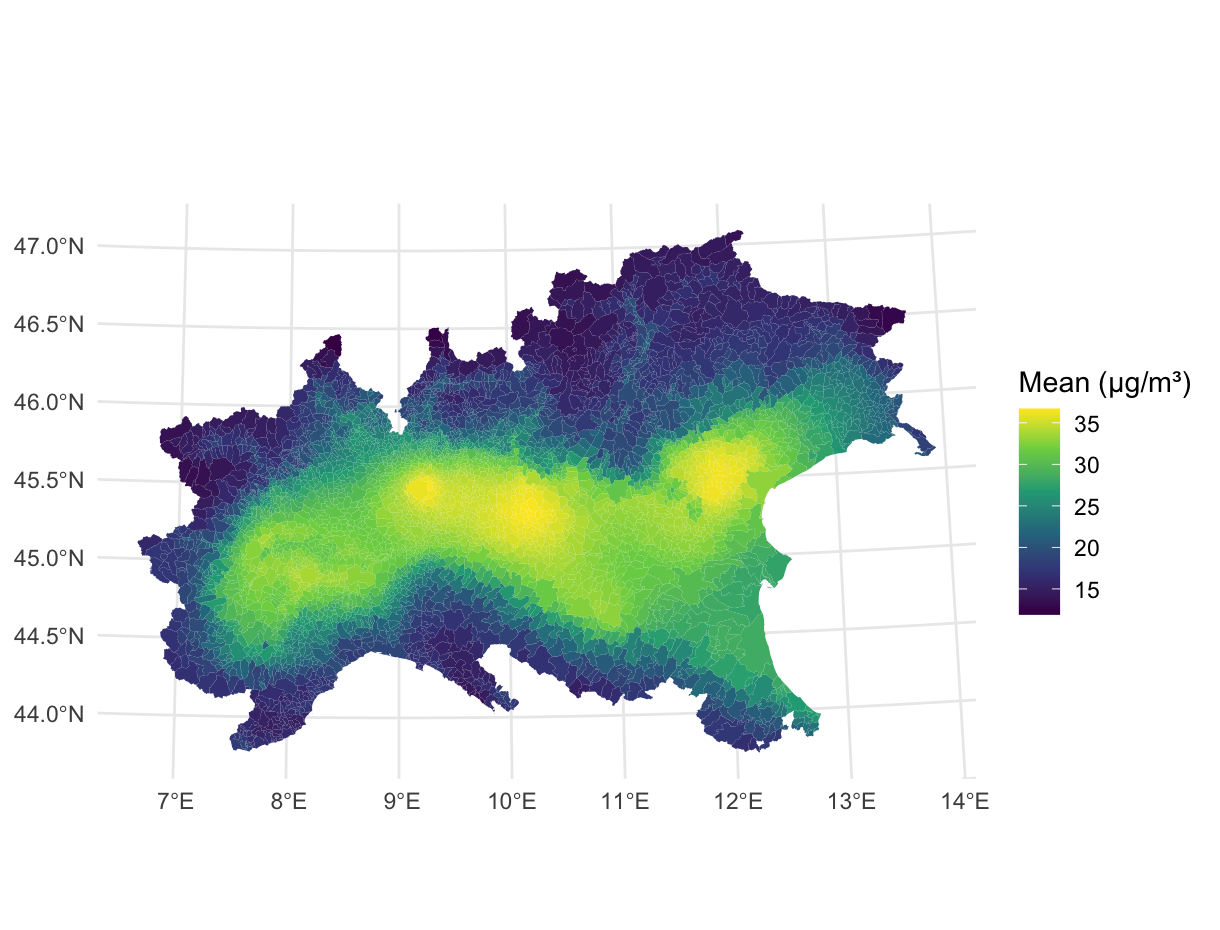}
        \caption*{(a) Mean PM\textsubscript{10} concentration}
    \end{minipage}
    \hfill
    \begin{minipage}[t]{0.46\textwidth}
        \centering
        \includegraphics[width=\linewidth]{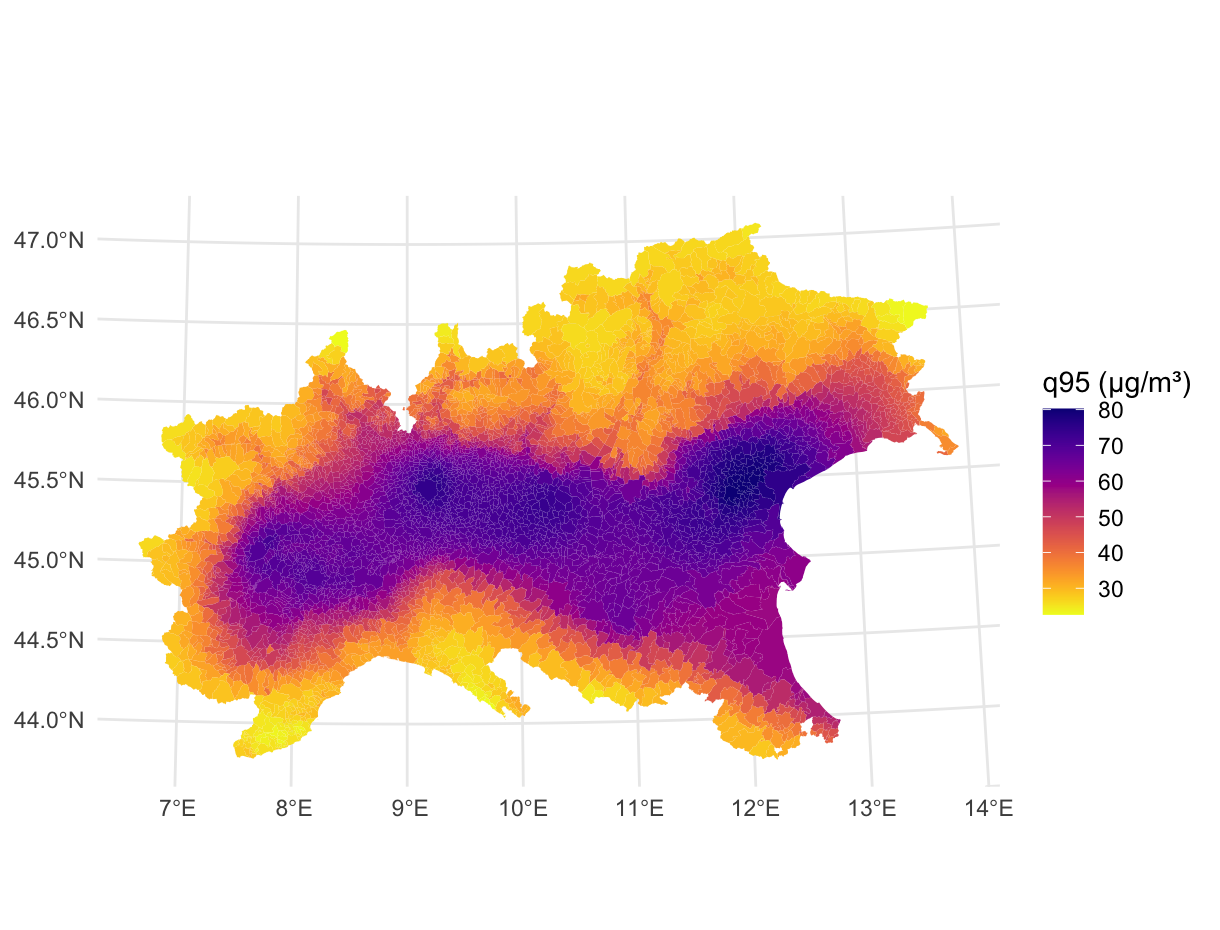}
        \caption*{(b) 95\textsuperscript{th} percentile (\(Q_{95}\))}
    \end{minipage}
    
    \caption{Spatial distribution of (a) the mean PM\textsubscript{10} concentration and (b) the 95\textsuperscript{th} percentile (\(Q_{95}\)) estimated from the reconstructed density functions.}
    \label{fig:mean_q95_maps}
\end{figure}

Panel (a) of Figure~\ref{fig:mean_q95_maps} displays the spatial distribution of the mean PM\textsubscript{10} concentration. 
A clear and well documented pattern emerges: the highest values are concentrated across the Po Valley, particularly around the metropolitan areas of Milan, the provinces of Brescia and Cremona and in the Veneto region. 
This result is consistent with the characteristics of the Po Valley, which is one of the most industrialized and densely populated regions in Europe. Industrial activity, together with emissions from road traffic and domestic heating in large urban centers such as Milan and Turin, contributes to consistently high emission levels throughout the year, with particularly elevated levels in winter.
In addition to this, the geographical configuration of the Po Valley plays a crucial role: the valley is enclosed by the Alps and the Apennines on three sides, which limits air circulation. This reduced ventilation, together with frequent thermal inversions in winter, favors the accumulation and persistence of pollutants near the ground.
Lower concentrations are observed in Trentino–Alto Adige and along the Ligurian coast. 
In the former case, the mountainous terrain is generally associated with greater air circulation and dispersion, while in coastal areas such as Liguria, the presence of sea breezes can contribute to improved ventilation. These factors, together with lower emission densities, help explain the comparatively lower levels of PM\textsubscript{10} estimated in these regions.
Panel (b) of Figure~\ref{fig:mean_q95_maps} shows the spatial distribution of the \(Q_{95}\) values. The overall pattern is similar to that observed for the mean concentration. In the Po Valley, not only are average PM\textsubscript{10} levels elevated, but also intense peak episodes occur more frequently and with greater intensity. In contrast, the lowest \(Q_{95}\) values are observed along the Ligurian coast, suggesting that these areas are less affected by high-concentration episodes.

We now focus on two indicators that summarize the severity of pollution episodes. 
The first is the exceedance probability \(p_i = P(\text{PM}_{10} > 50)\), which expresses the likelihood that daily concentrations exceed the European limit value of \(50\,\mu\text{g/m}^3\). 
The second is the expected number of exceedance days over one year.
This quantity is directly comparable to the regulatory threshold established by the European Air Quality Directive (2008/50/EC), which allows up to 35 exceedance days per year.

Formally, for each municipality \(i\), the exceedance probability is computed by numerically integrating the estimated PDF \(\hat{f}_i(y)\) above the regulatory limit \(y_0 = 50\). The expected number of exceedance days is subsequently obtained as \(365\times p_i\), representing the expected annual frequency of days exceeding the regulatory threshold.

\begin{figure}[H]
    \centering
    \begin{minipage}[t]{0.46\linewidth}
        \centering
        \includegraphics[width=\linewidth]{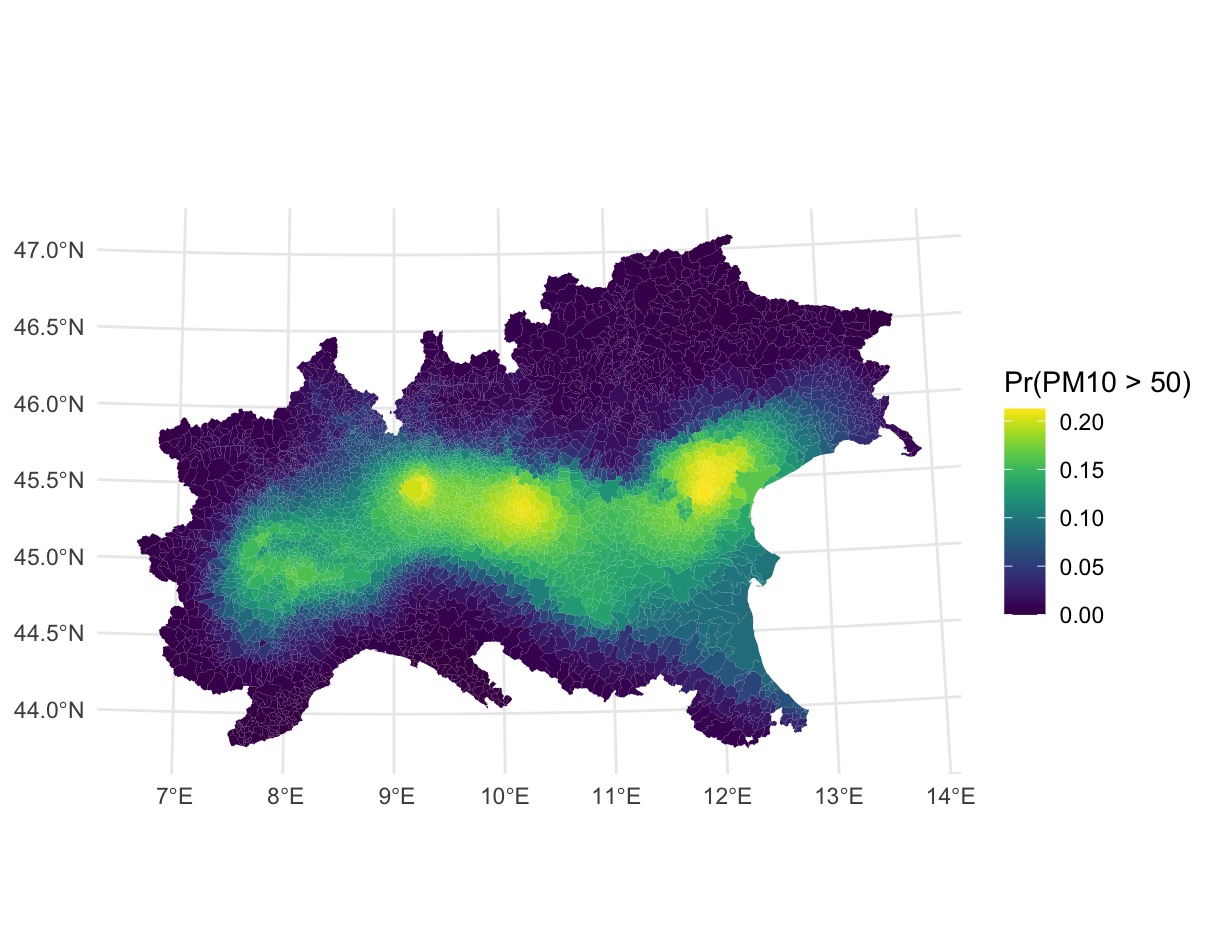}
        \caption*{(a) Exceedance probability \(p_i = P(\text{PM}_{10} > 50)\)}
    \end{minipage}
    \hfill
    \begin{minipage}[t]{0.38\linewidth}
        \centering
        \includegraphics[width=\linewidth]{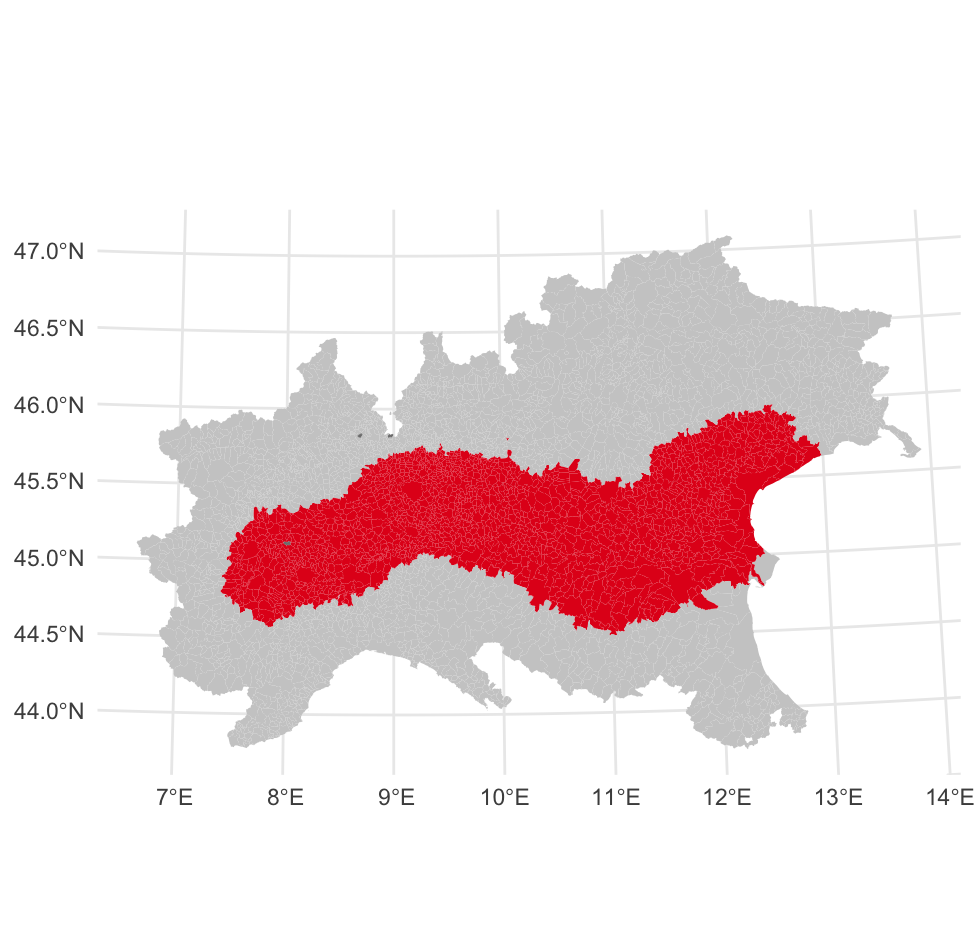}
        \caption*{(b) Municipalities exceeding the regulatory threshold of 35 days/year (red: above the limit)}
    \end{minipage}

    \caption{}
    \label{fig:exceedance_maps}
\end{figure}

Figure~\ref{fig:exceedance_maps} summarizes the frequency of high-pollution episodes. 
Panel (a) displays the exceedance probability map, while panel (b) highlights in red the municipalities where the expected number of exceedance days exceeds the regulatory limit of 35 days per year.
Once again, a clear spatial separation emerges: the Po Valley forms a continuous high-risk zone, while surrounding Alpine and coastal regions remain below the threshold. The transition between these two regimes is remarkably sharp, confirming the strong spatial gradient already suggested by the \(Q_{95}\) map. 

\begin{figure}[h]
    \centering
    \begin{minipage}[t]{0.4\textwidth}
        \centering
        \includegraphics[width=\linewidth]{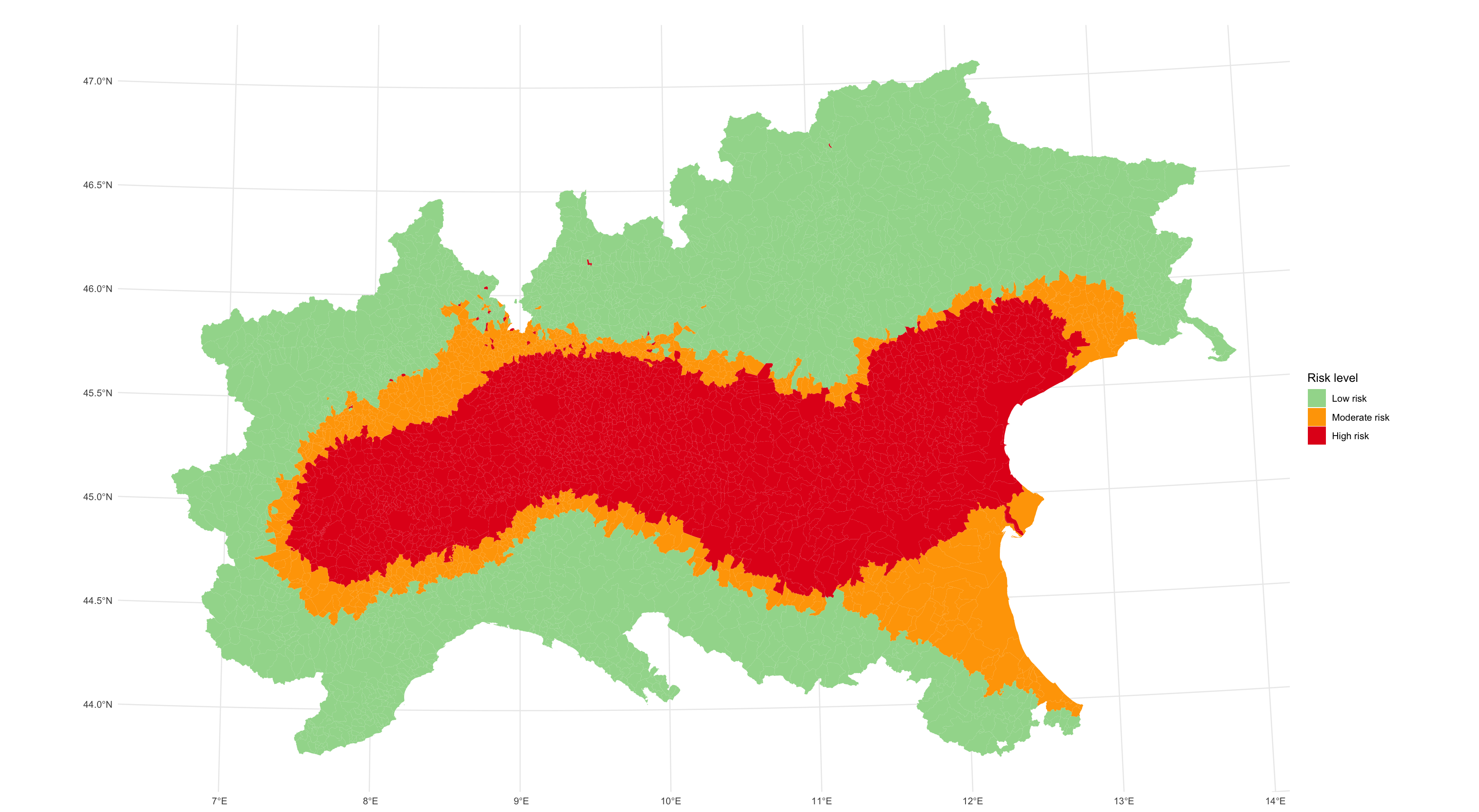}
        \caption*{(a) Spatial classification of municipalities into \emph{low}, \emph{moderate}, and \emph{high} risk zones.}
    \end{minipage}
    \hfill
    \begin{minipage}[t]{0.3\textwidth}
        \centering
        \includegraphics[width=\linewidth]{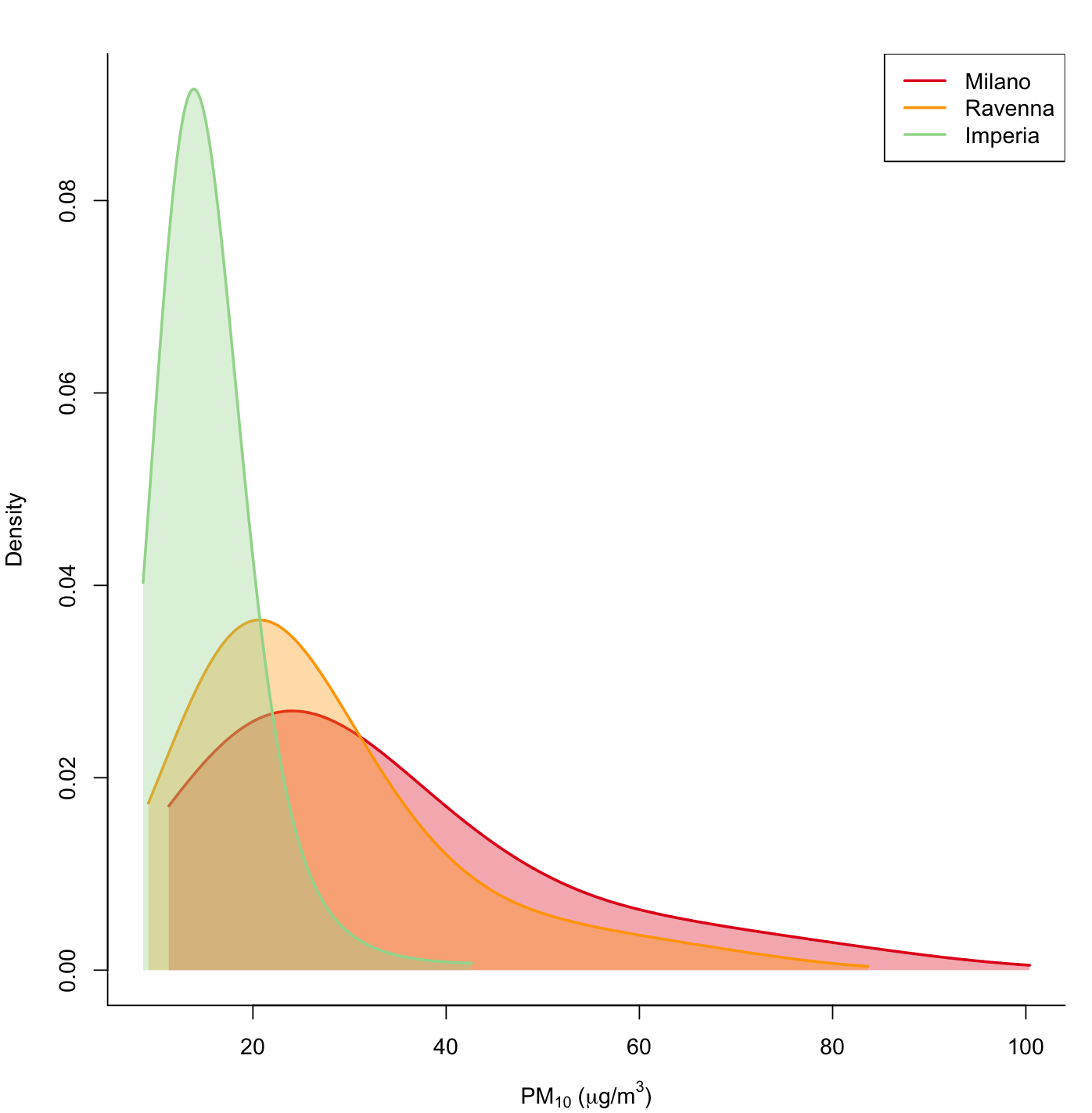}
        \caption*{(b) Reconstructed PM\textsubscript{10} density functions for three municipalities: Milan (high risk), Ravenna (moderate risk), and Imperia (low risk).}
    \end{minipage}

    \caption{}
    \label{fig:risk_zones}
\end{figure}

For interpretability, municipalities are classified into three risk levels according to their estimated exceedance probability. Specifically, we define three categories based on the value of \(p_{i}\): 
\emph{low risk} for \(p_i < 0.05\), \emph{moderate risk} for \(0.05 \le p_i < 0.10\), and \emph{high risk} for \(p_i \ge 0.10\). 
This classification reflects increasing likelihood of surpassing the European Air Quality Directive limit (2008/50/EC) of \(50\,\mu\text{g}/\text{m}^3\).  
Figure~\ref{fig:risk_zones} displays the reconstructed PM\textsubscript{10} density functions for three illustrative municipalities, each corresponding to one of the defined risk levels: 
Imperia (\emph{low risk}), Ravenna (\emph{moderate risk}), and Milan (\emph{high risk}). 
These examples highlight the distinct distributional characteristics associated with different pollution–risk profiles across Northern Italy.

\subsubsection{Comparing results}

To conclude the analysis, we present a comparison between the predictions obtained using coFRK and those produced in \cite{DeSanctis2025_DistributionalPM10}. Among the three approaches considered in that work, we focus on the Spatial Density Estimation (SDE) method, since it also models PM\textsubscript{10} concentrations in terms of their underlying probability distribution.

We compare the two models on the estimated probability of exceeding the regulatory threshold of $50~\mu\text{g}/\text{m}^3$. In this application, we do not expect coFRK to reveal substantially different spatial patterns compared to those obtained from the SDE approach, as this is not the setting under which coFRK would provide substantial additional gains in information. Instead, the aim of this comparison is to verify that the two formulations lead to coherent and consistent estimates. 

Similarity between the exceedance probabilities is evaluated using the Jensen–Shannon distance (\cite{Lin1991JSD}):
\[
JS(p_{\text{coFRK}}, p_{\text{SDE}}) =
\tfrac{1}{2} KL(p_{\text{coFRK}} \,\|\, M)
+ \tfrac{1}{2} KL(p_{\text{SDE}} \,\|\, M),
\qquad M = \tfrac{1}{2}(p_{\text{coFRK}} + p_{\text{SDE}}),
\]
where $KL(\cdot \| \cdot)$ denotes the Kullback--Leibler divergence and $p_{\text{coFRK}}$ and $p_{\text{SDE}}$ respectively denote the exceedance probabilities estimated by coFRK model and by SDE approach. We use the metric form of this quantity
\(
d_{JS}(p_{\text{coFRK}}, p_{\text{SDE}}) = \sqrt{JS(p_{\text{coFRK}}, p_{\text{SDE}})}
\).
In this setting, the Jensen--Shannon distance provides a measure of how closely the two models agree on the exceedance probability. Small values indicate that the estimates are very similar, while larger values highlight municipalities where the two models differ more noticeably. However, the Jensen--Shannon distance does not indicate the direction of the disagreement. To evaluate this aspect, we consider the difference in log-odds between the two exceedance probabilities:
\[
\Delta \text{log-odds} \;=\; 
\log\!\left(\frac{p_{\text{coFRK}}}{1 - p_{\text{coFRK}}}\right)
\;-\;
\log\!\left(\frac{p_{\text{SDE}}}{1 - p_{\text{SDE}}}\right).
\]
The log-odds scale is appropriate here because it enhances differences near $0$ and $1$, thus providing a more informative comparison in this setting where exceedance probabilities are generally low. Positive values of $\Delta \text{log-odds}$ indicate municipalities where coFRK predicts higher exceedance probabilities, while negative values indicate the opposite.

\begin{table}[H]
\centering
\begin{tabular}{lccc}
\toprule
 & Mean $\sqrt{JS}$ & Mean $|\Delta \text{log-odds}|$ & $\%(\Delta \text{log-odds} > 0)$ \\
\midrule
& 0.008069 & 0.0535457  & 71.85 \% \\
\bottomrule
\end{tabular}
\caption{Summary of comparison metrics between the proposed model and the SDE approach.}
\label{tab:comparison}
\end{table}

\begin{figure}[H]
\centering

\begin{minipage}{0.48\textwidth}
\centering
\includegraphics[width=\textwidth]{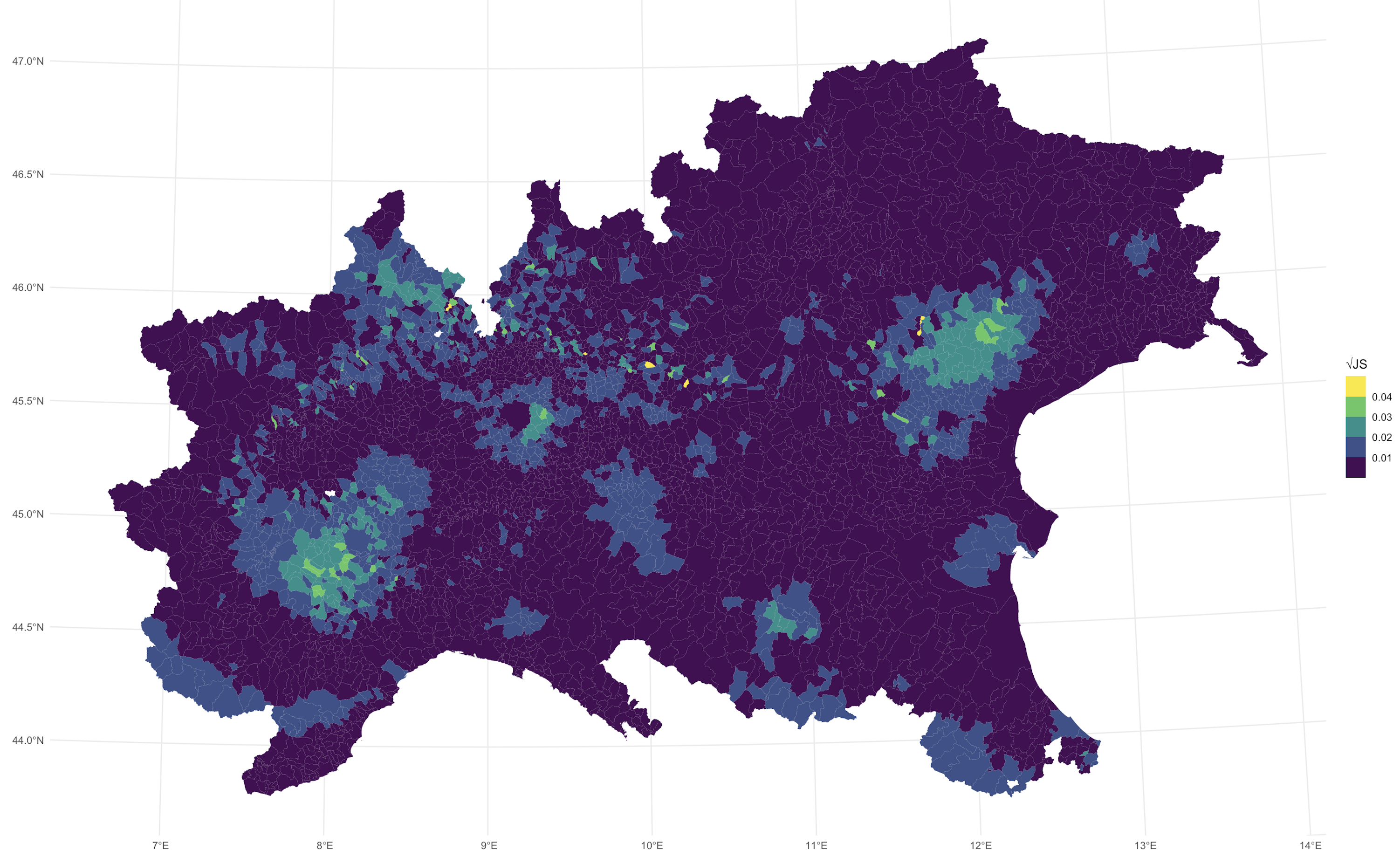}
\caption*{\textbf{(a)} Jensen--Shannon distance $\sqrt{JS}$}
\end{minipage}
\hfill
\begin{minipage}{0.48\textwidth}
\centering
\includegraphics[width=\textwidth]{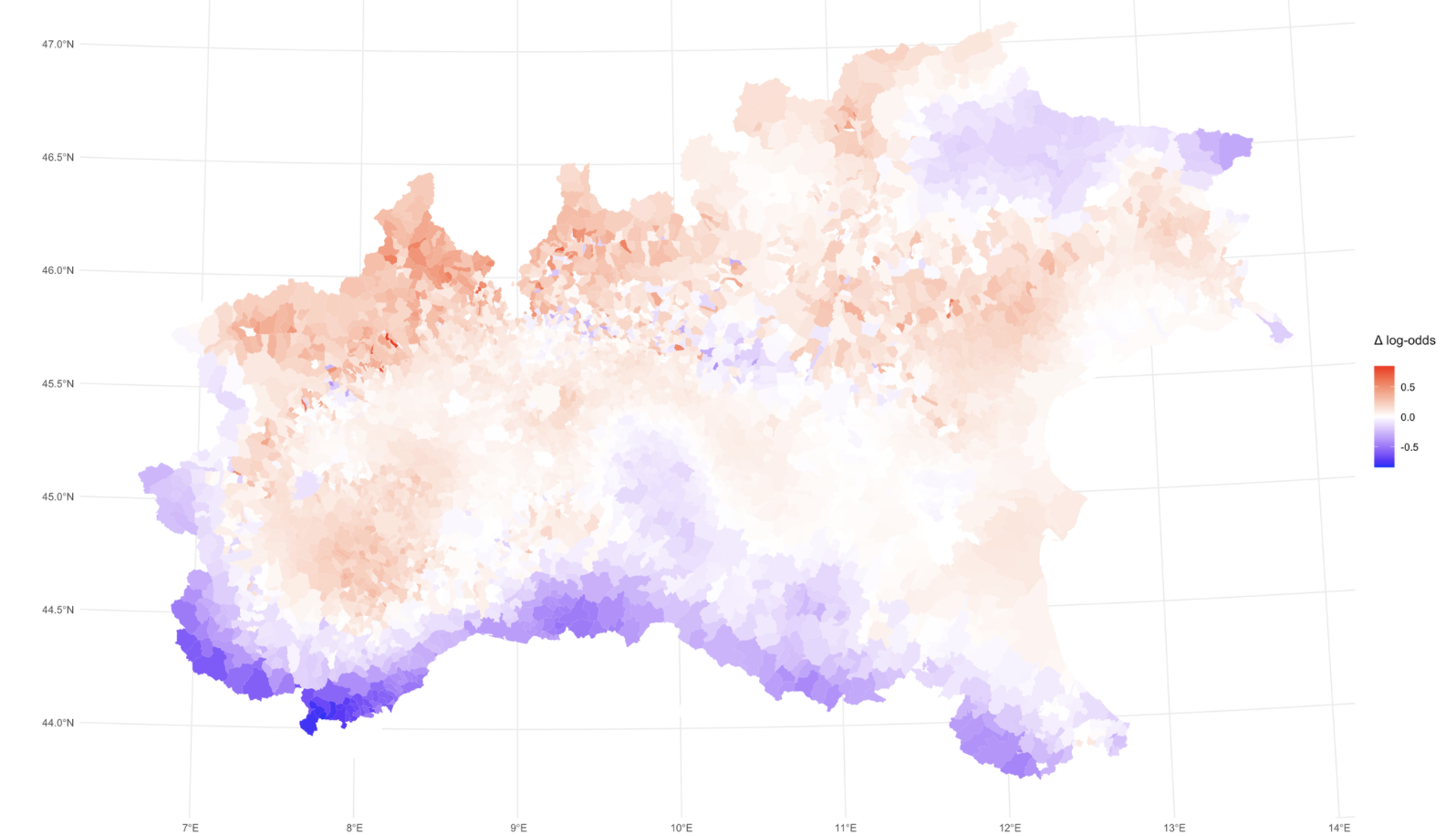}
\caption*{\textbf{(b)} $\Delta$ log-odds}
\end{minipage}

\caption{Comparison of exceedance probability estimates. Panel (a) shows the Jensen–Shannon distance between the two models, and panel (b) shows the log-odds difference indicating which model predicts higher exceedance probabilities.}
\label{fig:js_logodds}

\end{figure}

\begin{figure}[H]
\centering
\includegraphics[width=0.5\textwidth]{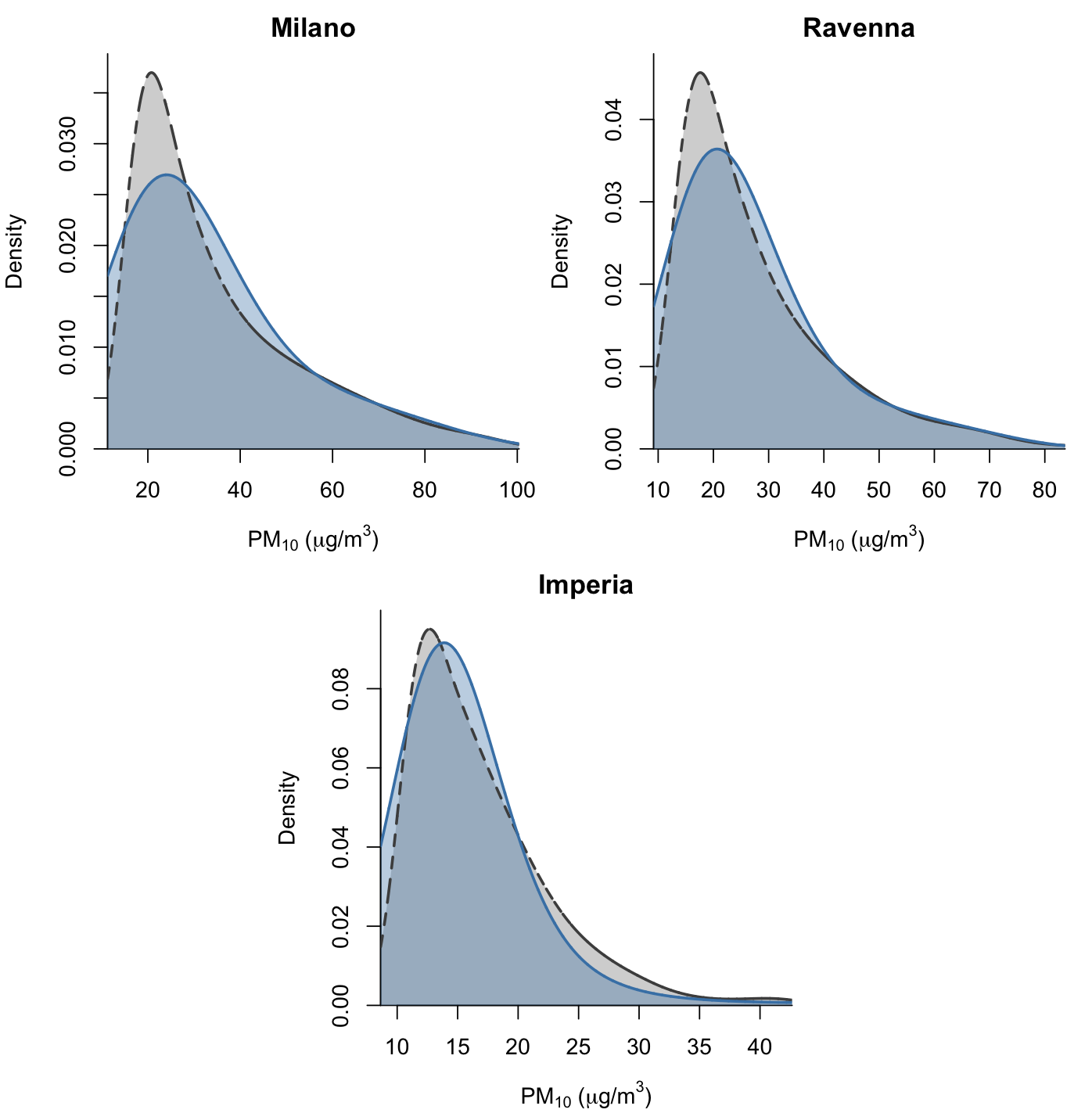}
\caption{Comparison of the predicted PM\textsubscript{10} densities for three municipalities (Milano, Ravenna and Imperia). The blue curve corresponds to the coFRK model, while the grey curve corresponds to the SDE approach.}
\label{fig:density_comparison_4municipalities}
\end{figure}

The results are summarized in Table~\ref{tab:comparison} and visually represented in Figure~\ref{fig:js_logodds}. As anticipated, the two models exhibit general agreement across most municipalities, with small values of $\sqrt{JS}$ and modest differences in log-odds. Nonetheless, some spatial patterns of divergence emerge. The largest values of $\sqrt{JS}$ 
are found in parts of Lombardy (particularly in Milan and in the surrounding area) and in the area south of Turin, as well as in the Veneto region. 
The sign of the difference provides further insight, indicating the areas where coFRK tends to be more pessimistic (i.e., where the difference is positive, highlighted in red) and those where the SDE model yields higher exceedance probabilities (in purple). For completeness, Figure~\ref{fig:density_comparison_4municipalities} reports the predicted PM\textsubscript{10} densities from the two models for the three municipalities previously analyzed, namely Milano, Ravenna and Imperia.

\vspace{1em}

\section{Discussion and Conclusions}

This work develops a multivariate extension of the Fixed Rank Kriging framework. This is achieved by incorporating a multiresolution formulation for cross-variable dependence directly into the FRK spatial random effects. Merging these two perspectives ensures a valid multivariate covariance structure while preserving the scalability and change-of-support properties of FRK. Importantly, the integration goes beyond simply solving the modeling constraint: the multiresolution component adds a structured and theoretically grounded way to distinguish shared from variable-specific spatial patterns across scales, providing a clearer and more informative characterization of cross-variable spatial dependence.

Alongside the modeling contribution, attention was devoted to its practical implementation. The EM algorithm was adapted to explicitly exploit the hierarchical multiresolution representation and the sparsity of the corresponding precision matrices. In addition, particular effort was devoted to the implementation of the estimation procedure. Several functions from the existing \texttt{FRK} R package were extended to handle the multivariate case, allowing the proposed model to be fitted within the same software environment and workflow used for univariate FRK. The corresponding implementation is available in a dedicated GitHub repository at
\url{https://github.com/gaiacaringi/coFRK}.

Beyond addressing the modeling question of how to represent cross-dependence within the FRK framework, this work also considered when the use of a multivariate model is justified, given the additional complexity it introduces. The simulation study in Section~4 speaks to this point: it shows that the model is particularly beneficial when information is unevenly distributed across variables or across space. In such settings, where one variable is observed more densely than another, or where some regions lack data entirely, the multivariate structure allows information to be shared coherently across variables and scales, improving predictions in data-sparse areas. These situations arise frequently in environmental monitoring and remote sensing, underscoring the practical relevance of the proposed approach. \\

Several aspects of the model suggest natural directions for further development. An important one concerns the specification of cross-dependence across variables. Although the parameterization adopted here is fairly general and can accommodate a wide range of dependence behaviors, different applications may call for different forms of cross-covariance structure. A systematic investigation of alternative parameterizations, and of criteria for selecting among them, would help clarify how to best exploit the model’s capacity to transfer information across variables and scales, ensuring that this feature is used effectively when the data structure allows it. 
Moreover, in this work, the number of resolution levels was kept intentionally limited to ensure computational feasibility. With greater computational resources, a finer resolution hierarchy could be employed, which would allow the model to capture more localized spatial variation while maintaining large-scale structure.

Another relevant direction concerns the use of the proposed framework in functional settings.
In the PM\textsubscript{10} application, each distribution is represented through a small set of spline coefficients. While this offers a compact numerical representation, the coefficients themselves do not have a clear spatial interpretation, and the structure of spatial dependence between them is not directly meaningful.
A more suitable approach would be to derive functional representations that are themselves spatially coherent (for example by identifying a small number of typical distributional shapes that recur across the region, and then represent each site as a combination of these shapes). Such representations would allow the multivariate spatial dependence structure to act directly on interpretable features of the distributions, rather than on abstract basis coefficients, strengthening both interpretability and predictive performance.

Finally, an interesting extension of the framework would be to incorporate a temporal dimension. Many environmental applications involve data collected repeatedly over time, where both spatial structure and cross-variable dependence evolve dynamically. Embedding the multiresolution representation within a spatio-temporal formulation would enable joint prediction in space and time. Spatio-temporal extensions of the univariate FRK framework already exist in the literature (\cite{ZammitMangionCressie2017}), suggesting that a multivariate analogue could be developed by embedding the proposed model within a dynamic state-space structure.

\section{Acknowledgements}

The authors acknowledge the project GRINS - Growing
Resilient, INclusive and Sustainable (GRINS PE00000018 – CUP D43C22003110001),
funded by the European Union - NextGenerationEU programme. The views and
opinions expressed are solely those of the authors and do not necessarily reflect those
of the European Union, nor can the European Union be held responsible for them.
The authors also acknowledge the research project Dipartimento di Eccellenza 2023-
2027, Dipartimento di Matematica, Politecnico di Milano, funded by the Italian
Ministry for University and Research.

\printbibliography

@article{CressieJohannesson2008,
  author    = {Noel Cressie and Gardar Johannesson},
  title     = {Fixed Rank Kriging for Very Large Spatial Data Sets},
  journal   = {Journal of the Royal Statistical Society: Series B (Statistical Methodology)},
  volume    = {70},
  number    = {1},
  pages     = {209--226},
  year      = {2008}
}

@article{Lin1991JSD,  author  = {Jianhua Lin},  title   = {Divergence Measures Based on the {Shannon} Entropy},  journal = {IEEE Transactions on Information Theory},  year    = {1991},  volume  = {37},  number  = {1},  pages   = {145--151}}

@article{LimWu2022_cokriging_vs_kriging,
  author = {Lim, Chae and Wu, Wei-Ying},
year = {2022},
month = {07},
pages = {105084},
title = {Conditions on which cokriging does not better than kriging},
volume = {192},
journal = {Journal of Multivariate Analysis}
}

@article{EldeiryGarcia2010_soilsalinity,
  author       = {Eldeiry, Ahmed A. and Garcia, Luis A.},
  title        = {Comparison of Ordinary Kriging, Regression Kriging, and Cokriging Techniques to Estimate Soil Salinity Using {LANDSAT} Images},
  journal      = {Journal of Irrigation and Drainage Engineering},
  year         = {2010},
  volume       = {136},
  number       = {6},
  pages        = {355--364}
}

@article{GentonKleiber2015,
  author    = {Marc G. Genton and William Kleiber},
  title     = {Cross-Covariance Functions for Multivariate Geostatistics},
  journal   = {Statistical Science},
  volume    = {30},
  number    = {2},
  pages     = {147--163},
  year      = {2015},
  doi       = {10.1214/14-STS487}
}

@article{Goovaerts2000,
author = {Goovaerts, Pierre},
year = {2000},
month = {02},
pages = {113-129},
title = {Geostatistical Approaches for Incorporating Elevation Into the Spatial Interpolation of Rainfall},
volume = {228},
journal = {Journal of Hydrology}
}

@article{Helbich,
author = {Helbich, Marco and Kuntz, Michael},
year = {2014},
month = {04},
pages = {},
title = {Geostatistical mapping of real estate prices: an empirical comparison of kriging and cokriging},
volume = {28},
journal = {International Journal of Geographical Information Science}
}

@article{DeSanctis2025_DistributionalPM10,
  title        = {Three Distributional Approaches for {PM\textsubscript{10}} Assessment in Northern Italy},
  author       = {De Sanctis, Marco F. and Gilardi, Andrea and Milan, Giacomo and Sangalli, Laura M. and Ieva, Francesca and Secchi, Piercesare},
   year={2025},
   eprint={2509.13886},
   archivePrefix={arXiv},
   primaryClass={stat.AP},
   note   = {arXiv:2509.13886}

}

@article{Castiglione2023_PDEQuantile,
author = {Castiglione, Cristian and Arnone, Eleonora and Bernardi, Mauro and Farcomeni, Alessio and Sangalli, Laura},
year = {2024},
month = {10},
pages = {105381},
title = {{PDE}-regularised spatial quantile regression},
volume = {205},
journal = {Journal of Multivariate Analysis}
}

@article{Sangalli2021_SR_PDE,
  title        = {Spatial Regression With Partial Differential Equation Regularisation},
  author       = {Sangalli, Laura M.},
  journal      = {International Statistical Review},
  volume       = {89},
  number       = {3},
  pages        = {505--531},
  year         = {2021}
}

@article{VanDenBoogaart2011_BayesLinearSpaces,
  author  = {Van Den Boogaart, K. G. and Egozcue, J. J. and Pawlowsky-Glahn, V.},
  title   = {Bayes linear spaces},
  journal = {SORT - Statistics and Operations Research Transactions},
  volume  = {34},
  number  = {2},
  pages   = {201--222},
  year    = {2011}
}

@article{Wackernagel1994,
  author       = {Wackernagel, Hans},
  title        = {Cokriging versus kriging in regionalized multivariate data analysis},
  journal = {Geoderma},
  volume = {62},
  number = {1},
  pages = {83-92},
  year = {1994}

}

@article{Dowd2023,
  author = {Dowd, Peter and Pardo-Iguzquiza, Eulogio},
year = {2023},
month = {10},
pages = {},
title = {The Many Forms of Co-kriging: A Diversity of Multivariate Spatial Estimators},
volume = {56},
journal = {Mathematical Geosciences}
}

@article{ZhangCai2015,
  author    = {Hao Zhang and Wenxiang Cai},
  title     = {When Doesn’t Cokriging Outperform Kriging?},
  journal   = {Statistical Science},
  year      = {2015},
  volume    = {30},
  number    = {2},
  pages     = {176--180}
}

@book{Cressie1993,
  author    = {Noel Cressie},
  title     = {Statistics for Spatial Data},
  year      = {1993},
  publisher = {Wiley}
}

@book{Wackernagel1998,
  author    = {Hans Wackernagel},
  title     = {Multivariate Geostatistics: An Introduction with Applications},
  edition   = {3rd},
  year      = {2003},
  publisher = {Springer}
}

@book{horn2012matrix,
  title     = {Matrix Analysis},
  author    = {Horn, Roger A. and Johnson, Charles R.},
  year      = {2012},
  edition   = {2nd},
  publisher = {Cambridge University Press},
  address   = {Cambridge, UK}
}

@misc{WHO2021_AQG,
  author       = {{World Health Organization}},
  title        = {Ambient (outdoor) air pollution},
  year         = {2024},
  howpublished = {\url{https://www.who.int/news-room/fact-sheets/detail/ambient-(outdoor)-air-quality-and-health}}
}

@misc{EEA2024website,
  author       = {{European Environment Agency}},
  title        = {European Environment Agency Website},
  howpublished = {\url{https://www.eea.europa.eu/it}},
  year         = {2024}
}

@misc{EEA2024_AirQualityStatus2024,
  author       = {{European Environment Agency}},
  title        = {Particulate matter (PM10) -- Annual limit value for the protection of human health},
  year         = {2024},
  howpublished = {\url{https://www.eea.europa.eu/en/analysis/maps-and-charts/particulate-matter-pm10-annual-limit-value-for-the-protection-of-human-health-3}}
}

@article{ZammitMangionCressie2017,
  author = {Zammit-Mangion, Andrew and Cressie, Noel},
year = {2017},
month = {05},
pages = {},
title = {{FRK}: An R Package for Spatial and Spatio-Temporal Prediction with Large Datasets},
volume = {98},
journal = {Journal of Statistical Software}
}

@article{GoulardVoltz1992,
  author = {Goulard, Michel and Voltz, Marc},
year = {1992},
month = {04},
pages = {269-286},
title = {Linear coregionalization model: Tools for estimation and choice of cross-variogram matrix},
volume = {24},
journal = {Mathematical Geology}
}

@article{CressieRonald,
author = {Cressie, Noel and Barry, Ronald},
year = {2004},
month = {06},
pages = {265-282},
title = {Flexible Spatial Models for Kriging and Cokriging Using Moving Averages and the Fast Fourier Transform (FFT)},
volume = {13},
journal = {Journal of Computational and Graphical Statistics - J COMPUT GRAPH STAT}
}

@article{Gneiting2010,
  author = {Gneiting, Tilmann and Kleiber, William and Schlather, Martin},
  title     = {Matérn Cross-Covariance Functions for Multivariate Random Fields},
  journal   = {Journal of the American Statistical Association},
  volume    = {105},
  number    = {491},
  pages     = {1167--1177},
  year      = {2010},
  publisher = {Taylor \& Francis},
  address   = {Washington, DC}
}

@article{Apanasovich2012,
  author = {Apanasovich, Tatiyana and Genton, Marc and Sun, Ying},
year = {2012},
month = {03},
pages = {180-193},
title = {A Valid Matérn Class of Cross-Covariance Functions for Multivariate Random Fields With Any Number of Components},
volume = {107},
journal = {Journal of The American Statistical Association}
}

@article{Nguyen2012,
  author    = {Hai Nguyen and Noel Cressie and Amy Braverman},
  title     = {Spatial Statistical Data Fusion for Remote Sensing Applications},
  journal   = {Journal of the American Statistical Association},
  year      = {2012},
  volume    = {107},
  number    = {499},
  pages     = {1004--1018}
}

@article{Kleiber2019,
  author    = {William Kleiber and Douglas Nychka and Soutir Bandyopadhyay},
  title     = {A Model for Large Multivariate Spatial Data Sets},
  journal   = {Statistica Sinica},
  year      = {2019},
  volume    = {29},
  number    = {3},
  pages     = {1085--1104},
  publisher = {Institute of Statistical Science, Academia Sinica}
}

@article{Nychka2002,
  author    = {Douglas Nychka and Christopher K. Wikle and J. Andrew Royle},
  title     = {Multiresolution models for nonstationary spatial covariance functions},
  journal   = {Statistical Modelling},
  year      = {2002},
  volume    = {2},
  number    = {4},
  pages     = {315--331}
}

@article{LindgrenLindstromRue2011,
  author    = {Finn Lindgren and Johan Lindström and H{\aa}vard Rue},
  title     = {An explicit link between Gaussian fields and Gaussian Markov random fields: The stochastic partial differential equation approach},
  journal   = {Journal of the Royal Statistical Society: Series B (Statistical Methodology)},
  year      = {2011},
  volume    = {73},
  number    = {4},
  pages     = {423--498}
}

@article{Nychka2015,
  author    = {Douglas Nychka and Soutir Bandyopadhyay and Dorit Hammerling and Finn Lindgren and Stephan Sain},
  title     = {A Multi-resolution Gaussian process model for the analysis of large spatial data sets},
  journal   = {Journal of Computational and Graphical Statistics},
  year      = {2015},
  volume    = {24},
  number    = {2},
  pages     = {579--599}
}

@article{ZammitMangion_Cressie_2024_FRKintro,
  author  = {Andrew Zammit-Mangion and Noel Cressie},
  title   = {Introduction to Fixed Rank Kriging: The R Package},
  journal = {Journal of Statistical Software},
  volume = {98},
  year    = {2024},
  month   = {July}
}

@article{PintadoRomo,
author = {Pintado, Sara and Romo, Juan},
year = {2009},
month = {06},
pages = {},
title = {On the Concept of Depth for Functional Data},
volume = {104},
journal = {Journal of the American Statistical Association}
}

@article{SunGenton,
author = {Sun, Ying and Genton, Marc},
year = {2010},
month = {10},
pages = {},
title = {Functional Boxplot},
volume = {20},
journal = {Journal of Computational and Graphical Statistics}
}

@article{ZammitMangion_2015_multivariateST,
  author = {Zammit-Mangion, Andrew and Rougier, Jonathan and Schoen, Nana and Lindgren, Finn and Bamber, Jonathan},
year = {2015},
month = {05},
pages = {},
title = {Multivariate spatio-temporal modelling for assessing Antarctica's present-day contribution to sea-level rise},
volume = {26},
journal = {Environmetrics}
}

@article{Zhang2007_MLcoregionalization,
  author = {Zhang, Hao},
title = {Maximum-likelihood estimation for multivariate spatial linear coregionalization models},
journal = {Environmetrics},
volume = {18},
number = {2},
pages = {125-139},
year = {2007}
}

@Manual{R-FRK,
  title        = {FRK: Fixed Rank Kriging},
  author       = {Andrew Zammit-Mangion and Matthew Sainsbury-Dale},
  year         = {2024},
  note         = {R package version 2.3.1}
}

@book{RasmussenWilliams2006_GPML,
  author    = {Carl Edward Rasmussen and Christopher K. I. Williams},
  title     = {Gaussian Processes for Machine Learning},
  year      = {2006},
  publisher = {The MIT Press},
  address   = {Cambridge, MA}
}

@book{ChilesDelfiner2012,
  author    = {Jean-Paul Chilès, Pierre Delfiner},
  title     = {Geostatistics: Modeling Spatial Uncertainty},
  edition   = {2nd},
  year      = {2012},
  publisher = {Wiley}
}

\appendix
\section{Appendix A: Univariate simulations results}
\label{AppendixA}

The spatial configuration (the domain, number of spatial locations, basis functions, and BAU construction) is identical to that of the bivariate setting illustrated in \ref{subsec:bivariate_simulation}. 
The simulation parameters are also kept unchanged, namely
$
\sigma_s^2 = 0.7, 
\sigma_\xi^2 = 0.01, 
\sigma_\varepsilon^2 = 10^{-4}, 
\kappa_0^2 = 0.05
$
As in the bivariate setting, we set \(\mathbf{V}_\xi = \mathbf{I}\) and \(\mathbf{V}_\varepsilon = \mathbf{I}\), corresponding to homoscedastic fine-scale variation and measurement error.

\subsection{Parameter recovery and confounding analysis}
\begin{figure}[H]
    \centering
   \includegraphics[width=0.45\textwidth]{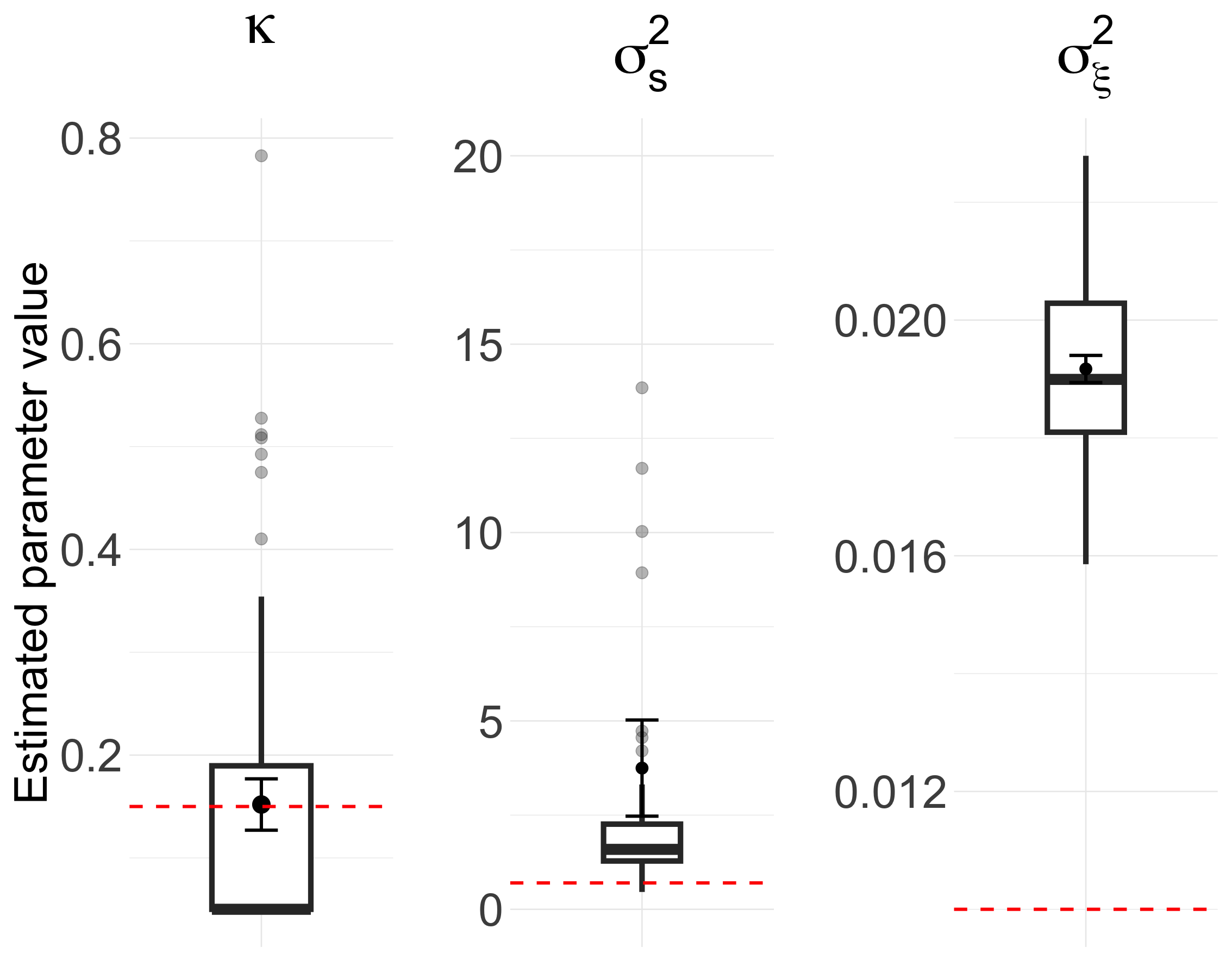}
    \caption{Parameter estimates of $\sigma^2_s$, $\sigma^2_\xi$, and $\kappa_0$ across the 50 Monte Carlo replications.
The boxplots summarize the empirical variability of the estimates across simulations, while the red dashed lines indicate the true parameter values used in the data-generating process.}

    \label{fig:param_box}
\end{figure}

Based on \(50\) Monte Carlo simulations, we summarized the distribution of the estimated parameters using boxplots in Figure~\ref{fig:param_box}.
While the estimates of $\sigma^2_\xi$ and $\kappa_0$ were generally accurate and stable across replicates, the spatial variance $\sigma^2_s$ exhibited a systematic tendency toward overestimation. 

To further investigate this behavior, we examined whether the bias in $\sigma^2_s$ could be attributed to confounding between the two variance components, $\sigma^2_s$ and $\sigma^2_\xi$. 
Specifically, we ran an additional experiment in which $\sigma^2_s$ was kept fixed at its true value during estimation, while $\sigma^2_\xi$ was allowed to vary in the set $\{0.001,\,0.01,\,0.1,\,0.5\}$. 
For each configuration, the model was fitted over $50$ Monte Carlo replications under the same spatial sampling scheme as before.
The results, reported in the first panel of Table~\ref{tab:sig2s_conf}, show that the estimated $\hat{\sigma}^2_s$ varies  systematically with the true value of $\sigma^2_\xi$, confirming the presence of confounding between these two components. 

Given the persistent overestimation of $\sigma^2_s$ and the evidence of confounding with $\sigma^2_\xi$, we introduced a penalization term on $\sigma^2_s$ in the estimation procedure. 
In particular, an $L_2$ (ridge) penalty was added to stabilize the estimation of $\sigma_s^2$. 
The penalty parameter $\lambda$ was selected by repeated 5-fold cross-validation over the grid $\{0.01,\,0.1,\,1,\,10,\,100\}$. 
For each $\lambda$, the model was re-estimated across $R = 20$ repetitions.
Two quantities were then computed for each $\lambda$:
(i) the average magnitude and stability of $\hat{\sigma}_s^2$, defined as
$B_\lambda = \mathrm{mean}(\hat{\sigma}_s^2) + \mathrm{sd}(\hat{\sigma}_s^2)$,
and (ii) the degree of confounding between $\hat{\sigma}_s^2$ and $\hat{\sigma}_\xi^2$, quantified as  
$C_\lambda = |\mathrm{cor}(\hat{\sigma}_s^2,\hat{\sigma}_\xi^2)|$.
The penalty parameter was selected as
\(
\lambda^\star = \arg\min_{\lambda} (B_\lambda + C_\lambda),
\)
which favors values of $\lambda$ that simultaneously reduce the systematic overestimation of $\sigma_s^2$ while mitigating its confounding with the fine-scale variance component.
The resulting estimates are reported in the second panel of Table~\ref{tab:sig2s_conf}.

\begin{table}[H]
\centering
\caption{Estimated spatial variance $\hat{\sigma}^2_s$ for different values of the true fine-scale variance $\sigma^2_\xi$, before (left panel) and after (right panel) applying ridge penalization.}
\label{tab:sig2s_conf}

\begin{minipage}{0.47\linewidth}
\centering
\textbf{(a) Without penalization}\\[4pt]
\begin{tabular}{cccc}
\toprule
$\sigma^2_{\xi,\text{true}}$ & $\mathrm{E}[\hat{\sigma}^2_s]$ & $\mathrm{SD}[\hat{\sigma}^2_s]$  \\
\midrule
0     & 1.99 & 0.59  \\
0.01  & 1.76 & 0.45  \\
0.10  & 1.55 & 0.72  \\
0.50  & 1.10 & 0.77  \\
\bottomrule
\end{tabular}
\end{minipage}
\hfill
\begin{minipage}{0.47\linewidth}
\centering
\textbf{(b) With ridge penalization}\\[4pt]
\begin{tabular}{cccc}
\toprule
$\sigma^2_{\xi,\text{true}}$ & $\mathrm{E}[\hat{\sigma}^2_s]$ & $\mathrm{SD}[\hat{\sigma}^2_s]$  \\
\midrule
0     & 0.888 & 0.0643  \\
0.01  & 0.807 & 0.088   \\
0.10  & 0.604 & 0.091   \\
0.50  & 0.396 & 0.0450  \\
\bottomrule
\end{tabular}
\end{minipage}

\end{table}

\subsection{Predictive performance}
To compare the predictive accuracy of the proposed model in the univariate setting and with the standard FRK formulation we report three commonly used predictive metrics, namely the Root Mean Squared Error (RMSE), the Mean Absolute Error (MAE), and the coefficient of determination ($R^2$), averaged over 50 Monte Carlo replications. The results, summarized in Table~\ref{tab:metrics_univariate}, indicate that coFRK achieves predictive accuracy comparable to standard FRK, with slightly higher error values on average. This behavior is expected, as the additional multiresolution structure introduced for multivariate dependence does not provide an advantage in the univariate setting.

\begin{figure}[H]
    \centering
    \includegraphics[width=0.9\columnwidth]{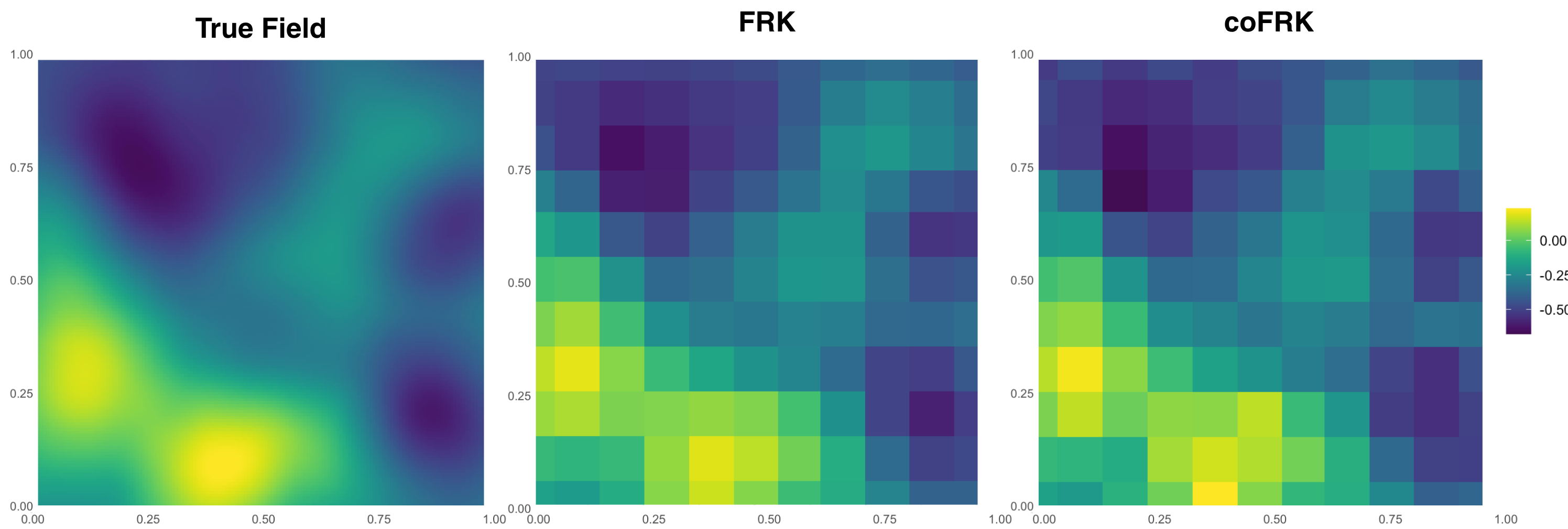}
    \caption{Comparison between the true field \( Z_1(s) \) and the predicted fields from FRK and coFRK. 
Predictions are shown at the BAU level.}
    \label{fig:true-frk-cofrk_UNI}
\end{figure}

Finally, in Figure~\ref{fig:true-frk-cofrk_UNI} we present a visual comparison between the true simulated field and the corresponding predicted field from a representative Monte Carlo replication.

\begin{table}[H]
\centering
\caption{Predictive performance metrics (mean $\pm$ standard deviation) across 50 Monte Carlo iterations for the univariate simulation.}
\label{tab:metrics_univariate}
\begin{tabular}{lccc}
\toprule
\textbf{Metric} & \textbf{coFRK} & \textbf{FRK} & \textbf{Difference} \\
\midrule
RMSE & $0.128 \pm 0.0011$ & $0.1226 \pm 0.0012$ & $+0.0054$ \\
MAE  & $0.121 \pm 0.006$ & $0.1025 \pm 0.0009$ & $+0.0185$ \\
$R^2$ & $0.78 \pm 0.0117$ & $0.80 \pm 0.01$ & $-0.02$ \\
\bottomrule
\end{tabular}
\end{table}

\subsection{Covariance matrix: a comparison with standard FRK model}

As remarked in \ref{subsec:coFRKFinalFormulation}, the main difference between the two models lies in the parametrization of the latent coefficients covariance matrix $\mathbf{K}$. 

\begin{figure}[H]
    \centering
    \includegraphics[width=0.7\columnwidth]{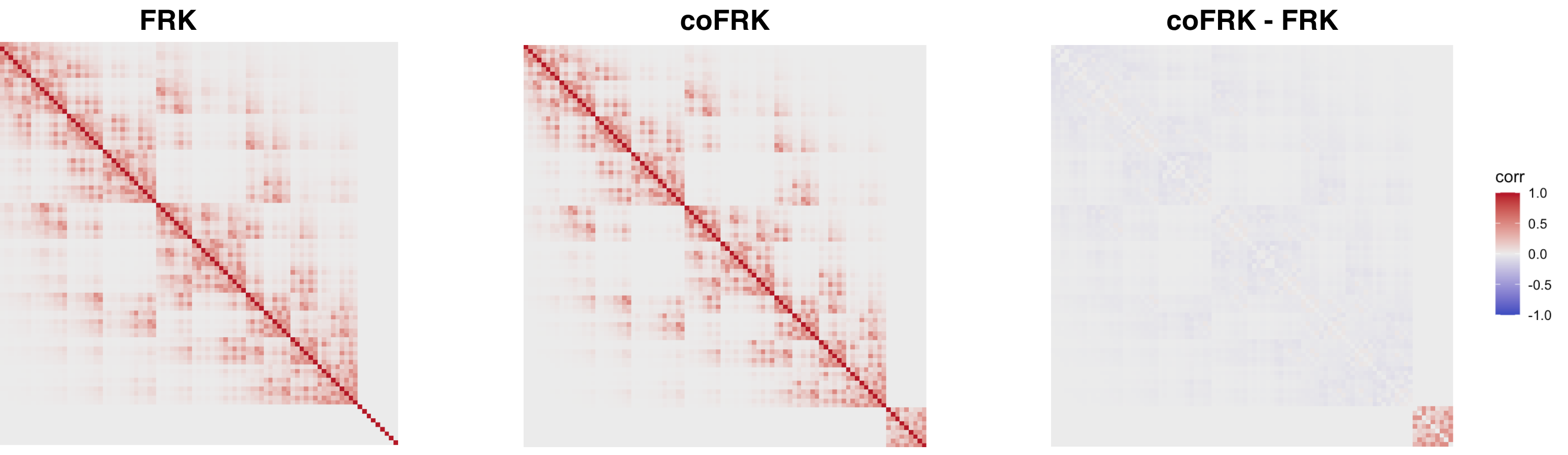}
    \caption{Correlation matrices of the latent coefficients for FRK (left), coFRK (center), and their difference (right).}
    \label{fig:heatmap}
\end{figure}

A qualitative comparison is performed by visualizing the corresponding correlation matrices. 
As shown in Figure~\ref{fig:heatmap}, both matrices exhibit two main blocks, corresponding to the two resolution levels. 
For the finer resolution level (the second), the correlation structures are very similar. 
A noticeable difference appears in the block associated with the coarser resolution level (the first). 
This difference is due to the parameterization typically adopted in practical implementations of FRK for the matrix \(\mathbf{K}\), 
where correlations between coefficients depend on the spatial distance between the centers of the corresponding basis functions 
(see the discussion on the structured \(\mathbf{K}\) matrix in \cite{R-FRK}). 
In contrast, our model enforces correlations among coefficients that belong to the same resolution level, 
even when the associated basis functions are spatially distant.

\section{Appendix B: Derivation of the EM Algorithm Updates}
\label{AppendixB}
In Section~\ref{sec:estimation}, we presented the EM algorithm used to estimate the model
parameters and reported only the resulting update expressions. In this appendix, we provide
the intermediate steps leading to those updates. In particular, we focus on the
maximization step (M--step) for the parameters governing the multiresolution covariance
structure of the latent coefficients, i.e., the parameters entering the precision matrix $\mathbf{Q}$. These updates differ from the
standard univariate FRK formulation of \cite{ZammitMangion_Cressie_2024_FRKintro}, because
here we exploit the specific parametrization imposed on $\mathbf{Q}$. Thanks to the level-wise organization and induced sparsity of this parametrization, the M--step reduces to operations involving significantly smaller matrices, yielding lower computational cost.

For the covariance parameters that enter the precision matrix $\mathbf{Q}$, 
we only need to consider the part of the EM objective that depends on $\mathbf{Q}$:
\[
\mathcal{Q}(\boldsymbol{\theta} \mid \boldsymbol{\theta}^{(t)}) 
= \tfrac{1}{2}\log|\mathbf{Q}|
- \tfrac12\operatorname{tr}\!\Big(
\mathbf{Q}\,
(\boldsymbol{\Sigma}_c^{(t)}+\boldsymbol{\mu}_c^{(t)}{\boldsymbol{\mu}_c^{(t)}}^{\top})
\Big) + \text{const.}
\]

The following matrix identities will be used repeatedly (see, e.g. \cite{horn2012matrix}):
\begin{align*}
\textit{(Kronecker determinant)} 
&\quad |A\otimes B| = |A|^{\dim(B)}\,|B|^{\dim(A)}, \\[4pt]
\textit{(Determinant multiplicativity)} 
&\quad |AB| = |A|\,|B|, \\[4pt]
\textit{(Cyclic property of the trace)} 
&\quad \operatorname{tr}(XYZ)=\operatorname{tr}(ZXY), \\[4pt]
\textit{(Block trace–Kronecker contraction)} 
&\quad \text{If } X=[X_{ij}]_{i,j=1}^p \text{ with } X_{ij}\in\mathbb{R}^{r\times r}, \text{ then}
\quad
\operatorname{tr}\!\big((A\otimes K)\,X\big)
=
\sum_{i,j=1}^p A_{ij}\,\operatorname{tr}(K\,X_{ij}).
\end{align*}

In all of the following derivations we will exploit two key facts:

A) Level-wise formulation of $\mathbf{Q}$ (via permutation matrix, see Section~\ref{sec:estimation}):
\[
\mathbf{Q}
= \mathbf{P}^{\top}
\Big( \bigoplus_{\ell=1}^{L} (\boldsymbol{\Sigma}_{\ell}^{-1} \otimes \mathbf{B}_{\ell}\mathbf{B}_{\ell}^{\top}) \Big)
\mathbf{P},
\]
which implies:
\[
\text{(i)}\quad \log|\mathbf{Q}|
= \sum_{\ell=1}^{L}\log|\boldsymbol{\Sigma}_{\ell}^{-1}\otimes \mathbf{B}_{\ell}\mathbf{B}_{\ell}^{\top}|,
\qquad
\text{(ii)}\quad 
\operatorname{tr}\!\big(\mathbf{Q}(\boldsymbol{\Sigma}_c^{(t)}+\boldsymbol{\mu}_c^{(t)}{\boldsymbol{\mu}_c^{(t)}}^{\top})\big)
=
\sum_{\ell=1}^{L}\operatorname{tr}\!\big(\mathbf{Q}_{\ell}\,\mathbf{S}_{c_ \ell}^{(t)}\big),
\]
where $\mathbf{Q}_\ell=\boldsymbol{\Sigma}_\ell^{-1}\otimes \mathbf{B}_{\ell}\mathbf{B}_{\ell}^{\top}$ and 
$\mathbf{S}_{c_\ell}^{(t)}$ denotes the level-$\ell$ block of
$\mathbf{P}(\boldsymbol{\Sigma}_c^{(t)}+\boldsymbol{\mu}_c^{(t)}{\boldsymbol{\mu}_c^{(t)}}^{\top})\mathbf{P}^{\top}$.

B) Decomposition of the cross–process covariance at each level:
\[
\boldsymbol{\Sigma}_\ell = \mathbf{D}_\ell\,\mathbf{R}_\ell\,\mathbf{D}_\ell,
\]
where \(
\mathbf{D}_\ell=\operatorname{diag}\!\big(\sqrt{\alpha_{\ell 1}\sigma_{s_1}^2},\ldots,\sqrt{\alpha_{\ell p}\sigma_{s_p}^2}\big),
\)
and $\mathbf{R}_\ell$ the $p\times p$ equicorrelation matrix at level $\ell$. The inverse equicorrelation matrix is denoted as $\mathbf{C}_\ell$. \\

Before presenting the parameter–specific updates, we make one clarification regarding the
trace term
\[
\operatorname{tr}\!\big(\mathbf{Q}\,(\boldsymbol{\Sigma}_c^{(t)}+\boldsymbol{\mu}_c^{(t)}{\boldsymbol{\mu}_c^{(t)}}^{\top})\big),
\]
which appears in every maximization step.
Throughout the derivations, our strategy is to rewrite this quantity so that only the
portion of the precision matrix that actually depends on the parameter being updated
remains explicit.
Although this may at first look like a notational complication, it is in fact a computational
trick: by isolating the parameter–dependent component of $\mathbf{Q}$, we avoid working
with the full precision matrix at each EM iteration and instead manipulate only small,
structured matrices.

\subsubsection*{Update for $\sigma_s^2$}
For the update of $\sigma_s^2$, the contribution of level $\ell$ to $\mathcal{Q}(\sigma_s^2)$ is given by its determinant and trace components. 
Using the Kronecker determinant identity,
\[
\log|\mathbf{Q}|
=
\sum_{\ell=1}^{L} \log|\mathbf{Q}_\ell|
=
\sum_{\ell=1}^{L} \Big(R_\ell \log|\boldsymbol{\Sigma}_\ell^{-1}| + p\,\log|\mathbf{B}_\ell\mathbf{B}_\ell^{\top}|\Big)
=
-\,2\sum_{\ell=1}^{L} R_\ell \sum_{i=1}^{p} \log d_{\ell_i} + \text{const.},
\]
where $d_{\ell_i}=\sqrt{\alpha_{\ell i}\sigma_{s_i}^2}$ and all terms independent of $\sigma_s^2$ have been
absorbed into the constant. \\

Next, partitioning $\mathbf{S}_{c_\ell}$ into $p\times p$ sub-blocks $\mathbf{S}_{c_\ell}^{(i,j)}\in\mathbb{R}^{R_\ell\times R_\ell}$, we define
\(
(\mathbf{G}_\ell)_{ij} = \operatorname{tr}\!\big((\mathbf{B}_\ell\mathbf{B}_\ell^{\top})\,\mathbf{S}_{c_\ell}^{(i,j)}\big)\), \(
\mathbf{G}_\ell\in\mathbb{R}^{p\times p}.
\)
Thus, the trace term can be expressed as 
\[
\operatorname{tr}\!\big(\mathbf{Q}_{\ell}\mathbf{S}^{(\ell)}\big)
=
\operatorname{tr}\!\big(\mathbf{C}_\ell\,\mathbf{D}_\ell^{-1}\,\mathbf{G}_\ell\,\mathbf{D}_\ell^{-1}\big).
\]
Since $\mathbf{C}_\ell$ and $\mathbf{G}_\ell$ do not depend on $\sigma_s^2$, all dependence enters
through $\mathbf{D}_\ell$, and thus the update for $\sigma_s^2$ reduces to optimizing the
diagonal scaling in $\mathbf{D}_\ell$. Collecting terms over levels gives
\[
\mathcal{Q}(\sigma_s^2)
=
\sum_{\ell=1}^L\left(
-2\,R_\ell\sum_{i=1}^p\log d_{\ell_i}
-
\operatorname{tr}\!\big(\mathbf{C}_\ell\,\mathbf{D}_\ell^{-1}\,\mathbf{G}_\ell\,\mathbf{D}_\ell^{-1}\big)
\right).
\]
which is maximized numerically. 

\subsubsection*{Update for $\kappa_0$}

Analogously, the contribution of level $\ell$ to $\mathcal{Q}(\kappa_0)$ is obtained from the determinant
and trace components. Using the Kronecker determinant identity and noting that
$\boldsymbol{\Sigma}_\ell$ does not depend on $\kappa_0$,
\[
\log|\mathbf{Q}|
=
\sum_{\ell=1}^{L} \log|\mathbf{Q}_\ell(\kappa_0)|
=
2p \sum_{\ell=1}^{L} \log|\mathbf{B}_\ell(\kappa_0)|+\text{const.},
\]
up to an additive constant independent of $\kappa_0$.

Next, define
\(
\big(\mathbf{T}_\ell(\kappa_0)\big)_{ij}
=
\operatorname{tr}\!\big(\mathbf{B}_\ell(\kappa_0)\mathbf{B}_\ell(\kappa_0)^{\top}\,\mathbf{S}_{c_\ell}^{(i,j)}\big)
\).
Then the trace term becomes
\[
\operatorname{tr}\!\big(\mathbf{Q}_\ell\mathbf{S}^{(\ell)}\big)
=
\operatorname{tr}\!\big(\boldsymbol{\Sigma}_\ell^{-1}\,\mathbf{T}_\ell\big),
\]
with $\boldsymbol{\Sigma}_\ell$ fixed in this step.

Collecting terms over levels gives
\[
\mathcal{Q}(\kappa_0)
=
\sum_{\ell=1}^L\left(
2p\,\log|\mathbf{B}_\ell|
-
\operatorname{tr}\!\big(\boldsymbol{\Sigma}_\ell^{-1}\,\mathbf{T}_\ell\big)
\right),
\]
which is maximized numerically.

\subsubsection*{Update for $r_0$ and $r_1$}

The parameters $r_0$ and $r_1$ determine cross–process dependence across levels through
\(
\rho_\ell = r_0\,\exp(-r_1(\ell-1)) \).
In this step, $\mathbf{D}_\ell$ and $\mathbf{B}_\ell$ are fixed, and only $\mathbf{C}_\ell$
depends on $(r_0,r_1)$.

Using the formula for the determinant of an equi-correlation matrix,  $|\mathbf{R}(\rho)| = (1-\rho)^{p-1}(1+(p-1)\rho)\,$,
\[
\log|\mathbf{Q}_\ell|
=
R_\ell \log|\mathbf{C}_\ell| + \text{const.}
=
-\,R_\ell\big[(p-1)\log(1-\rho_\ell)+\log(1+(p-1)\rho_\ell)\big] + \text{const.},
\]
up to a constant independent of $(r_0,r_1)$.

As in the previous case, let
\(
(\mathbf{T}_\ell)_{ij} = \operatorname{tr}\!\big(\mathbf{B}_\ell\mathbf{B}_\ell^{\top}\,\mathbf{S}_{c_\ell}^{(i,j)}\big).
\)
Moreover, since \(\boldsymbol{\Sigma}_\ell^{-1} = \mathbf{D}_\ell^{-1}\mathbf{C}_\ell\mathbf{D}_\ell^{-1}\),
it is convenient to introduce
\(
\mathbf{M}_\ell := \mathbf{D}_\ell^{-1} \mathbf{T}_\ell \mathbf{D}_\ell^{-1}.
\)
Then, using the cyclic property of the trace, the trace term becomes
\[
\operatorname{tr}\!\big(\mathbf{Q}_\ell\mathbf{S}^{(\ell)}\big)
=
\operatorname{tr}(\mathbf{C}_\ell \mathbf{M}_\ell).
\]
where only \(\mathbf{C}_\ell\) depends on \((r_0,r_1)\), while 
\(\mathbf{M}_\ell\) is fixed in this step.

Collecting the level contributions yields
\[
\mathcal{Q}(r_0,r_1)
=
\sum_{\ell=1}^L\left(
-\,R_\ell\big[(p-1)\log(1-\rho_\ell)+\log(1+(p-1)\rho_\ell)\big]
-
\operatorname{tr}(\mathbf{C}_\ell \mathbf{M}_\ell)\right)
\]

\section{Appendix C: Code Implementation}
\label{AppendixC}
The implementation of the proposed coFRK model was developed entirely in~R (version 4.4.2). 
The aim was to preserve the overall modeling workflow and user interface introduced in
the univariate \texttt{FRK} package (\cite{R-FRK}), while extending it to support a multiresolution
GMRF representation of the latent spatial effects and to allow for cross-process dependence.
While the workflow design mirrors that of the univariate setting, the code used here is
original.

A single model object is first constructed, collecting the data, covariates, BAUs, and basis
functions, and assigning initial values to all parameters, including those governing the
spatial structure and cross-process dependence. Model fitting is then carried out through the
EM algorithm, with parameter updates written directly back to the same object. Predictions
may be obtained either at the BAU resolution or on any user-defined spatial support by
aggregating BAU-level estimates using the appropriate mapping matrix.
All functions developed for this work are available in a dedicated GitHub repository at
\url{https://github.com/gaiacaringi/coFRK}.

Table~\ref{tab:functions_overview} summarizes the main functions implemented and their
respective roles within the modeling workflow.

\begin{table}[H]
\centering
\renewcommand{\arraystretch}{1.2}
\begin{tabularx}{\textwidth}{>{\raggedright\arraybackslash}p{3.3cm} X}
\toprule
\textbf{Function} & \textbf{Description} \\
\midrule
\multicolumn{2}{l}{\textbf{(A) User-facing high-level functions}} \\

\texttt{SRE\_mv} &
Constructs the multivariate spatial random effects model object. Gathers data and covariates
for all processes, attaches BAUs and multiresolution basis matrices, and initializes all model
parameters ($\boldsymbol{\sigma}_s^2$, $\boldsymbol{\sigma}_\xi^2$, $\nu$, $r_0$, $r_1$, $\kappa_0$, $\boldsymbol{\beta}$). This object
stores all quantities used during estimation and prediction.
\\[0.5em]

\texttt{build\_BAUs\_basis} &
Constructs the BAU grid and multiresolution basis system. The basis functions are evaluated at
BAU centroids and grouped by resolution level for use in \texttt{SRE\_mv}.
\\[0.5em]

\texttt{SRE\_mv.fit} &
Main fitting routine implementing the EM algorithm. The estimation proceeds through an
iterative loop composed of three internal functions:
\texttt{E\_step\_mv} (updates the conditional mean and covariance of $\mathbf{c}$),
\texttt{M\_step\_mv} (updates the model parameters),
and \texttt{logLik\_mv} (evaluates the log-likelihood for convergence monitoring). Parameter values and posterior quantities are written directly back
to the \texttt{SRE\_mv} object.
\\[0.5em]

\texttt{SRE\_mv.predict} &
Produces posterior means and variances at the BAU level, and aggregates them to user-specified
prediction supports using the aggregation matrix $\mathbf{C}_P$.
\\[0.8em]

\multicolumn{2}{l}{\textbf{(B) Internal functions: GMRF precision construction}} \\

\texttt{build\_B} & Constructs the local adjacency matrices $\mathbf{B}_\ell$ for each resolution level. \\
\texttt{build\_Qlist} & Builds the level-specific precision blocks $\mathbf{Q}_\ell = \boldsymbol{\Sigma}_\ell^{-1} \otimes \mathbf{B}_\ell \mathbf{B}_\ell^\top$. \\
\texttt{build\_Sigma\_list} & Constructs $\boldsymbol{\Sigma}_\ell$ encoding cross-process dependence at each level. \\
\texttt{build\_Q\_total} & Assembles the full sparse multiresolution precision matrix $\mathbf{Q}$. \\[0.8em]

\multicolumn{2}{l}{\textbf{(C) Internal functions: M-step parameter updates}} \\

\texttt{update\_beta} & Updates regression coefficients $\boldsymbol{\beta}$ via generalized least squares. \\
\texttt{update\_sigma2\_s} & Updates spatial scale variances $\sigma_{s,i}^2$. \\
\texttt{update\_kappa} & Updates $\kappa_0$ controlling decay across resolutions. \\
\texttt{update\_r0\_r1} & Updates cross-process correlation parameters $(r_0, r_1)$. \\
\texttt{update\_sigma2\_xi} & Updates process-specific fine-scale variances $\sigma_{\xi,j}^2$. \\[0.8em]

\multicolumn{2}{l}{\textbf{(D) Internal functions: Prediction aggregation}} \\

\texttt{map\_data\_to\_BAUs} & Assigns observations to BAUs. \\
\texttt{buildC} & Identifies BAU-to-region membership. \\
\texttt{make\_CP} & Constructs and normalizes the sparse aggregation matrix $\mathbf{C}_P$. \\
\bottomrule
\end{tabularx}
\caption{Summary of all functions implemented for the multivariate FRK model. 
Only the high-level functions in Section (A) are intended to be called by the user; 
sections (B)--(D) are internal routines automatically executed within \texttt{SRE\_mv.fit()} and \texttt{SRE\_mv.predict()} }.
\label{tab:functions_overview}
\end{table}

\end{document}